\documentclass{aa}

 \pdfoutput=1 
 
\usepackage{txfonts}
\usepackage{natbib,twoopt}  
\usepackage{graphicx}

\bibpunct{(}{)}{;}{a}{}{,} 

\newcommand{\msun}{\,\hbox{$M_{\odot}$}}
\newcommand{\lsun}{\,\hbox{$L_{\odot}$}}
\newcommand{\kms}{\,\hbox{\hbox{km}\,\hbox{s}$^{-1}$}}

\newcommand{\lir}{\,\hbox{$L_{\rm IR}$}}
\newcommand{\htwo}{\,\hbox{$\rm{H_ 2}$}}
\newcommand{\ha}{\,\hbox{$H_{\rm \alpha}$}}
\newcommand{\hbeta}{\,\hbox{$H_{\rm \beta}$}}
\newcommand{\ang}{\,\hbox{\AA}}

\newcommand{\oiii}{\,\hbox{[\ion{O}{III}}]}
\newcommand{\oii}{\,\hbox{[\ion{O}{II}}]}

\newcommand{\um}{\,\hbox{$\mu$m}}

\newcommand{\cooz}{\hbox{$\rm CO(1$-$0)$}}

\newcommand{\cott}{\hbox{$\rm CO(3$-$2)$}}

\newcommand{\coft}{\hbox{$\rm CO(4$-$3)$}}

\newcommand{\hcop}{\hbox{$\rm HCO^+(5$-$4)$}}
\newcommand{\water}{\hbox{$\rm H_ 2O$}}

\newcommand{\hst}{\hbox{\it Hubble Space Telescope}}
\newcommand{\spi}{{\it Spitzer}}

\begin{document}

\textheight=24.8cm
\addtolength{\topmargin}{-.2cm}
\setlength{\parskip}{0.5mm plus 0.0mm minus0.0mm}
        
\title{Complex molecular gas kinematics in the inner 5\,kpc of 4C12.50 \\ as seen by ALMA}

\subtitle{}     
      
\author{C. M. Fotopoulou\inst{1,2},
                K. M. Dasyra\inst{1,3}, 
                F.  Combes\inst{4},
                P. Salom\'e\inst{4}
                \and
                M. Papachristou\inst{1,3} 
        }
        
\institute{
                Department of Astrophysics, Astronomy \& Mechanics, Faculty of Physics, National and Kapodistrian University of Athens, Panepistimiopolis Zografos 15784, Greece
                \and
                Max-Planck Institut f\"ur Astrophysik, Karl-Schwarzschild-Str. 1, D-85741 Garching, Germany
                \and
                Institute for Astronomy, Astrophysics, Space Applications, and Remote Sensing, National Observatory of Athens, 15236 Penteli, Greece
                \and    
                Observatoire de Paris, LERMA, 61 Av.\,de l'Observatoire, 75014, Paris, France
        }
        
\date{}

\abstract {
	The nearby system 4C12.50, also known as IRAS\,13451+1217 and PKS\,1345+12, is a merger of gas-rich galaxies with infrared and radio activity. It has a perturbed interstellar medium (ISM) and a dense configuration of gas and dust around the nucleus. The radio emission at small ($\sim$100 pc) and large ($\sim$100 kpc) scales, as well as the large X-ray cavity in which the system is embedded, are indicative of a jet that could have affected the ISM. We carried out observations of the CO(1-0), (3-2), and (4-3) lines with the Atacama Large Millimeter Array (ALMA) to determine basic properties (i.e., extent, mass, and excitation) of the cold molecular gas in this system, including its already-known wind. The CO emission reveals the presence of gaseous streams related to the merger, which result in a small ($\sim$4kpc-wide) disk around the western nucleus. The disk reaches a rotational velocity of 200 \kms , and has a mass of  3.8($\pm$0.4)$\times$10${^9}$\msun . It is truncated at a gaseous ridge north of the nucleus that is bright in \oiii . Regions with high-velocity CO emission are seen at signal-to-noise ratios of between 3 and 5 along filaments that radially extend from the nucleus to the ridge and that are bright in \oiii\ and stellar emission.  
	A tentative wind detection is also reported in the nucleus and in the disk. The molecular gas speed could be as high as 2200\kms and the total wind mass could be as high as 1.5($\pm$0.1)$\times$10$^9$\msun . Energetically, it is possible that the jet, assisted by the radiation pressure of the active nucleus or the stars, accelerated clouds inside an expanding bubble. 
	}
        
        
\titlerunning{Complex molecular gas kinematics in 4C12.50}
\authorrunning{C. M. Fotopoulou et al.}
        
\maketitle
            
\section{Introduction}
\label{sec:intro}
Galactic winds that are driven by the feedback of active galactic nuclei (AGNs) or young stars and that are detected in molecular gas tracers are now considered common \citep{sakamoto06,sakamoto09,leon07,feruglio10,fischer10, fluetsch19,   alatalo11,rangwala11,sturm11,krips11,dasyra_combes11,dasyra_combes12,aalto12,tsai12,morganti13a,combes13,spoon13,veilleux13,cicone14,garciaburillo14,george14,sakamoto14,alatalo15,tombesi15,aalto16,stone16,gonzalez-alfonso17,pereira18}. 
In galaxies with active nuclei, the momentum rate of the molecular winds is often considerably higher, that is, about 20 times higher, than the pressure exerted by the AGN radiation \citep[e.g.,][]{cicone14,carniani15}. Multiple photon scatterings  \citep{ishibashi_fabian15} and, more frequently, an energy-conserving expansion of an ionized gas bubble that leads to momentum boosting \citep{king11,faucher12,zubovas14} have been evoked to justify the high momentum rates of molecular winds. Radiation pressure can drive such winds during specific phases of the ionized medium expansion: when the expansion happens so rapidly compared to the radiative cooling that it is nearly adiabatic (as in the Sedov-Taylor phase of supernovae). Radio jets, when powerful or nearly relativistic, can efficiently drive adiabatically expanding bubbles as they rapidly deposit energy in the interstellar medium (ISM) for most expansion phases \citep{wagner16}. Indeed, several of the molecular winds in the above-mentioned studies were detected in galaxies with AGN jets.
        
\citet{dasyra16} and \citet{oosterloo17} examined what happens in the molecular gas when a radio jet impacts clouds. Focusing on the nearby galaxy IC5063, they found that the flux ratios of CO lines reveal the presence of highly excited and optically thin gas in the wind. This result indicated that jet--ISM interactions can leave traces in the kinematics and/or in the excitation of the impacted gas. In previous work, \citet{dasyra14} reported heating of the molecular gas in the wind of another radio galaxy, 4C12.50. This result emerged from the comparison of two molecular gas probes: the fraction of accelerated cold ($<$25K) gas in CO data was less than one third of the total reservoir, which corresponded to the fraction of accelerated warm ($\sim$400K) gas in \htwo\ data \citep[from the \spi\ Space Telescope;][]{dasyra14}.  
        
\begin{figure*}
        \begin{center}
                \includegraphics[width=7.8cm]{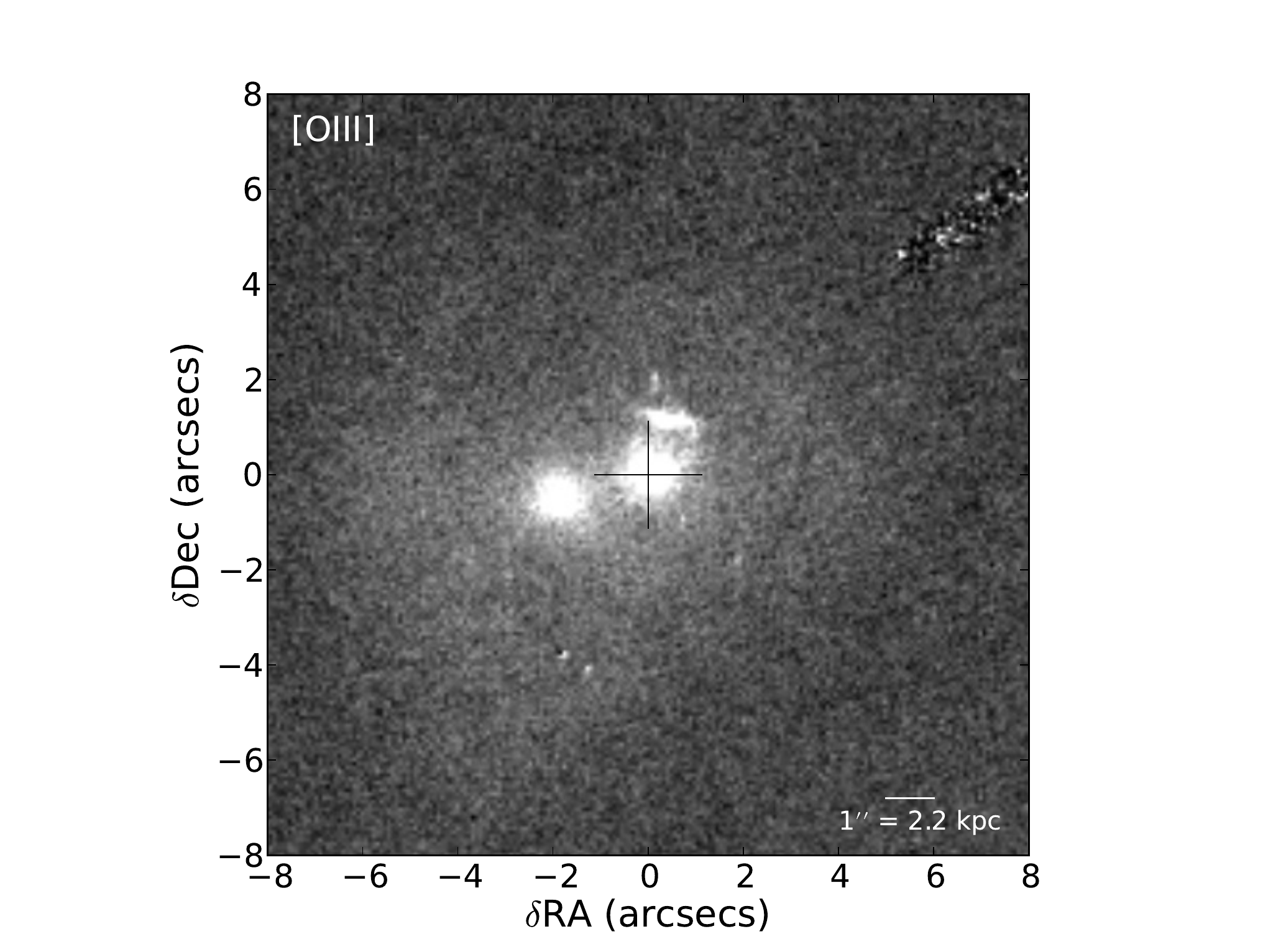}
                \includegraphics[width=7.8cm]{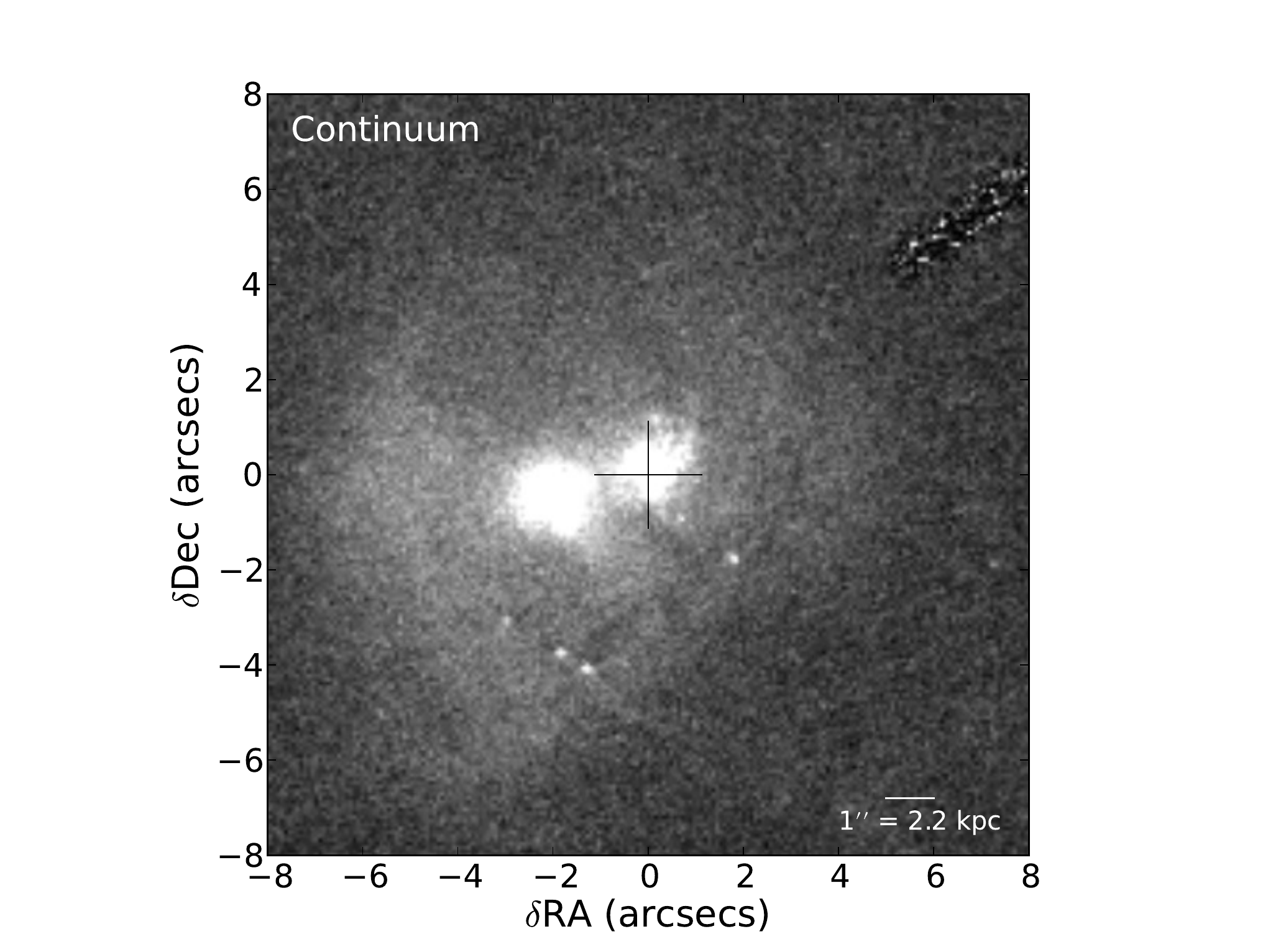}

                \includegraphics[width=7.8cm]{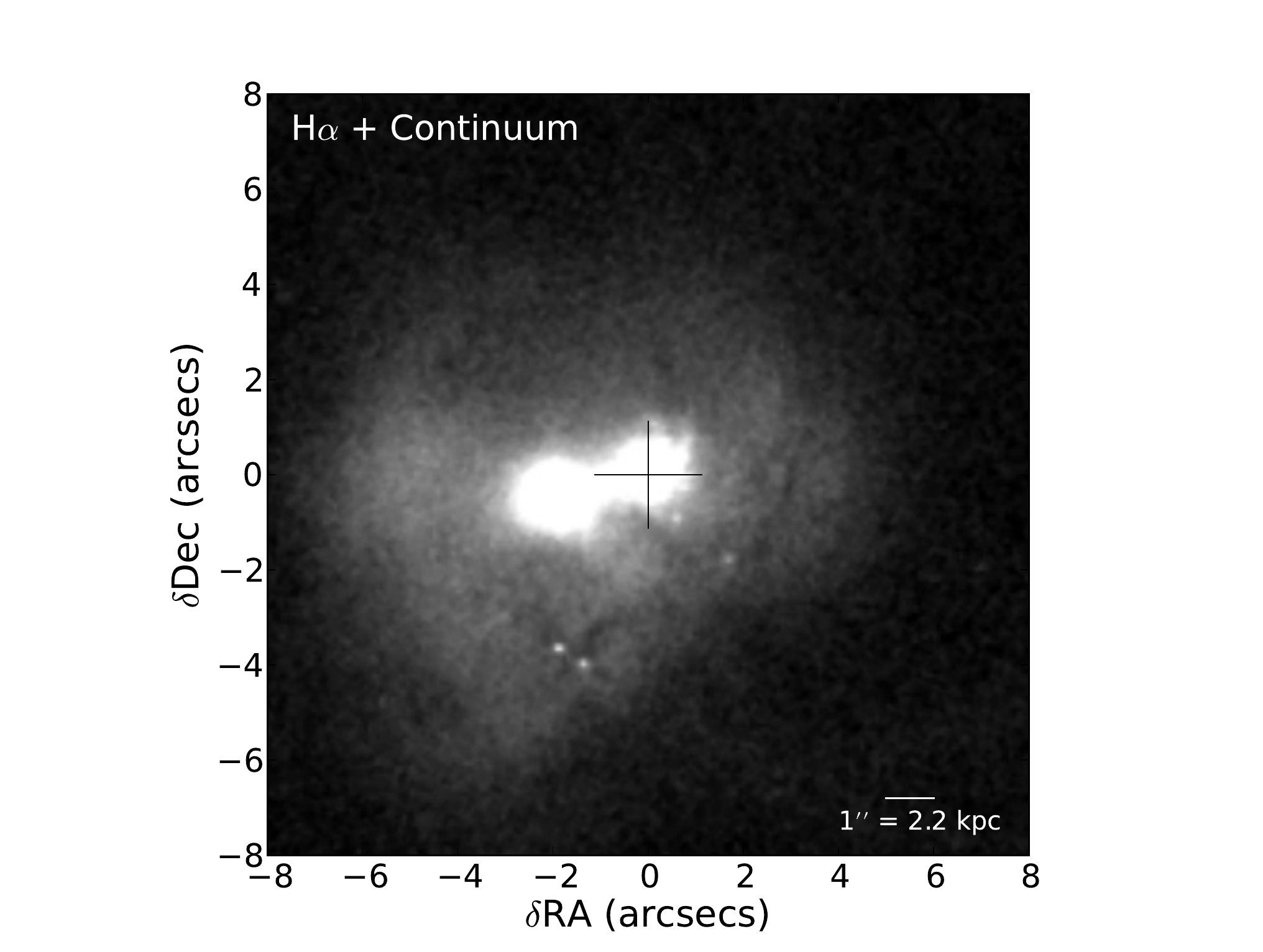}           
                \includegraphics[width=7.8cm]{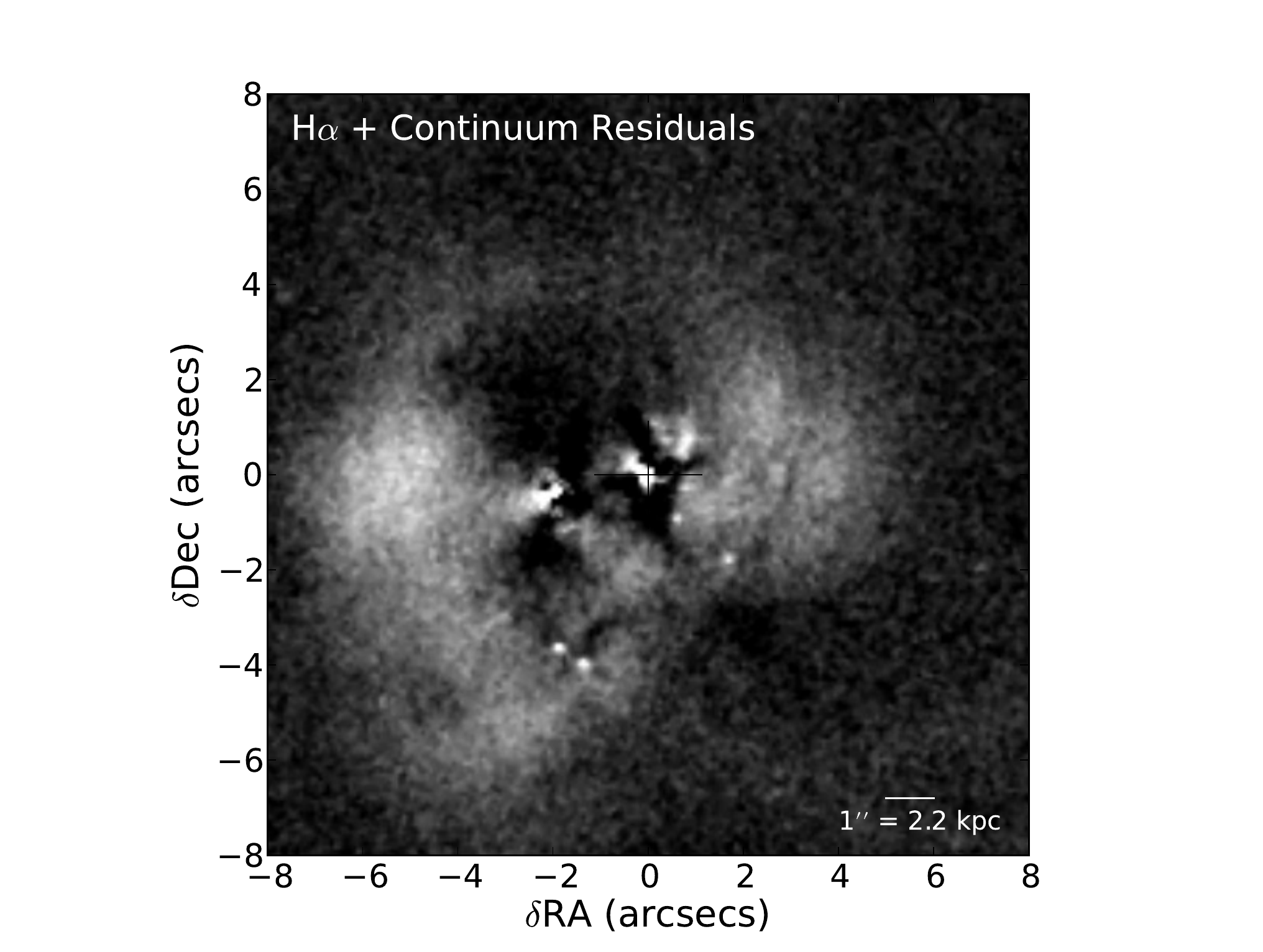}   
                \caption[]{Optical images of 4C12.50 from the \hst , previously presented by \citet{batcheldor07} and \citet{dasyra11}. The crosses mark the location of the peak continuum emission in the ALMA band 3 data. The same applies to all following figures with the crosses.
                        {\it Upper left}: Gas emission comprising primarily \oiii\ emission with potential contribution from \hbeta . The image was created from the subtraction of a \hst\ ACS image at 5092\ang\ (filter FR459M) from an ACS image at 5580\ang\ (filter F550M).
                        {\it Upper right}: Continuum emission at 5092\ang\ (filter FR459M).
                        {\it Lower left}: \ha\ and continuum emission (filters FR647M at 6616\ang). 
                        {\it Lower right}: Residual image originating from the subtraction of galactic component models (i.e., a bulge and two disks) from an optical \hst\ image of \ha\ and 5900\ang\ continuum emission \citep{batcheldor07} using GALFIT \citep{peng02}. The subtraction method was described in detail in \citet{dasyra11}. Its purpose was to bring up any structures underlying the bright regular emission. } 
                \label{fig:optical}
        \end{center}
\end{figure*}

The system 4C12.50, also known as IRAS\,13451+1217 and PKS\,1345+12, is a good candidate for further studies of the impact of AGN feedback on the ISM. This system is an ultraluminous infrared galaxy (ULIRG) in the local Universe, which originated from the merger of other galaxies (Fig. \ref{fig:optical}). It has two nuclei 4.4 kpc away: a western nucleus, from which the radio emission emerges, and an eastern nucleus. We refer to these nuclei as main or primary, and secondary, respectively. In optical wavelengths, the continuum emission of the primary nucleus is comparable to (i.e., lower by a factor of 1.3 than) that of the secondary nucleus within a radius of 1$\arcsec$. The K- and L-band images with Subaru show that the primary galaxy is roughly twice as massive \citep{imanishi14}. Contrarily, the primary nucleus has a significantly greater \oiii\ emission than the secondary nucleus (i.e., greater by a factor of three within the same radius), as derived from the subtraction of two \hst\ images (Fig. \ref{fig:optical}). This indicates that the gas transfer toward the main nucleus has significantly progressed. The \oiii\ emission reveals nonregular structures, such as a filamentary ridge to the northwest of the main nucleus. Traces of this ridge can also be observed in the continuum emission, indicating the presence of stars. Several shells and tidal tails that are caused by the merger are seen at distances greater than 2\arcsec. Some of them are visualized in another optical, H$\alpha,$ and continuum image, from which we subtracted a model of one bulge and two disks to bring up underlying structures \citep{dasyra11}. Further larger-scale tails are seen by \cite{emonts16}. \citet{stanghellini93} suggested that the merging system is part of a poor cluster. A third, smaller galaxy is seen 18 kpc northwest of the main nucleus, which could potentially be part of a merging group. 
        
\begin{figure*}
        \begin{center}
                \includegraphics[width=19cm]{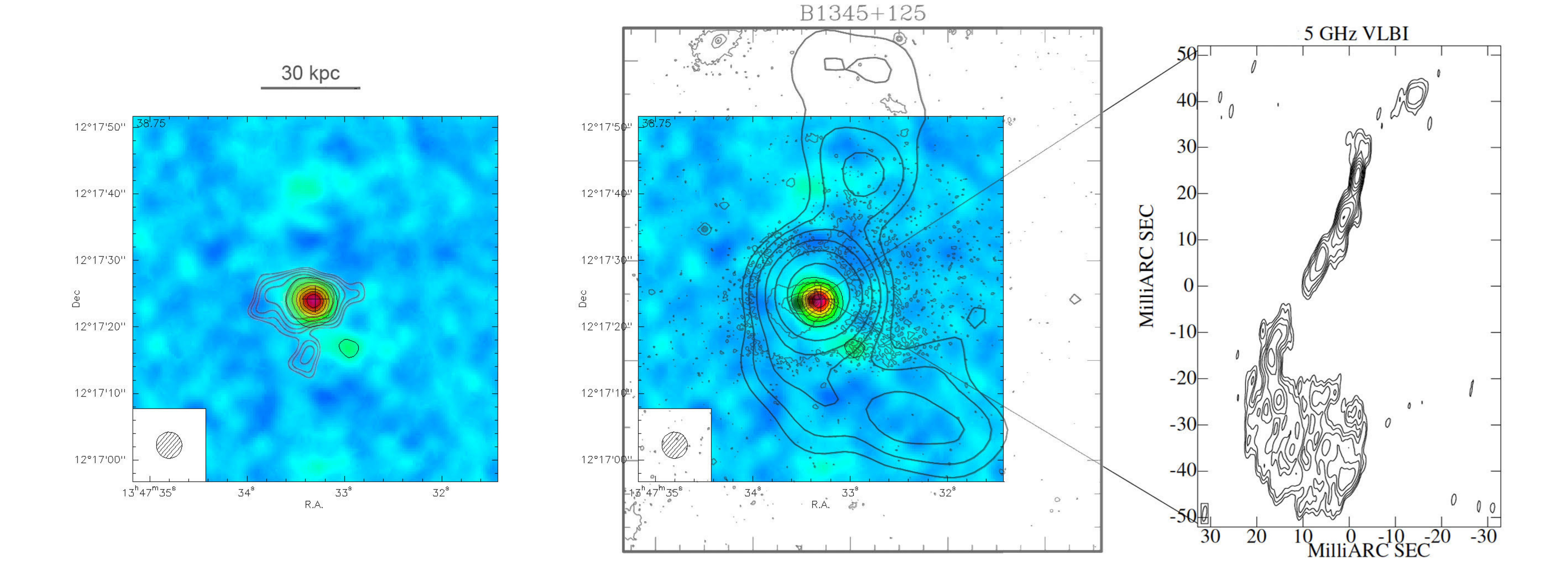}     
                \caption[]{{\it Left:} X-ray emission \citep{siemiginowska08} plotted over an IRAM Plateau de Bure \cooz\ image \citep{dasyra14}. {\it Middle, right:} Large and small scale radio emission at $\sim$1.4 and 5 GHz \citep{stanghellini05} compared to the same \cooz\ image.}
                \label{fig:radio_emission}
        \end{center}
\end{figure*}

The two unambiguously merging nuclei are embedded in an X-ray bubble of about 30$\times$15 kpc (Fig. \ref{fig:radio_emission}; \citealt{siemiginowska08}). Another, discrete X-ray component is detected along the radio jet axis further to the south. Radio emission is detected in two very different scales. A superluminal jet is seen within 100 pc from the main nucleus at 2 to 6 cm wavelengths \citep{stanghellini97, lister03}. More diffuse emission, extending out to 100 kpc, is seen  at 1.36 and 1.66 GHz (see Fig. \ref{fig:radio_emission}; \citealt{stanghellini05}). This indicates the presence of a jet that restarted recently - as recently as a few thousand years \citep{lister03}. Tilts of the radio emission in all scales indicate that the jet is precessing and that it could impact the ISM at various locations. Indeed, an outflow of atomic gas had been seen in absorption by \citet{morganti04,morganti13b} in clouds at the tip of the small-scale radio emission. The outflow had long been studied in its ionized phase \citep{holt08,holt09,holt11}. \citet{dasyra_combes11}, \citet{guillard12}, and \citet{dasyra_combes12} reported the outflow detection in rotational \htwo\ lines and in CO(2$-$3) absorption. The outflow had yet to be seen in CO emission.
        
To further investigate the physical conditions of the low-temperature molecular gas in the wind of 4C12.50 (including its location, mass, and excitation), we acquired new millimeter interferometric data of CO (1$-$0), (3$-$2), and (4$-$3). 
For simplicity, we adopted a $\Lambda$CDM cosmology with H$_0$=70 \kms\  Mpc$^{-1}$, $\Omega_{M}$=0.3, and $\Omega_{\Lambda}$=0.7, which yields 2.2kpc per arcsecond and a luminosity distance of $D_L=570$ Mpc \citep{wright06} at the redshift of the source.

\section{The ALMA data and their reduction}
\label{sec:data}
The Atacama Large Millimeter Array (ALMA) observed 4C12.50 for the cycle 2 program 2013.1.00180.S (PI Dasyra) in band 3 (July 2015), and in bands 7 and 8 (May 2015). At the time, the array had 37 operational antennas. In all bands, four spectral windows of 1.875\,GHz bandwidth were employed. In band 3, we used one spectral window to cover the velocity range -3300\kms\ to 2100\kms\ around the \cooz\ line, and one spectral window to cover the velocity range -3800\kms\ to 1100\kms\ around the HCS$^+$(3-2) line. The other two spectral windows targeted the continuum at rest frame 2.641($\pm$0.026)\,mm and 2.364($\pm$0.021)\,mm. In band 7, we used two partially overlapping spectral windows to cover the velocity range -2600\kms\ to 900\kms\ around the \cott\ line, and two spectral windows to obtain the restframe 896($\pm3$)\um\ and 902($\pm3$)\um\ continuum. In band 8, we used two spectral windows to cover the velocity range \hbox{-1350\kms} to 840\kms\ around the \coft\ line, one spectral window to target the \hcop\ line, and one spectral window to obtain the continuum at restframe 668.4($\pm$1.7)\um . For the line observations, we typically chose a velocity bin of 5($\pm$1)\kms , which is well suited for the detection of lines originating from small cloud ensembles (either in emission or in absorption in front of a background continuum). For the continuum observations, we chose coarser velocity bins of $\sim$20 up to $\sim$90\kms .

To create the cubes, we reran the Common Astronomy Software Applications (CASA) reduction routines delivered by ESO. The calibration routines were run with CASA version 4.3.1, 4.2.2, and 4.3.1 for bands 3, 7, and 8, respectively. The automated pipeline results were restored for bands 3 and 7. For band 8, the calibrations were fine tuned: the data from the  antennas DA34 and DV02 were flagged due to high system temperatures caused by an error in the antenna position database. After a first basic flagging, the calibration pipeline computed the water vapor, system temperature, and antenna position calibration solutions and applied them to the measurement set (MS), that is, to the visibilities of all baselines, spectral windows, and target fields. It then extracted the pertinent spectral windows and continued with the masking of antenna-shadowed baselines and edge channels. The assignment of a flux model for the calibrator, and the computation of analytic expressions for the bandpass (phase and amplitude vs. frequency)  and gain (phase and amplitude vs. time) calibrations were then done. The flux calibration solution was found using 3C273 in band 7, and using Titan in bands 3 and 8. Some band 8 baselines had to be excluded due to the extent of Titan. In all bands, the bandpass calibration solution was found using the QSO J1256-0547 (3C279). The phase was self-calibrated on 4C12.50 using additional observations dedicated for this purpose.

Once the calibrations were applied, the visibilities of the selected fields were converted into image cubes. For the imaging part of the pipeline, CASA version 5.4 was used. The image reconstruction was executed with the routine tclean that also performs the beam cleaning. The science-intended data were merged with the calibration-intended data of 4C12.50 during the image reconstruction phase for the sake of maximum uv plane coverage, even though the addition of the calibration-intended data had a small contribution to the total exposure time. As science target, 4C12.50 was observed for 24.4, 5.2, and 24 minutes, and as phase calibrator, 4C12.50 was observed for 2.0, 0.8, and 2.0 minutes in bands 3, 7, and 8, respectively. For the image reconstruction, we used Briggs weighting of the visibilities with robustness parameter of 2 for band 3, and 1 for bands 7 and 8.  A robustness parameter of 2 corresponds to 
natural weighting, most appropriate for high-sensitivity detection experiments. Indeed, a robustness parameter below 2 led to considerable flux losses in band 3. In bands 7 and 8, a robustness parameter of either 1 or 2 led to identical flux measurements. In these cases, the lower value was selected, as it leads to a narrower beam with smaller sidelobes. During the image reconstruction, a primary beam correction was also applied.  This procedure was repeated for continuum-free cubes, produced by the uvcontsub routine. First-order polynomials were employed for the modeling and the subtraction of the continuum in all spectral windows with one exception: in band 8, we had to use a zero-order polynomial as the main spectral window extended from -400\kms\ to 840\kms\ of the line center, and only the right-hand side of the continuum could be sampled. Contrarily, in band 3 
more than 1000\kms\ were available on the left
side of the line and a few hundred\kms\ on the
right side of the line enabling the continuum determination.  
Access to the continuum flux on both sides of the line reduces the uncertainty in the continuum slope. The small number of continuum-free channels on the right side of the line and the dynamic range of the observations add to the uncertainty in the continuum level.  

In all bands, the configuration of the  antennas led to a sub-arcsecond beam. The beam was  0.62\arcsec\ $\times$ 0.56\arcsec\ at a position angle (PA) of 14$\degr$ in band 3, 0.61\arcsec\ $\times$ 0.47\arcsec\ at a PA of 53$\degr$ in band 7, and 0.43\arcsec\ $\times$ 0.36\arcsec\ at a PA of -1$\degr$ in band 8. At a common pixel scale of 0.08\arcsec\ and at a common velocity resolution of 20\kms , the noise level is at 0.40 mJy/beam for \cooz , at 1.1 mJy/beam for \cott , and at 1.2 mJy/beam for \coft . For the examination of flux ratios, we produced a second set of datacubes, by convolving the data to a common beam of 0.63\arcsec $\times$0.63\arcsec\ using the routine ia.convolve2d. In this fully homogenized dataset the noise level is at 0.41 mJy/beam, 1.2 mJy/beam, and 1.4 mJy/beam for \cooz , \cott, and \coft , respectively.

To create the deepest possible cubes for each line, we also looked into the archive for additional CO line data from observations of 4C12.50 as a calibrator for other programs. Indeed, the \cooz\ line was observed with 38 antennas for 13.6 mins as a calibrator for the program 2013.1.00976.S. We calibrated the measurement set using the delivered calibration pipeline with CASA 4.7.0, and then we merged the calibrated measurement set with ours using the routine concat in CASA 5.4 to create a final cube with $\sim$40 mins of on-source exposure. As before, we used tclean to reconstruct this cube, which has a new (common) beam of 0.81\arcsec $\times$0.74\arcsec at a PA of -22$\degr$. The common pixel scale of 0.08\arcsec\ was selected for comparison purposes. This \cooz\ cube reached noise levels of 0.17 mJy at a spectral resolution of 100\kms . Likewise, the \cott\ line was observed with 27 antennas for 5.8 mins as a calibrator for the program 2012.1.00797.S. We calibrated the data in CASA 4.1, and we merged the calibrated measurement set with ours in CASA 5.4. We again adopted the common pixel scale of 0.08\arcsec\ for comparison purposes. The beam was 0.63\arcsec $\times$0.48\arcsec\ at a PA of 51$\degr$. This \cott\ cube reached noise levels of 1.1 mJy/beam at a spectral resolution of 20\kms  and 0.57 mJy at a spectral resolution of $\sim$100 \kms . No further data were found for \coft . From here onwards, we refer to the concatenated low-spatial-resolution cubes as deep cubes.

\section{Results}
\label{sec:results}

\subsection{Line fluxes}
\label{sec:basic}
\begin{figure*}
        \begin{center}
                \includegraphics[width=5.5cm,height=2.9cm]{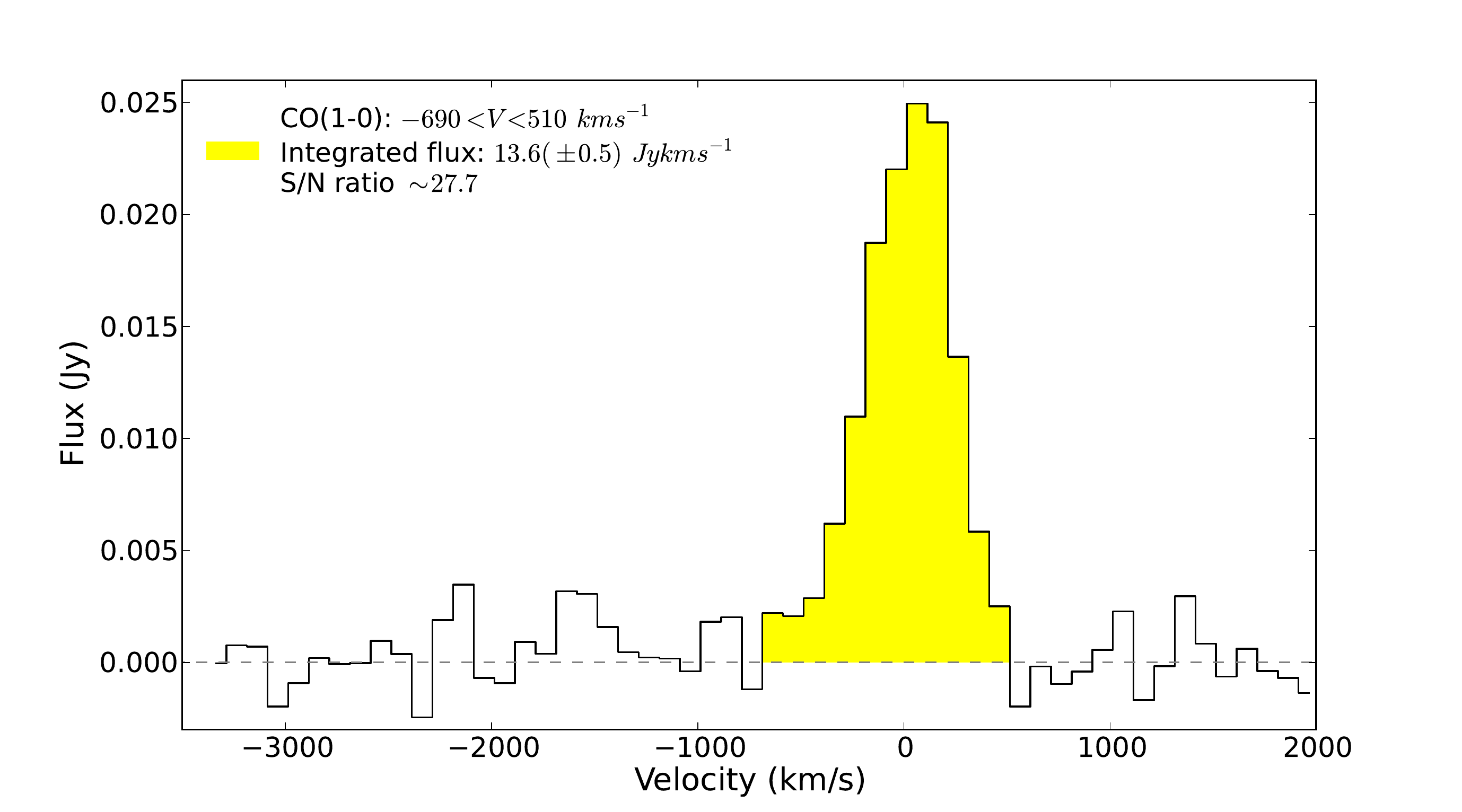}
                \includegraphics[width=5.5cm,height=2.9cm]{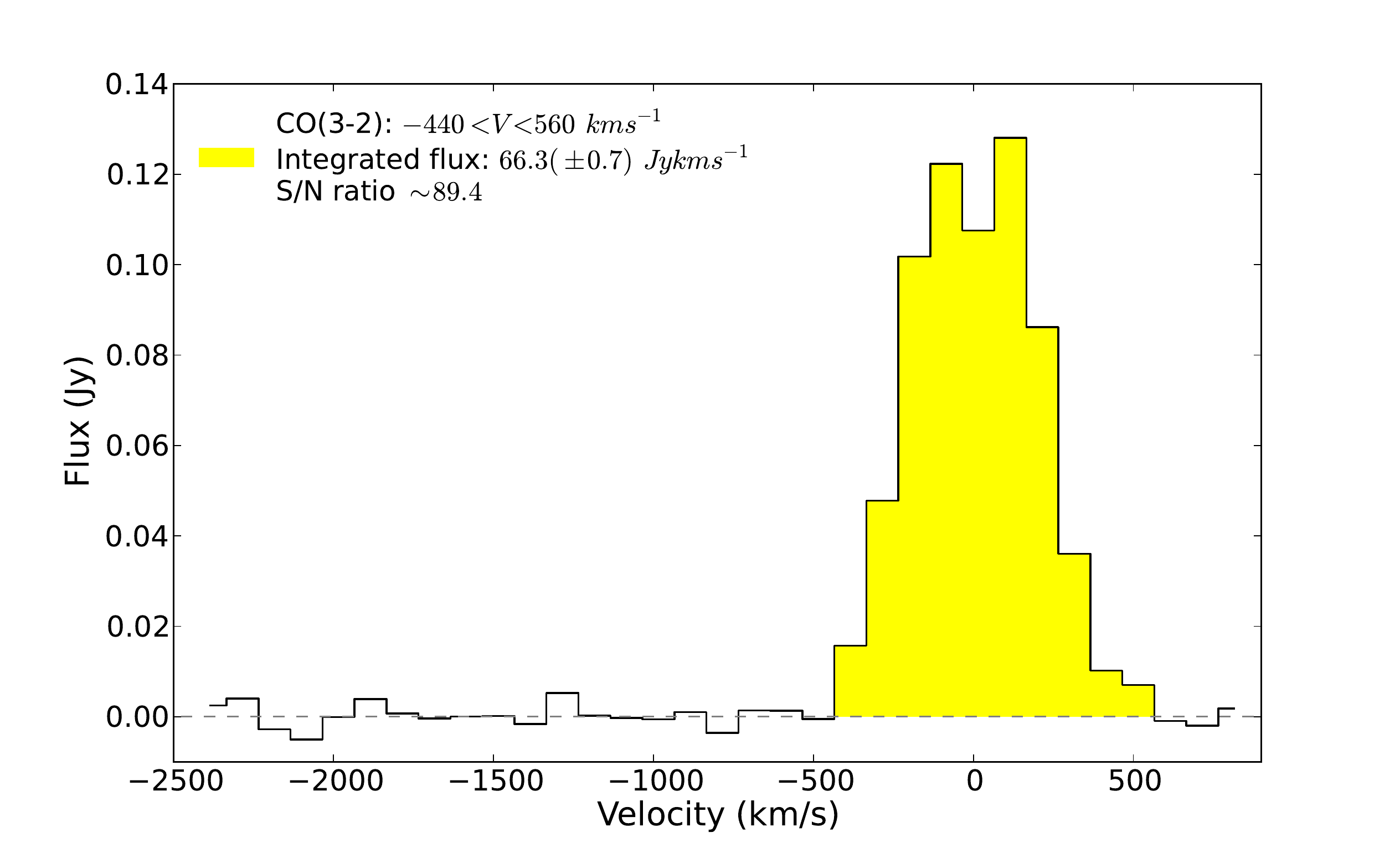}
                \includegraphics[width=5.5cm,height=2.9cm]{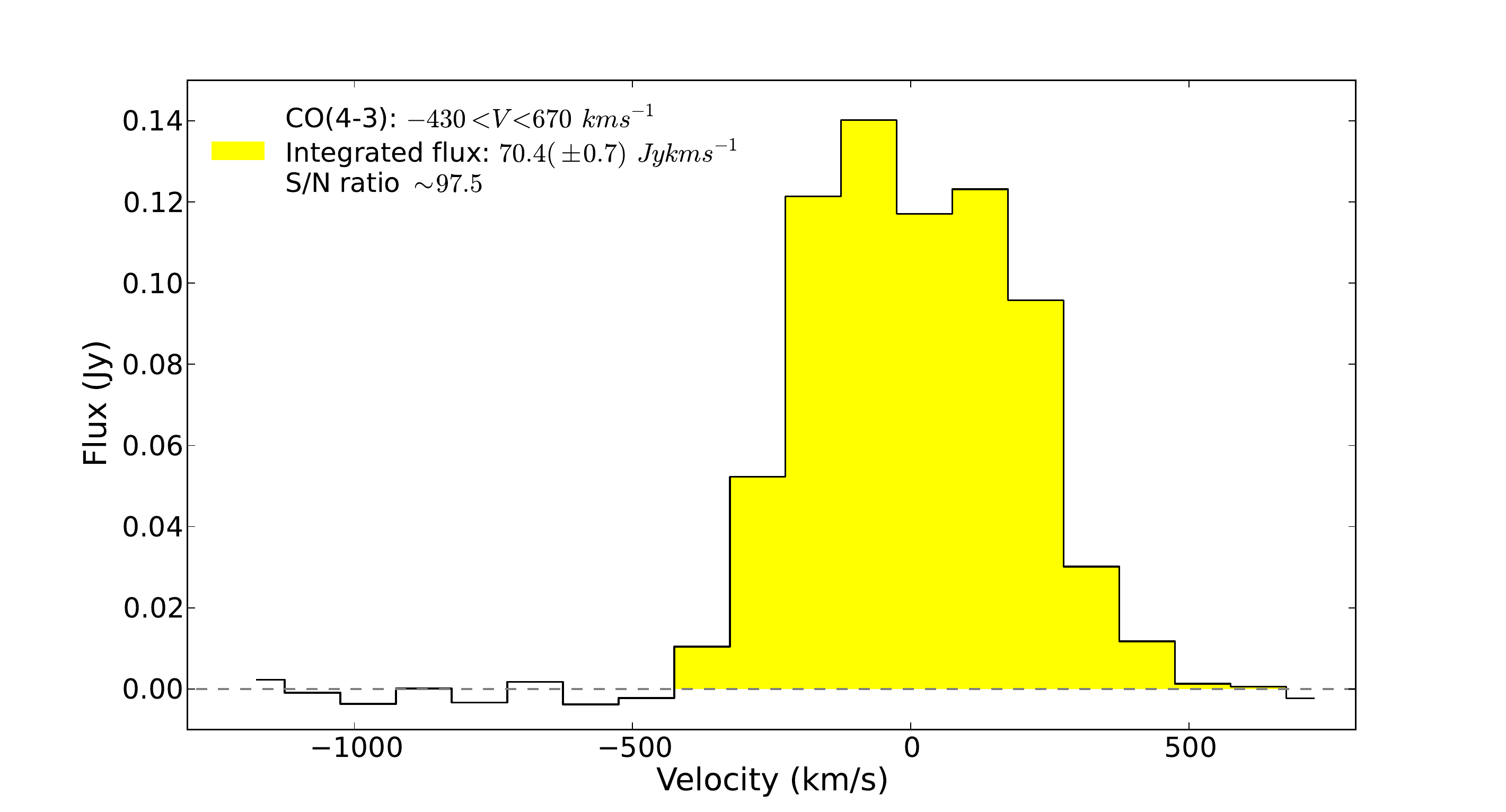}

                \includegraphics[width=5.5cm,height=2.9cm]{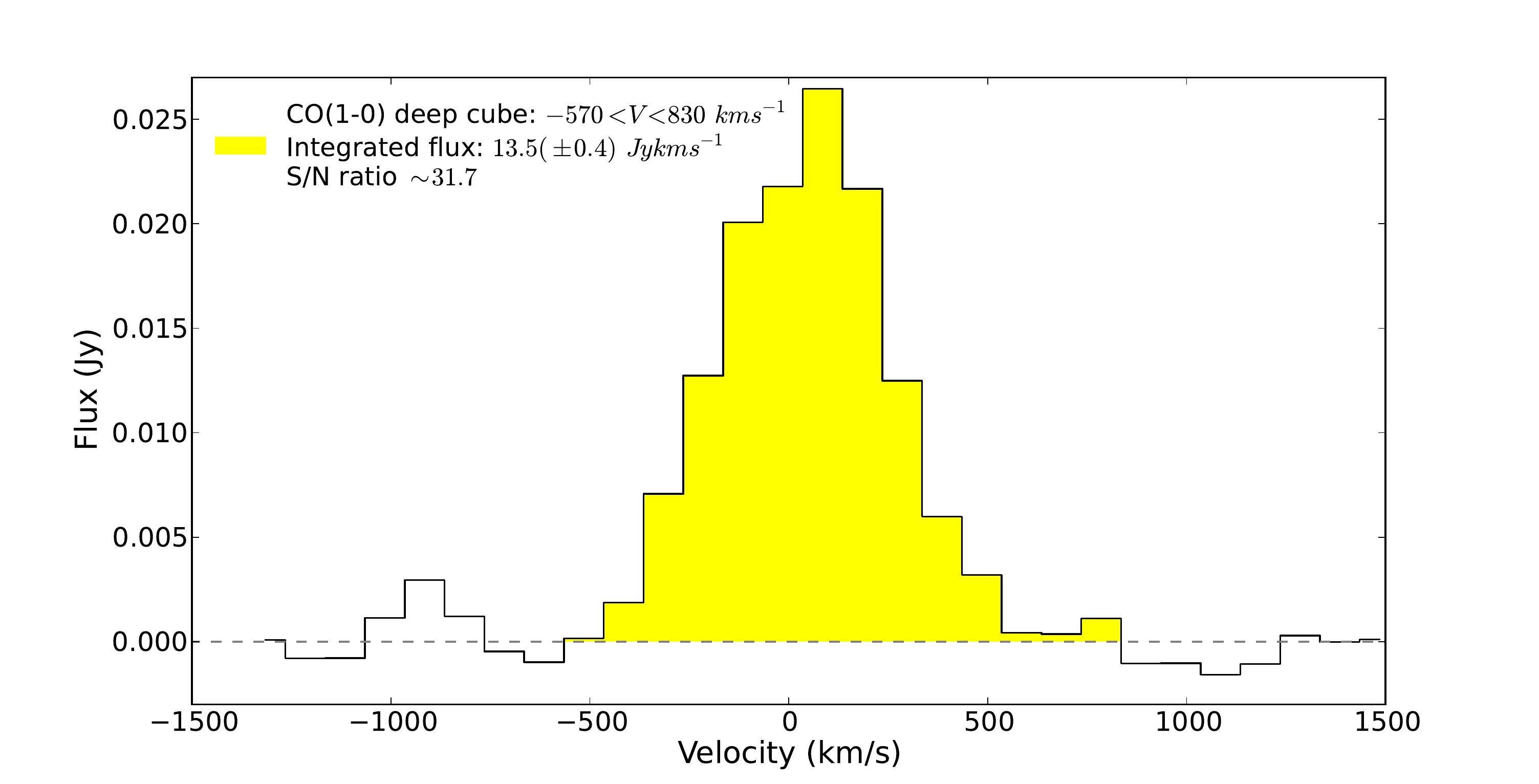}
                \includegraphics[width=5.5cm,height=2.9cm]{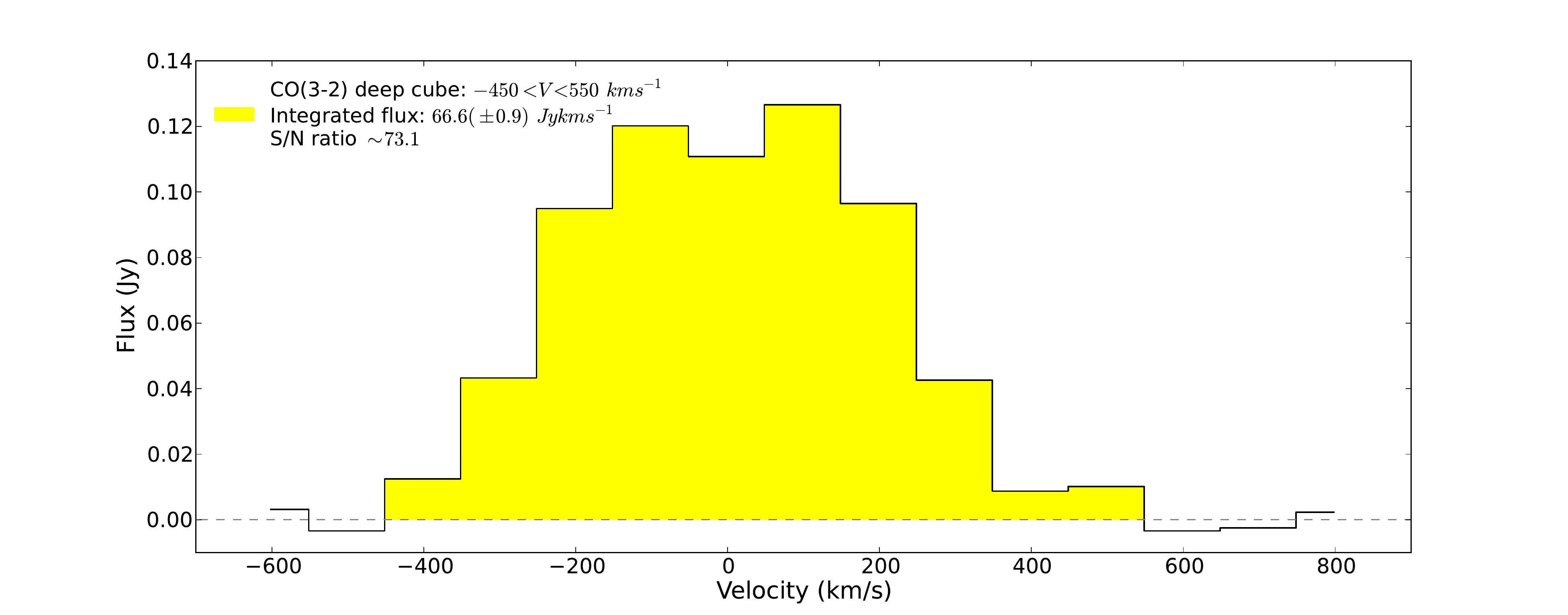}    
                \caption[]{\cooz\ , \cott\ and \coft\ galaxy spectra derived from our high-resolution data cube (upper panels) and the deep data (lower panels) that originated from the merging of the measurements of two programs.
                }
                \label{fig:galaxy_spectra}
        \end{center}
\end{figure*}

\begin{figure*}
        \begin{center}
                \includegraphics[width=5.5cm,height=2.9cm]{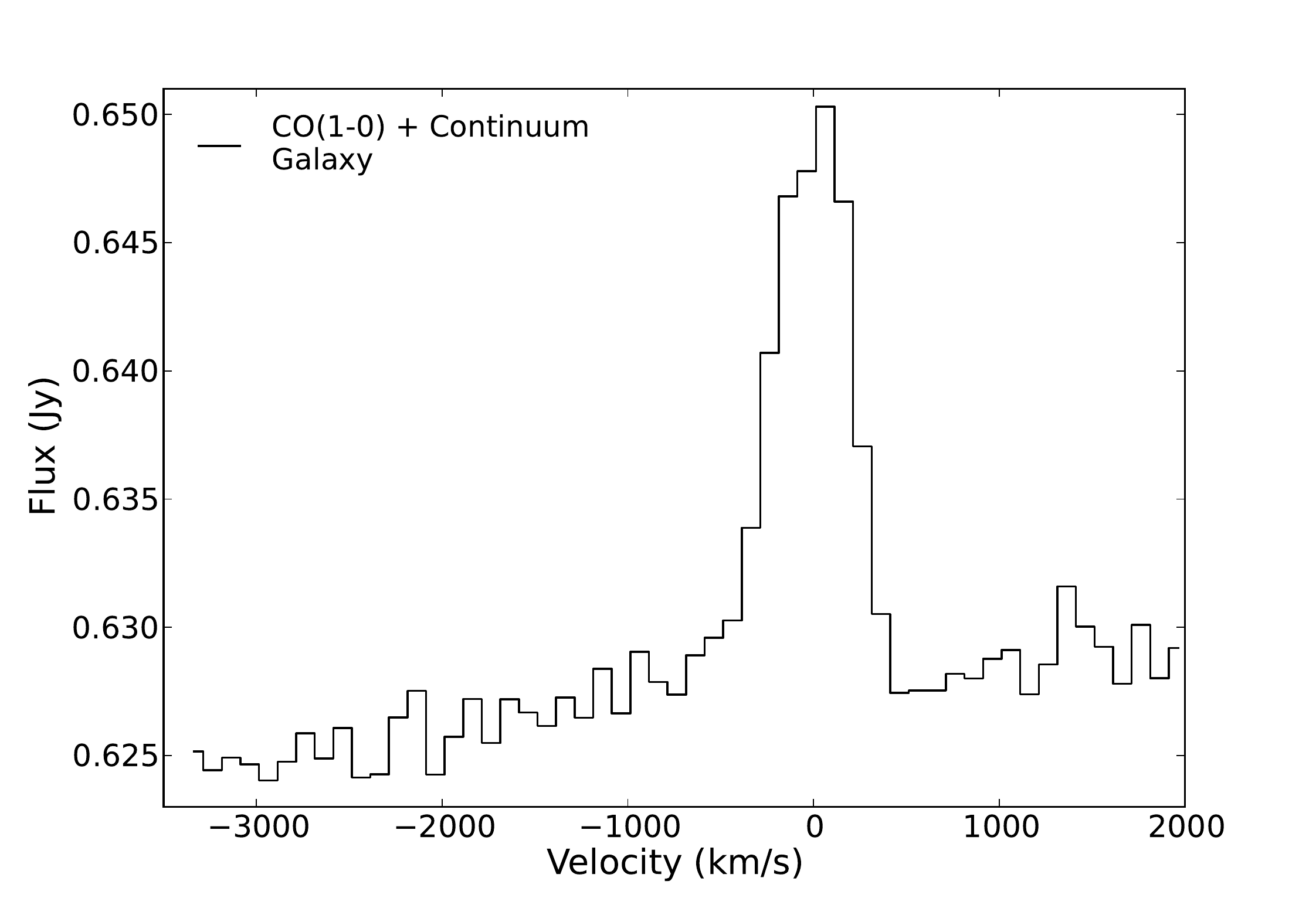}
                \includegraphics[width=5.5cm,height=2.9cm]{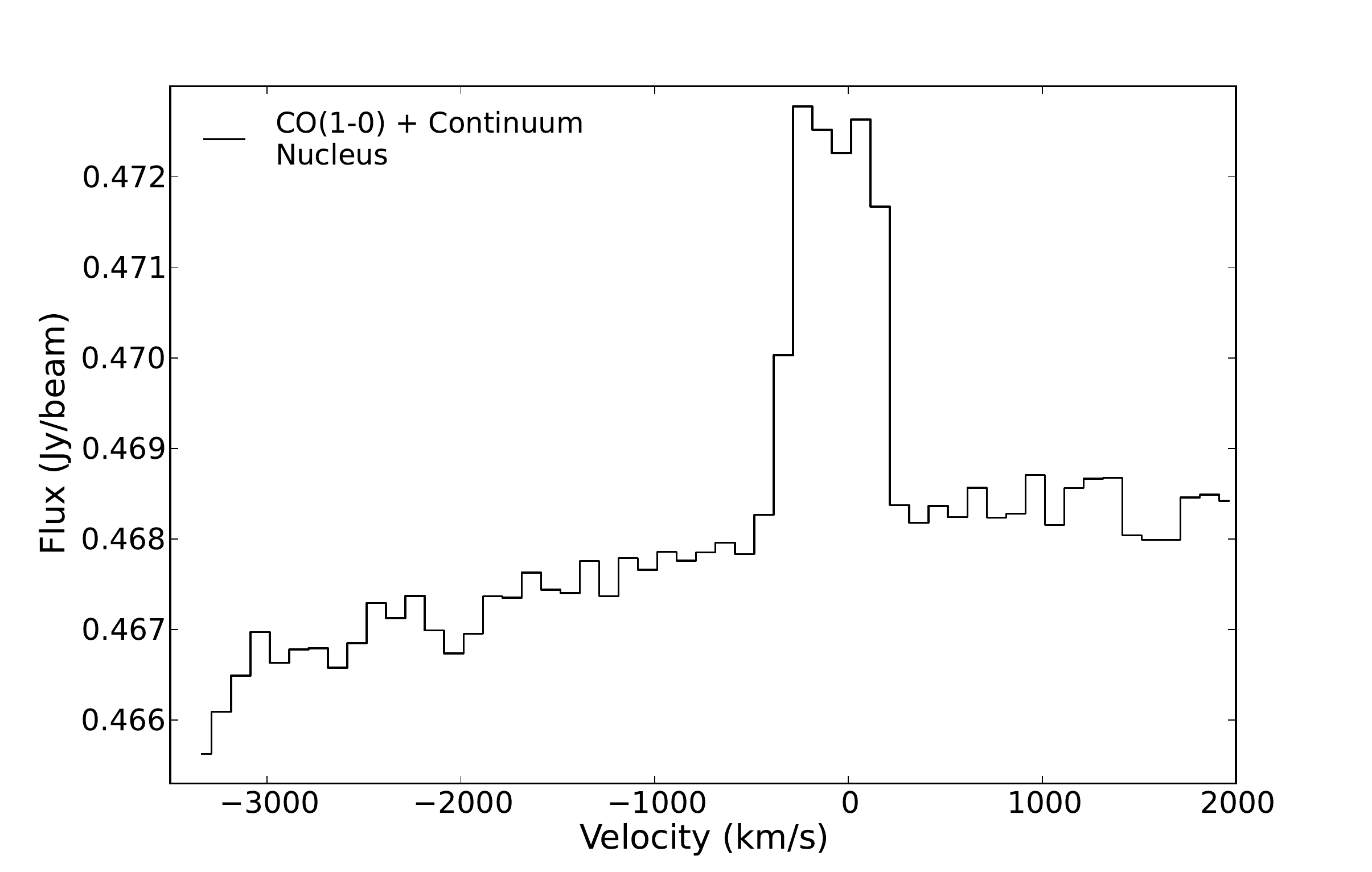}

                \includegraphics[width=5.5cm,height=2.9cm]{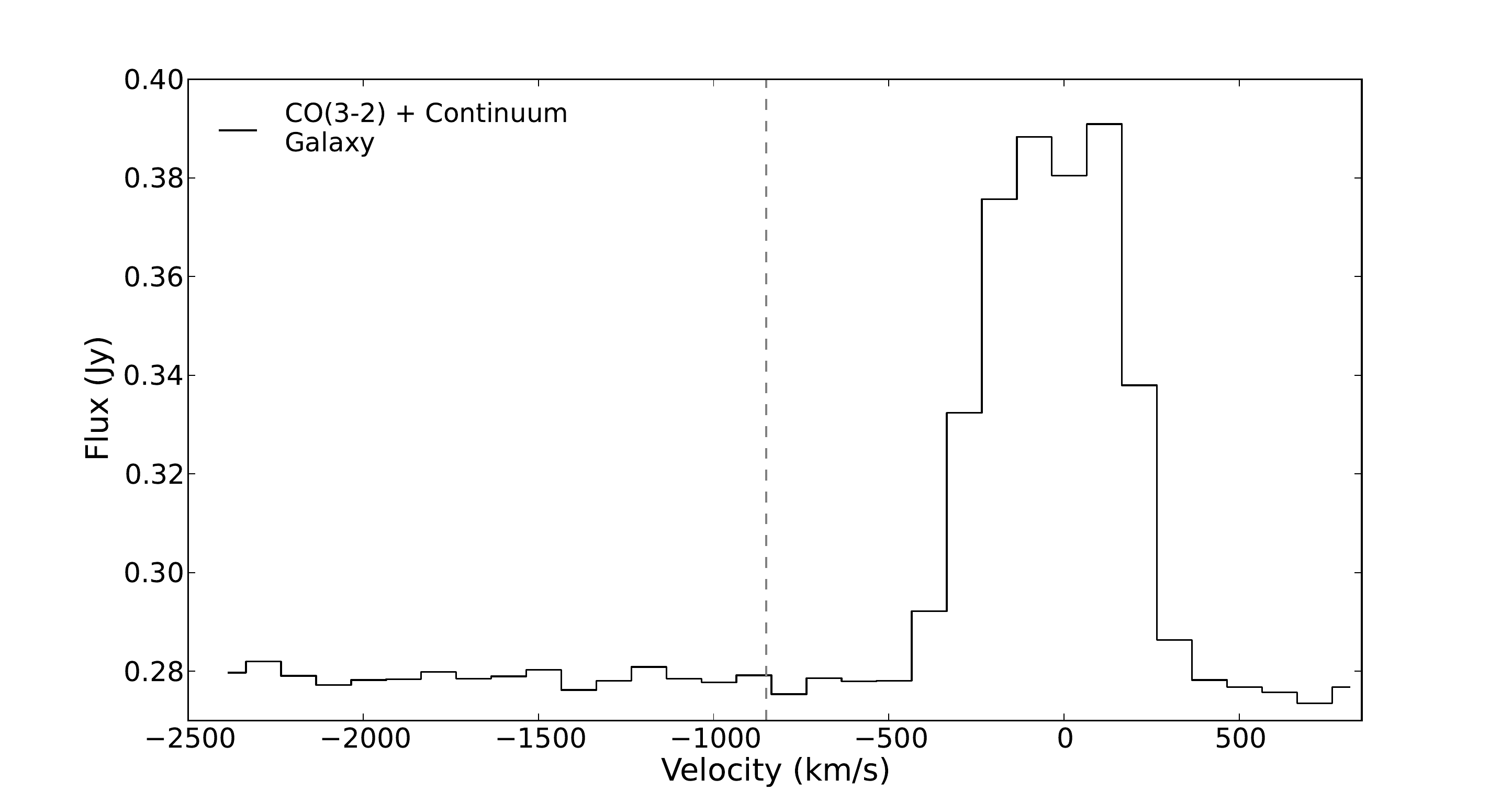}
                \includegraphics[width=5.5cm,height=2.9cm]{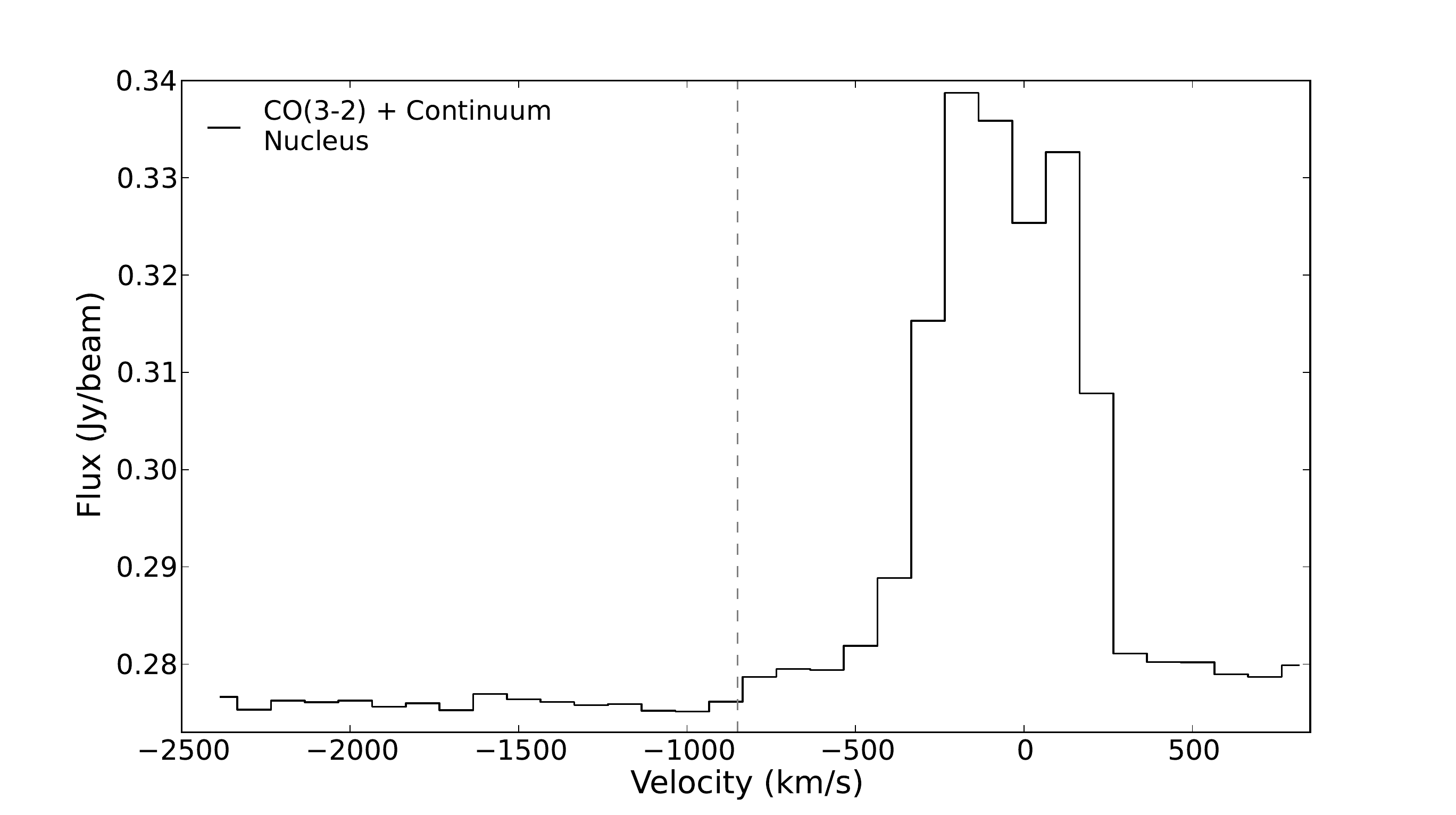}

                \includegraphics[width=5.5cm,height=2.9cm]{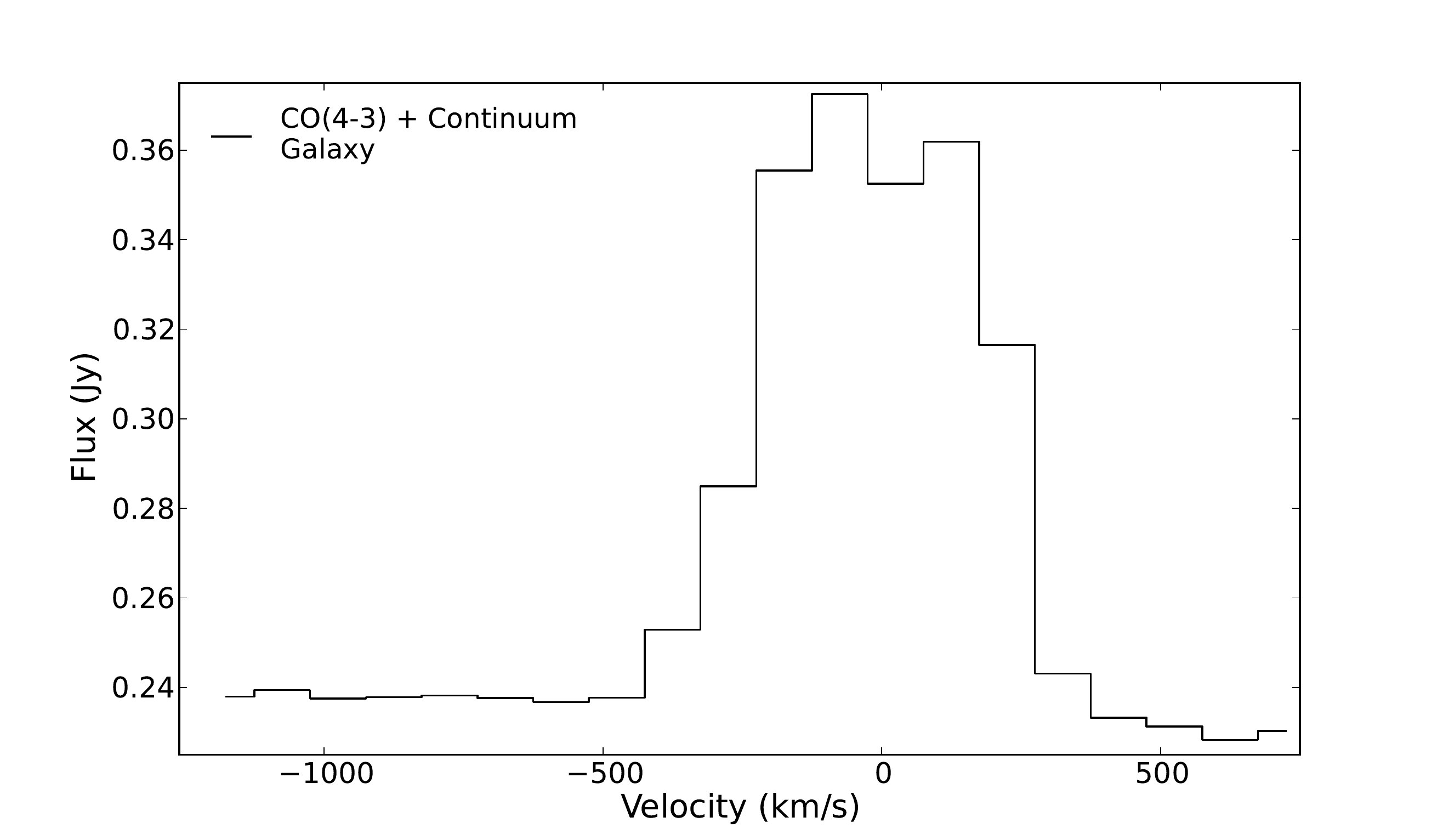}
                \includegraphics[width=5.5cm,height=2.9cm]{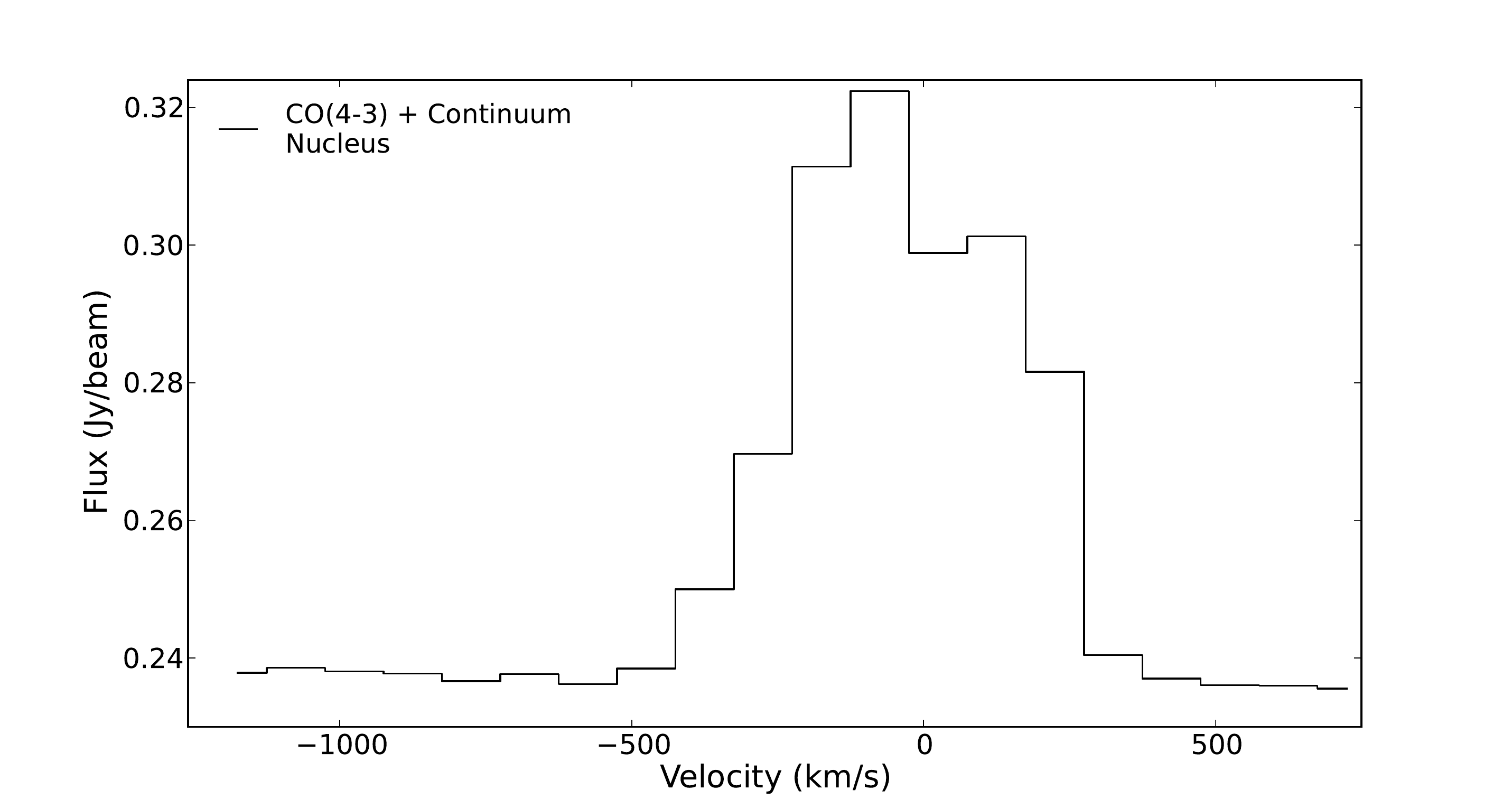}

                \includegraphics[width=5.5cm,height=2.9cm]{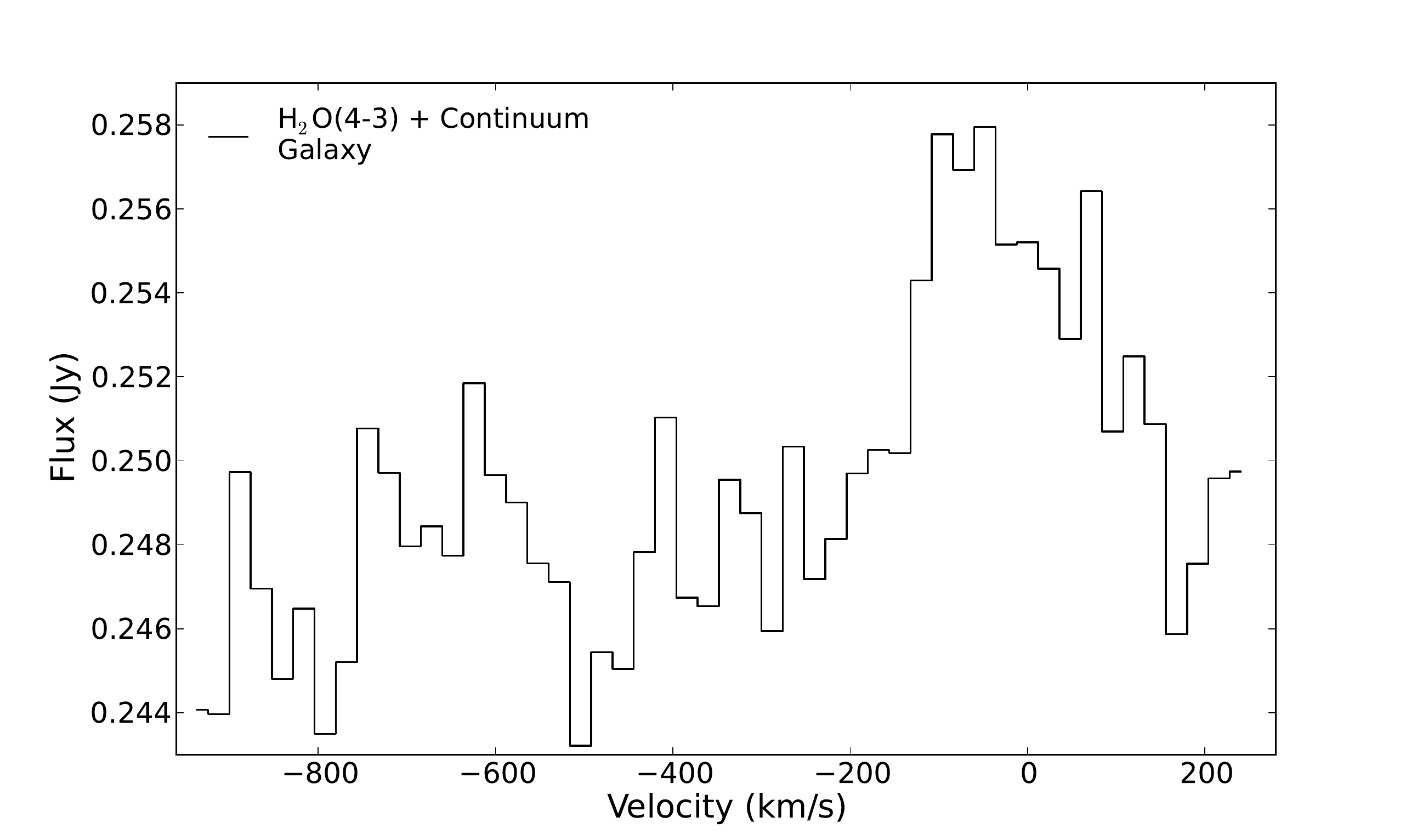}
                \includegraphics[width=5.5cm,height=2.9cm]{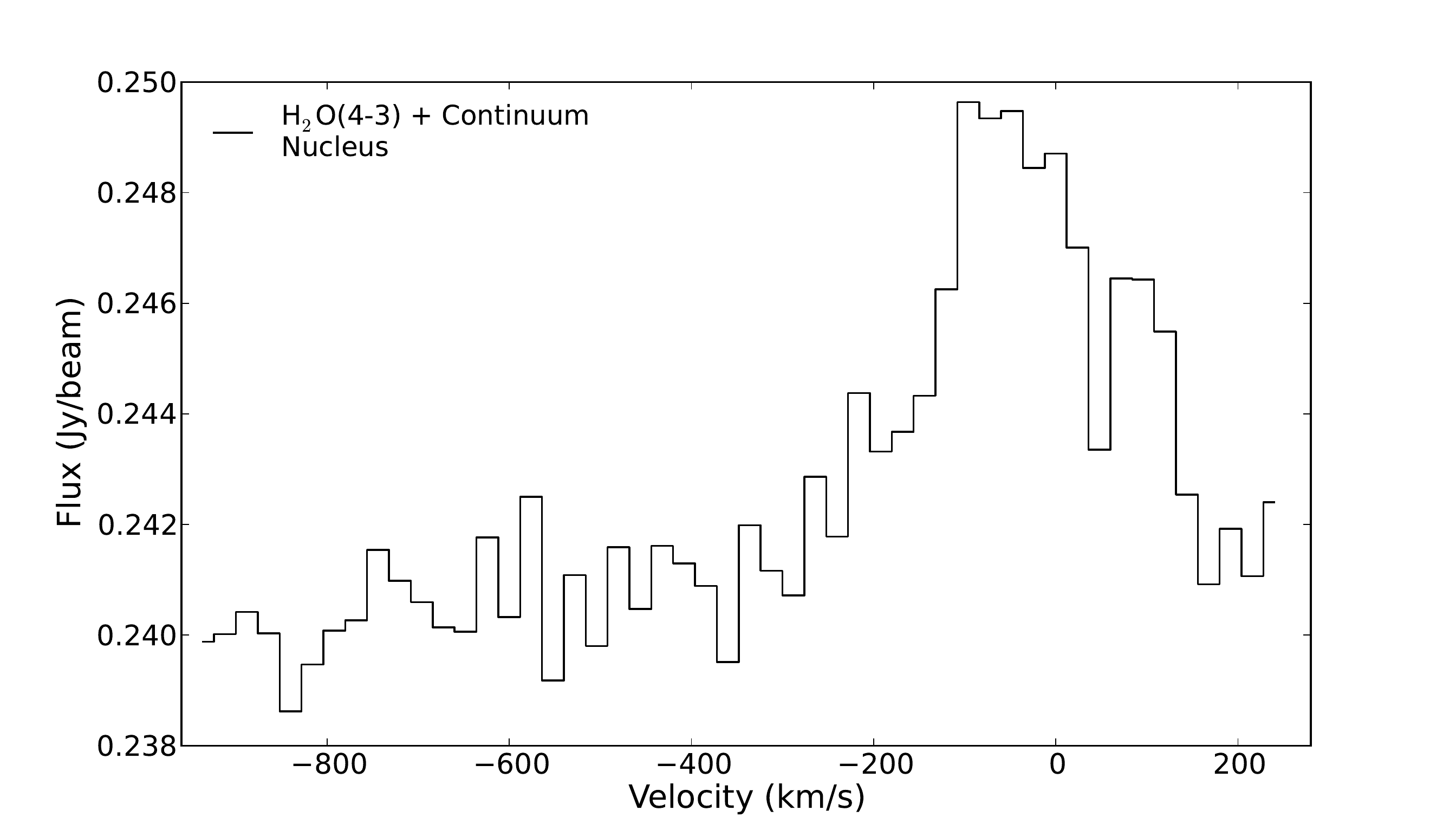}

                \includegraphics[width=5.5cm,height=2.9cm]{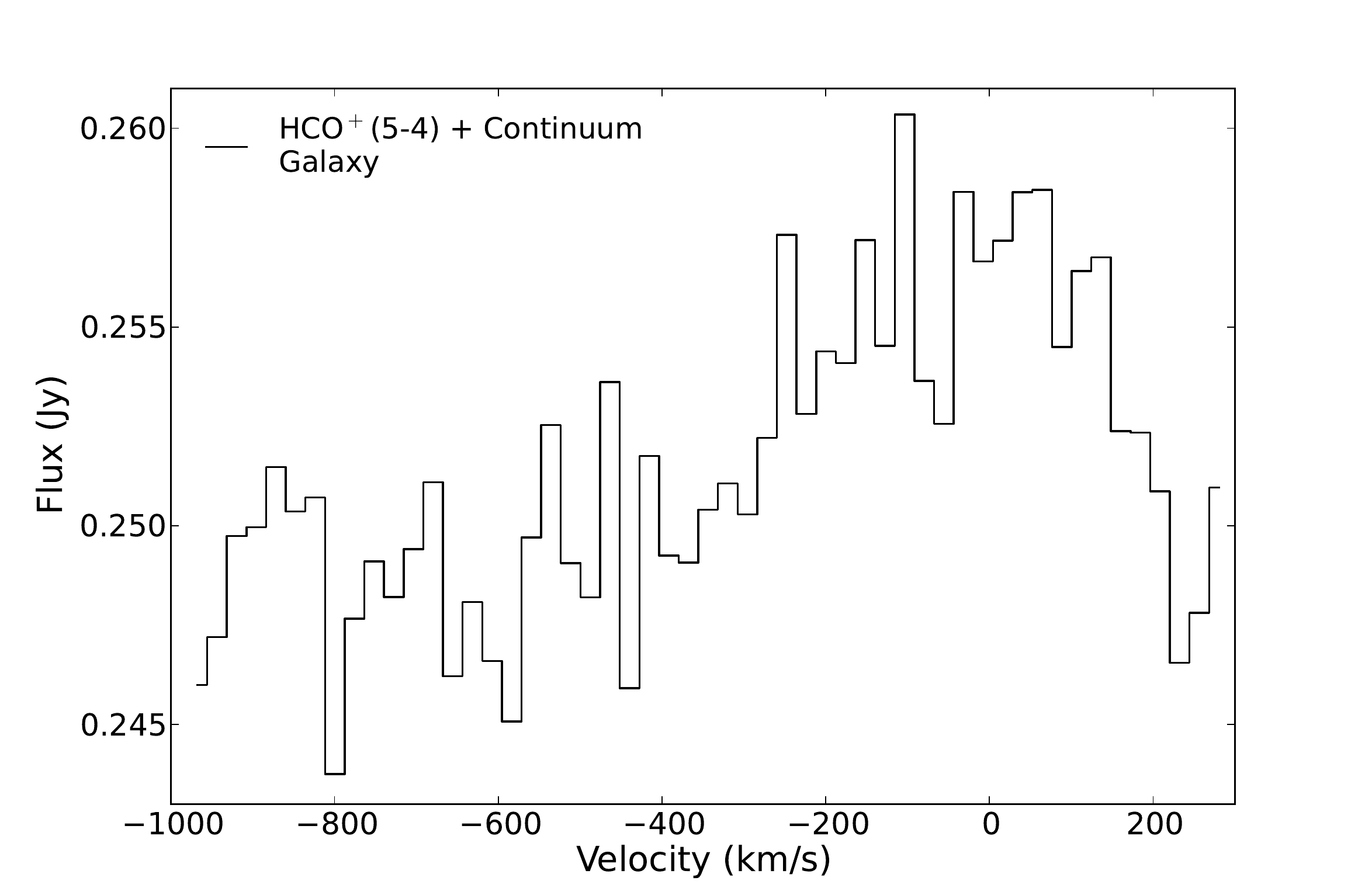}
                \includegraphics[width=5.5cm,height=2.9cm]{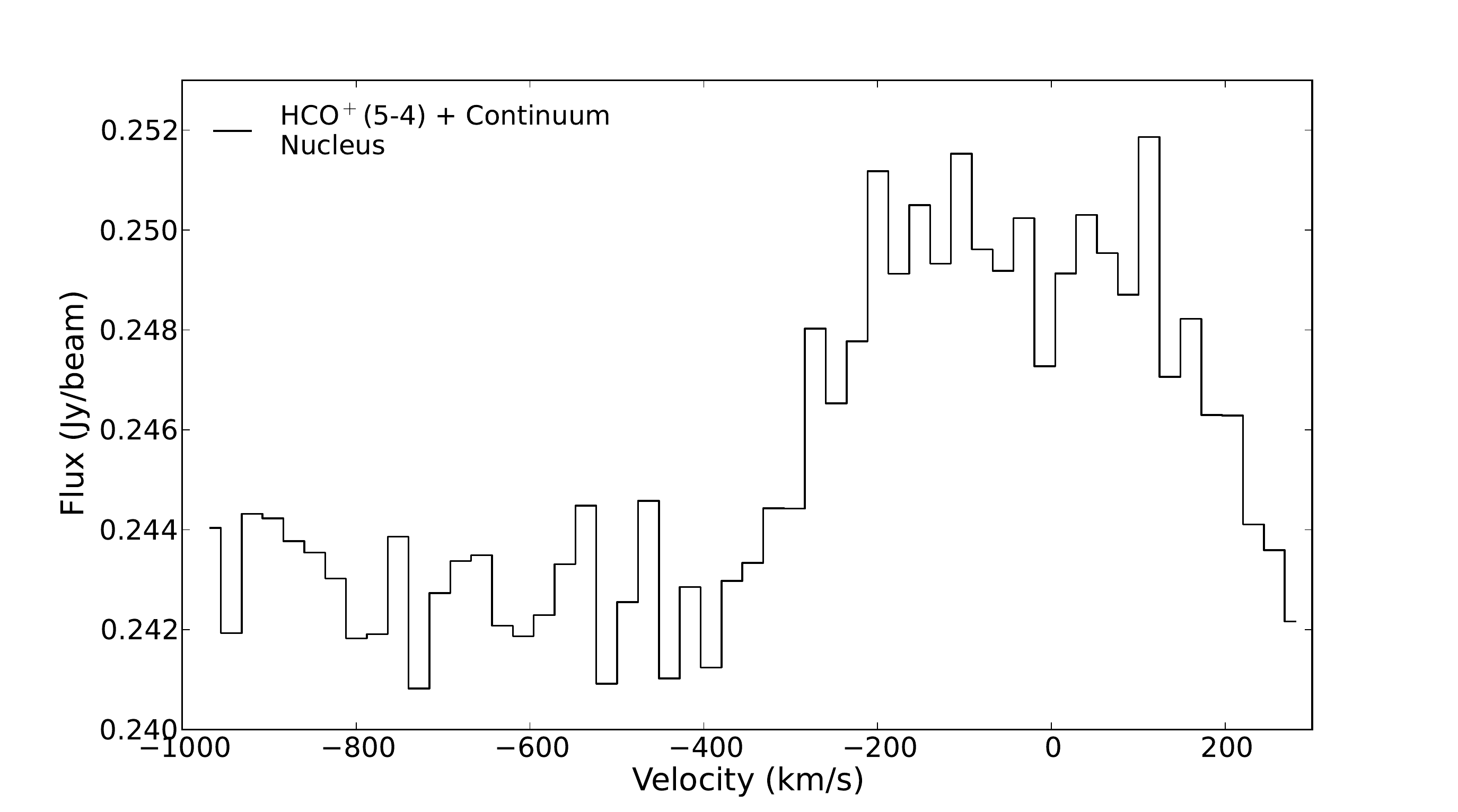}      
                \caption[]{Spectra of all molecular gas lines and continuum emission detected in the ALMA data of the program 2013.1.00180.S. {\it Left}: Spectra of the entire galaxy. {\it Right}: Spectra of the nuclear pixel. The dashed line in \cott\ spectrum corresponds to the frequency at which the stitching of the data of the two spectral windows, in which the continuum level and slope are not known with sufficient accuracy for the determination of the presence of a wind, took place.
                }
                \label{fig:spectra}
        \end{center}
\end{figure*}

Spectra of all detected lines are shown in Figs. \ref{fig:galaxy_spectra} and \ref{fig:spectra}. The CO (1$-$0), (3$-$2), and (4$-$3) lines that we detected provided an average redshift $z$ of 0.122($\pm$0.001) when fitted by Gaussian profiles. At that redshift, 1\arcsec\ corresponds to 2.2kpc. 
The integrated CO line fluxes from the high-spatial-resolution cubes are 
 13.6($\pm$1.5) Jy\kms , 66($\pm$2) Jy\kms\, and  70($\pm$2) Jy\kms\ 
for CO (1$-$0), (3$-$2), and (4$-$3), respectively.  The integrated CO(1$-$0) flux from the deep concatenated data, which have a high coverage of the uv-plane in short baselines, is  13.5($\pm$1.2) Jy\kms .
 The results from the \cooz\ data are in good agreement with the results of \cite{evans99}  yielding a \cooz\ flux of 14($\pm$4) Jy\kms .
Previous \cooz\ interferometric observations taken with the IRAM Plateau de Bure (PdB) array yielded a \cooz\ flux of 18.2($\pm$1.5) Jy\kms\ at a resolution of 4.0\arcsec$\times$ 3.8\arcsec\ \citep{dasyra14}. The integrated CO(3$-$2) flux from the deep concatenated cube is 67($\pm$2) Jy\kms . For comparison, the \cott\ line flux is comparable to the value we had previously measured with the IRAM 30m single dish telescope: 50($\pm$8) Jy\kms\  \citep{dasyra_combes12}. This means that not only did the band 7 data not miss extended emission, but that they were more sensitive than the 30m telescope data. As in both bands the high-resolution data reach comparable flux levels to the deep and the ancillary data, we perform all of our calculations with them.

Additionally, \hcop\ and a \water\ line ($\nu$=0 4[2,3] - 3[3,0]) at restframe 448.00108 GHz were detected in band 8. The \water\ emission, a shock tracer, is nearly unresolved, originating from the vicinity of the main nucleus of 4C12.50. The \hcop\ ion emission tracing the AGN radiation is also circumnuclear with a low-signal extension towards the northwest. The overall fluxes of the \hcop\ and \water\ lines are  3.5($\pm$0.2) and  2.1($\pm$0.2) Jy\kms, respectively.  HCS$^+$(3$-$2) was not detected in band 3.
Unfortunately, we were unable to test the CO(2$-$3) absorption previously reported by \citet{dasyra_combes12}. The concatenation of the two spectral windows of the band 7 observations at about -900\kms\ (indicated in Fig.~\ref{fig:spectra}), together with bandpass calibration and continuum subtraction uncertainties, led to significant differences in the continuum level and slope between the spectral window edges. These differences can mimic absorption or emission features or remove real absorption or emission features near the concatenation region. In the nuclear spectrum of \cott \  for example, the continuum to the right of the line is higher than the continuum to the left of the line (Fig.~\ref{fig:spectra}).

\subsection{ Disk and ambient gas emission}
\label{sec:disk}

\begin{figure*}
        \begin{center}
                \includegraphics[width=6cm]{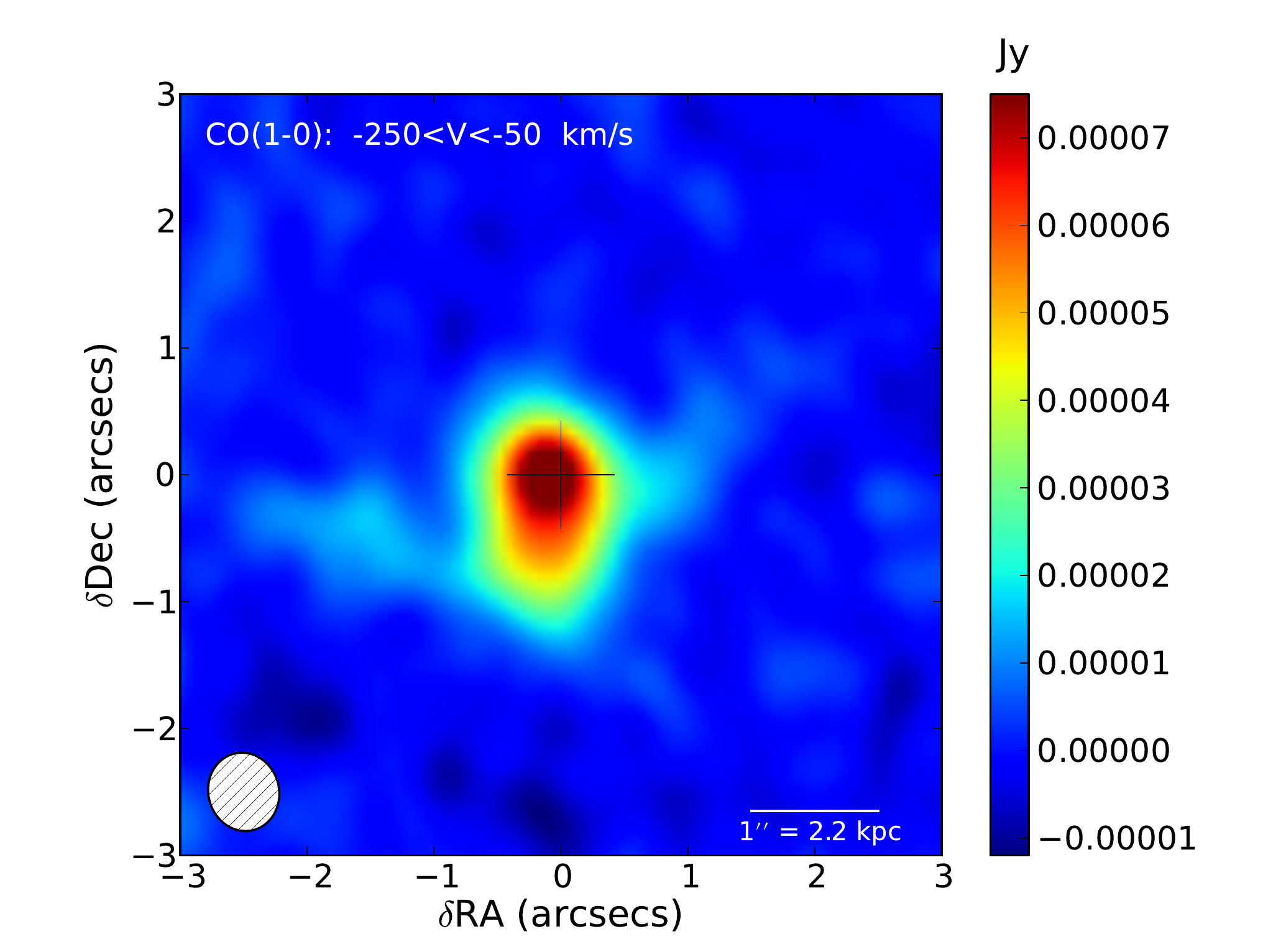}
                \includegraphics[width=6cm]{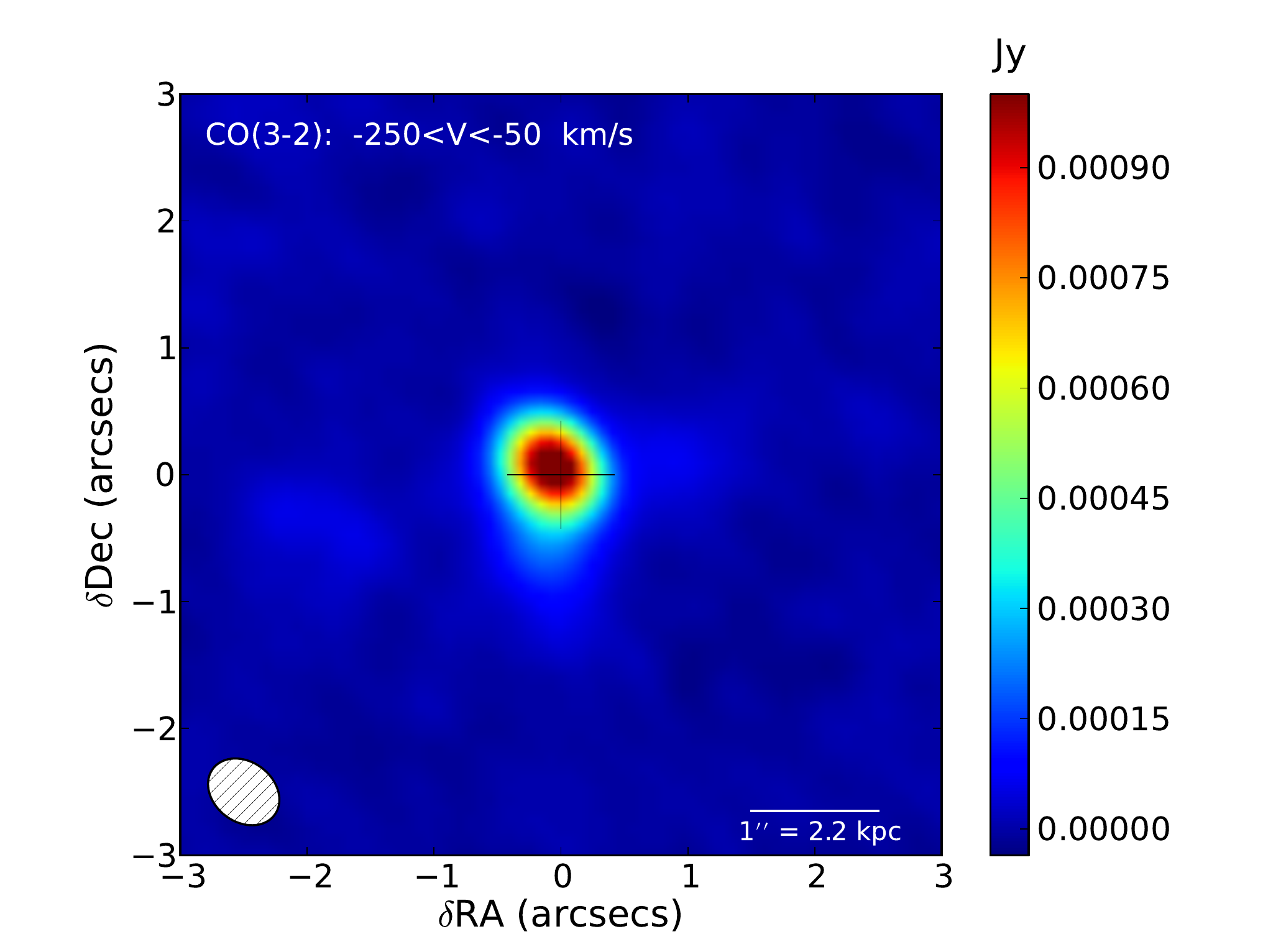}
                \includegraphics[width=6cm]{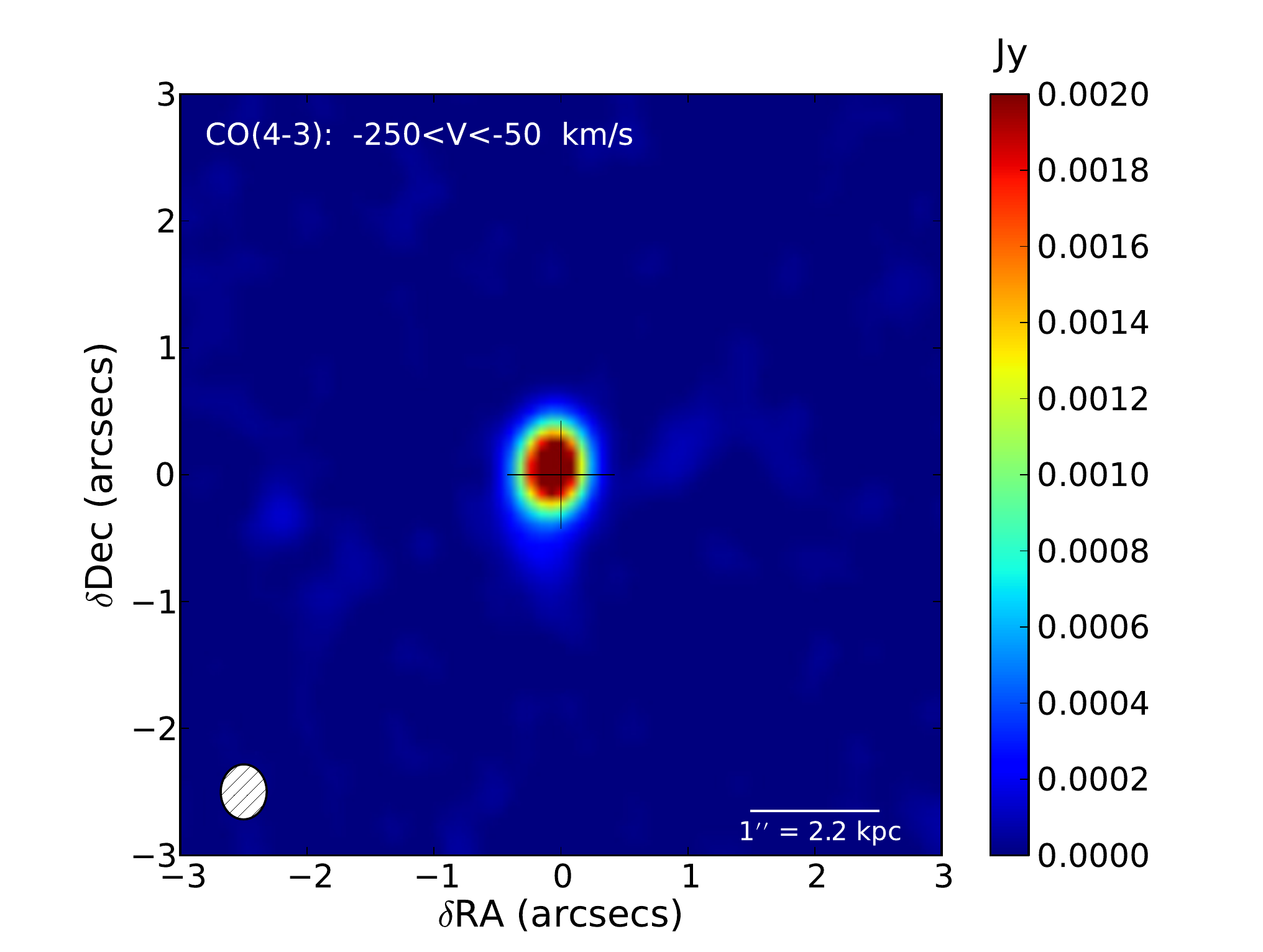}

                \includegraphics[width=6cm]{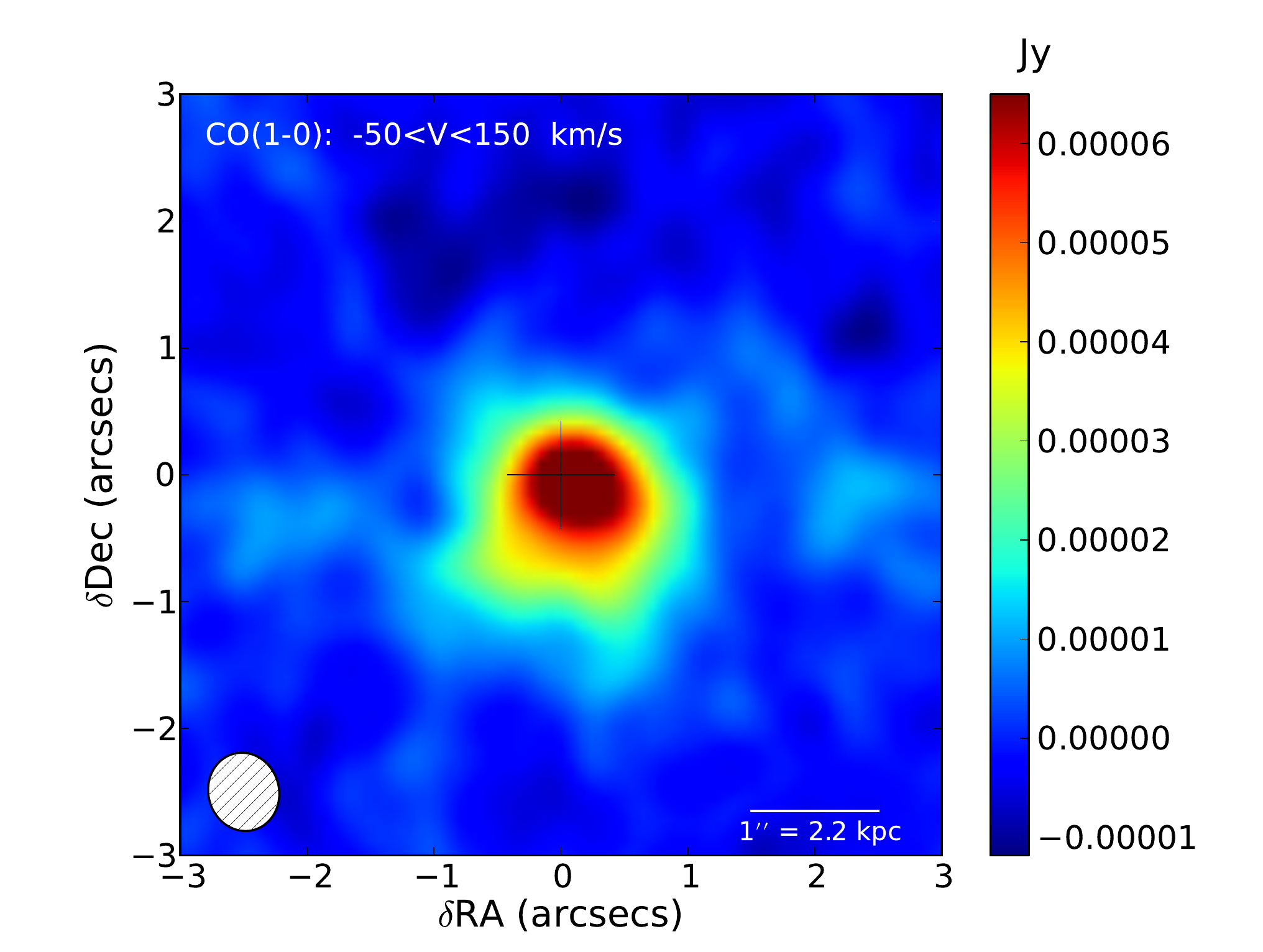}
                \includegraphics[width=6cm]{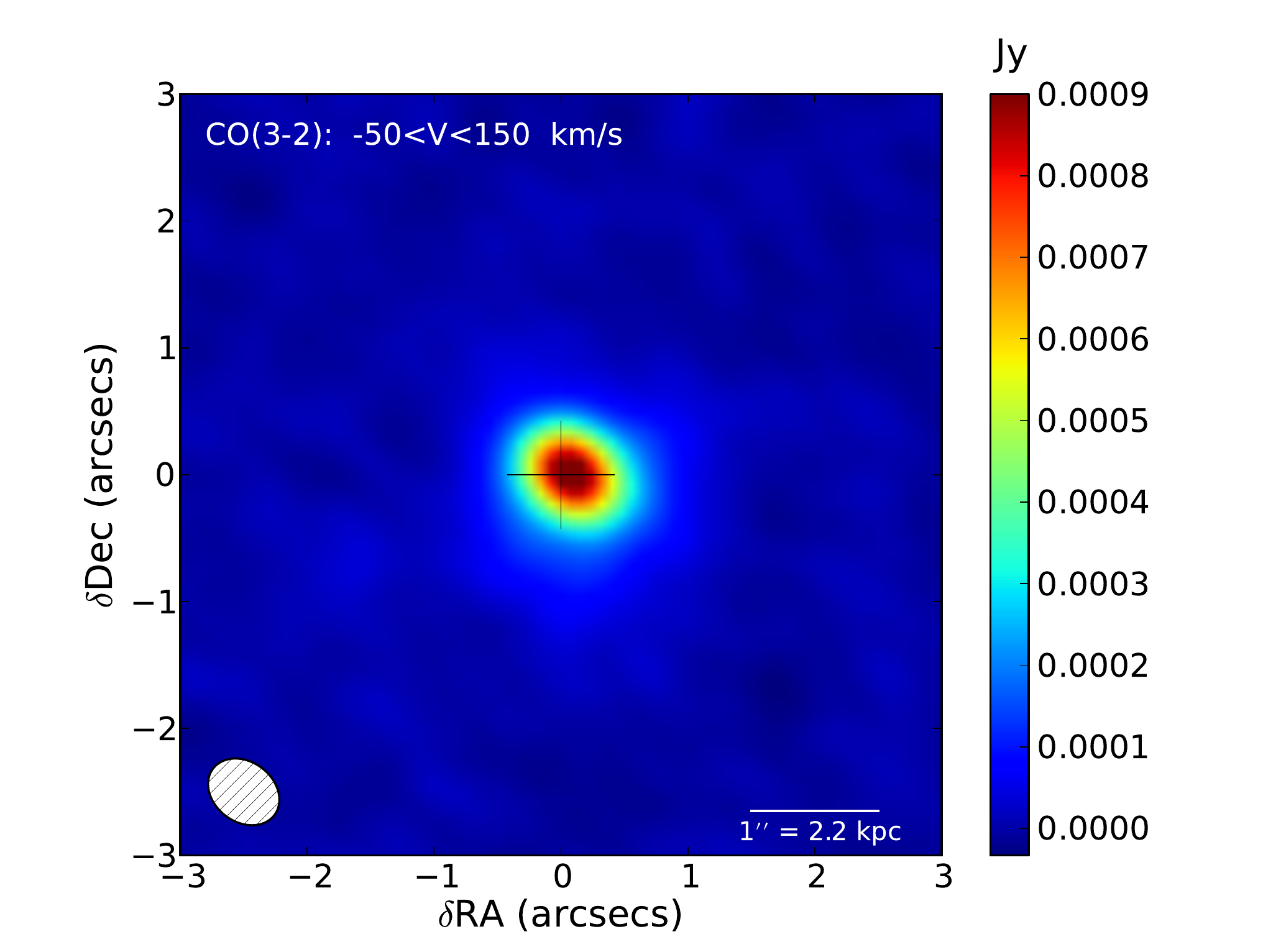}
                \includegraphics[width=6cm]{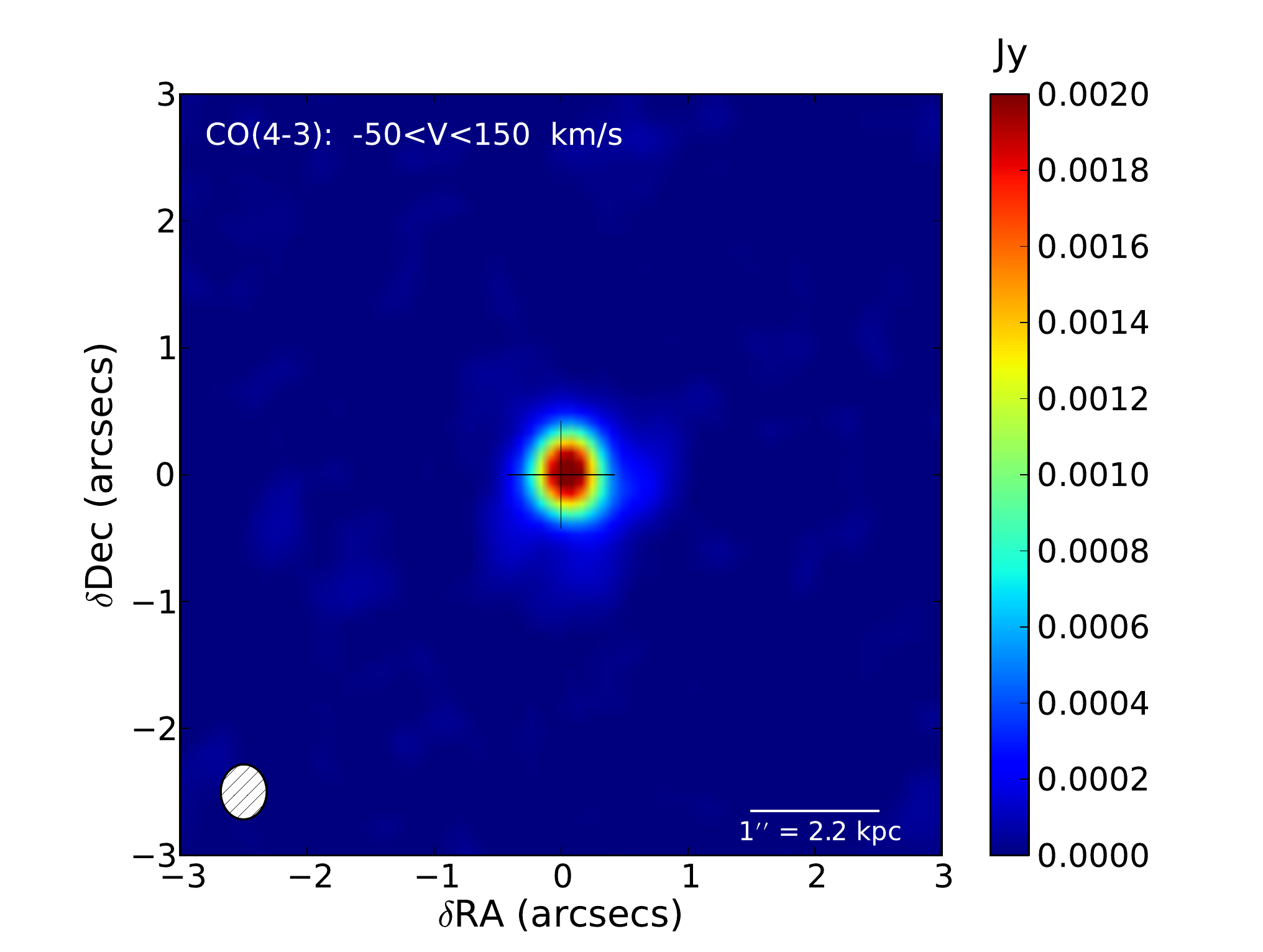}

                \includegraphics[width=6cm]{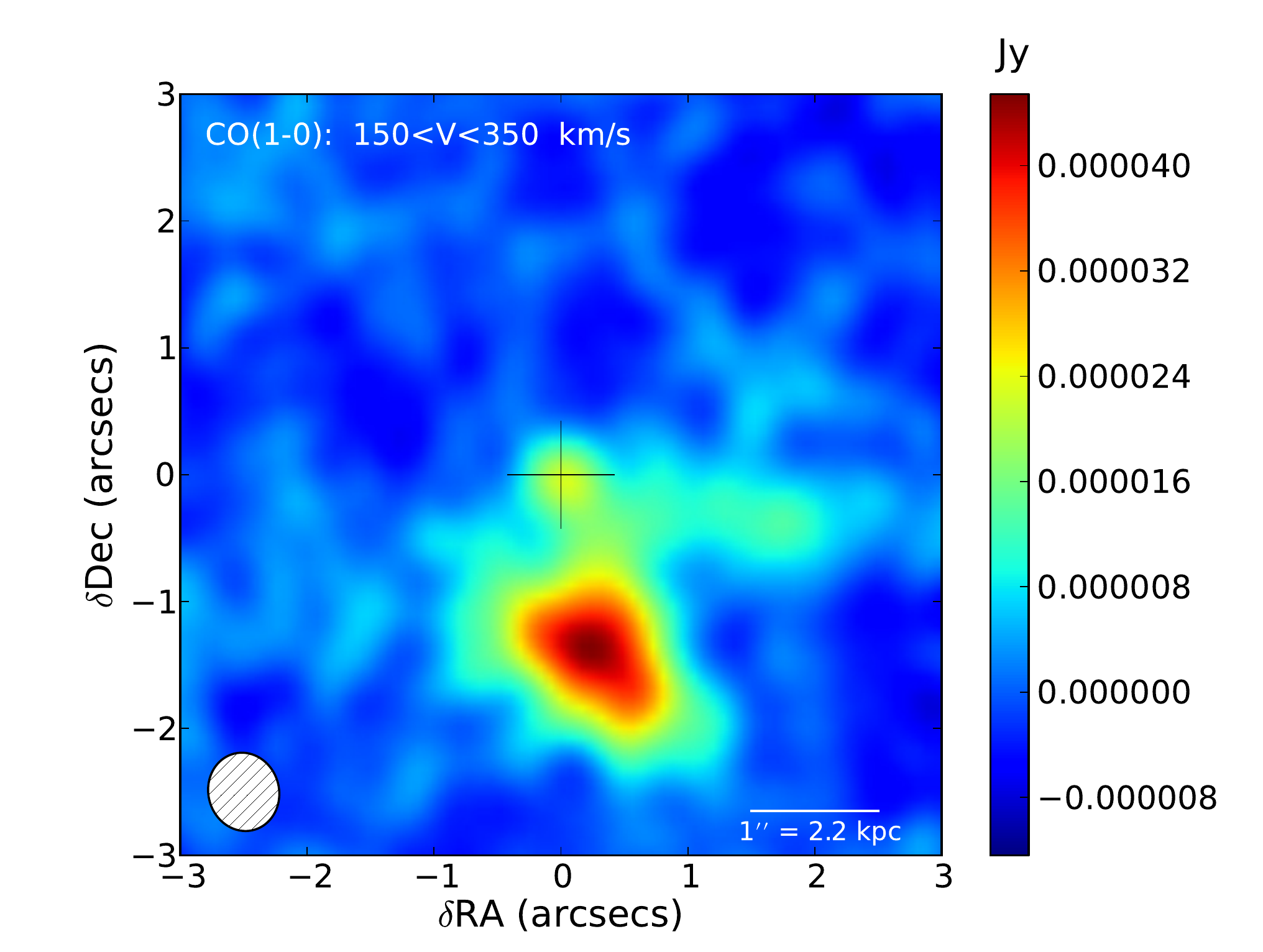}
                \includegraphics[width=6cm]{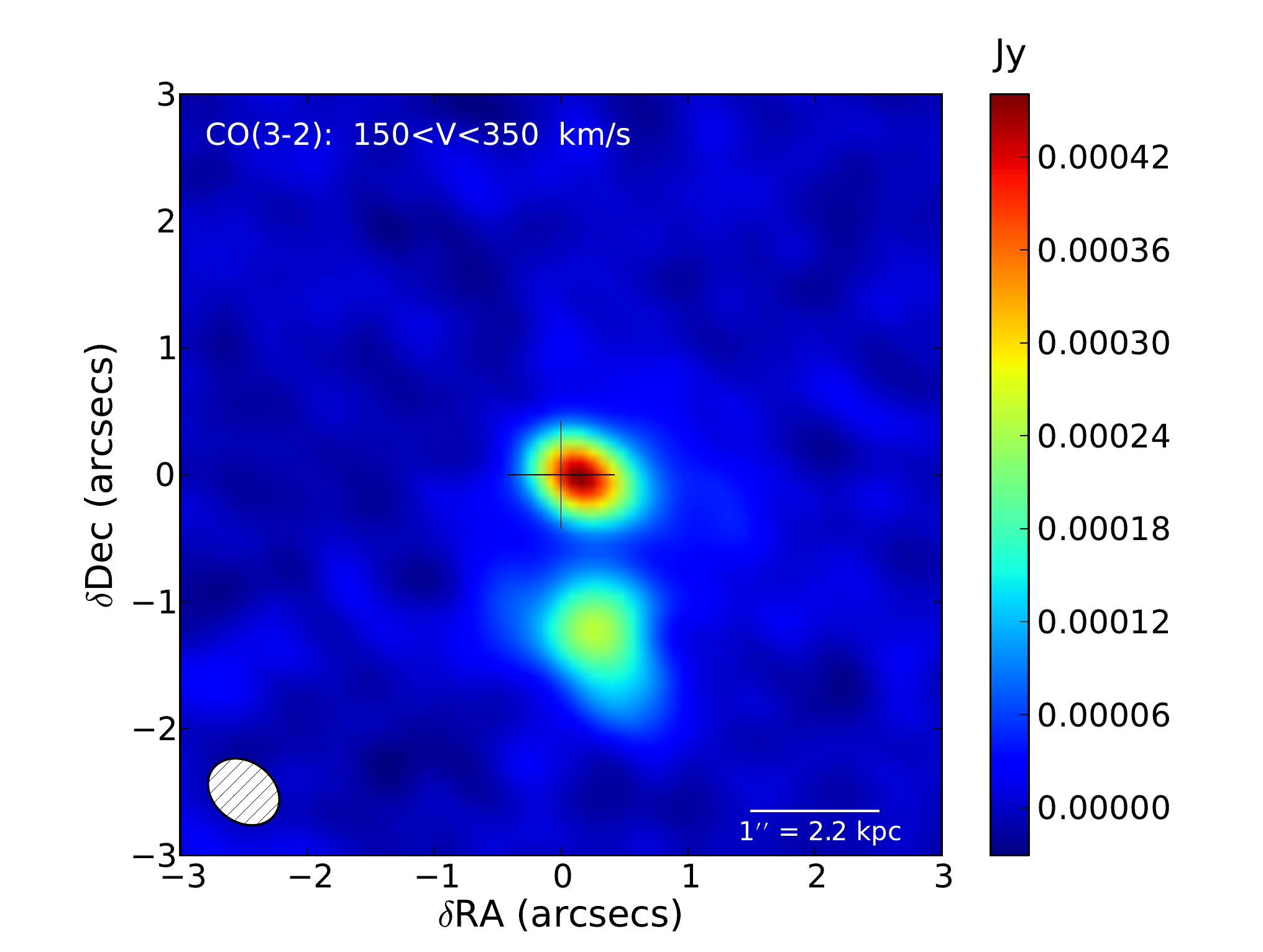}
                \includegraphics[width=6cm]{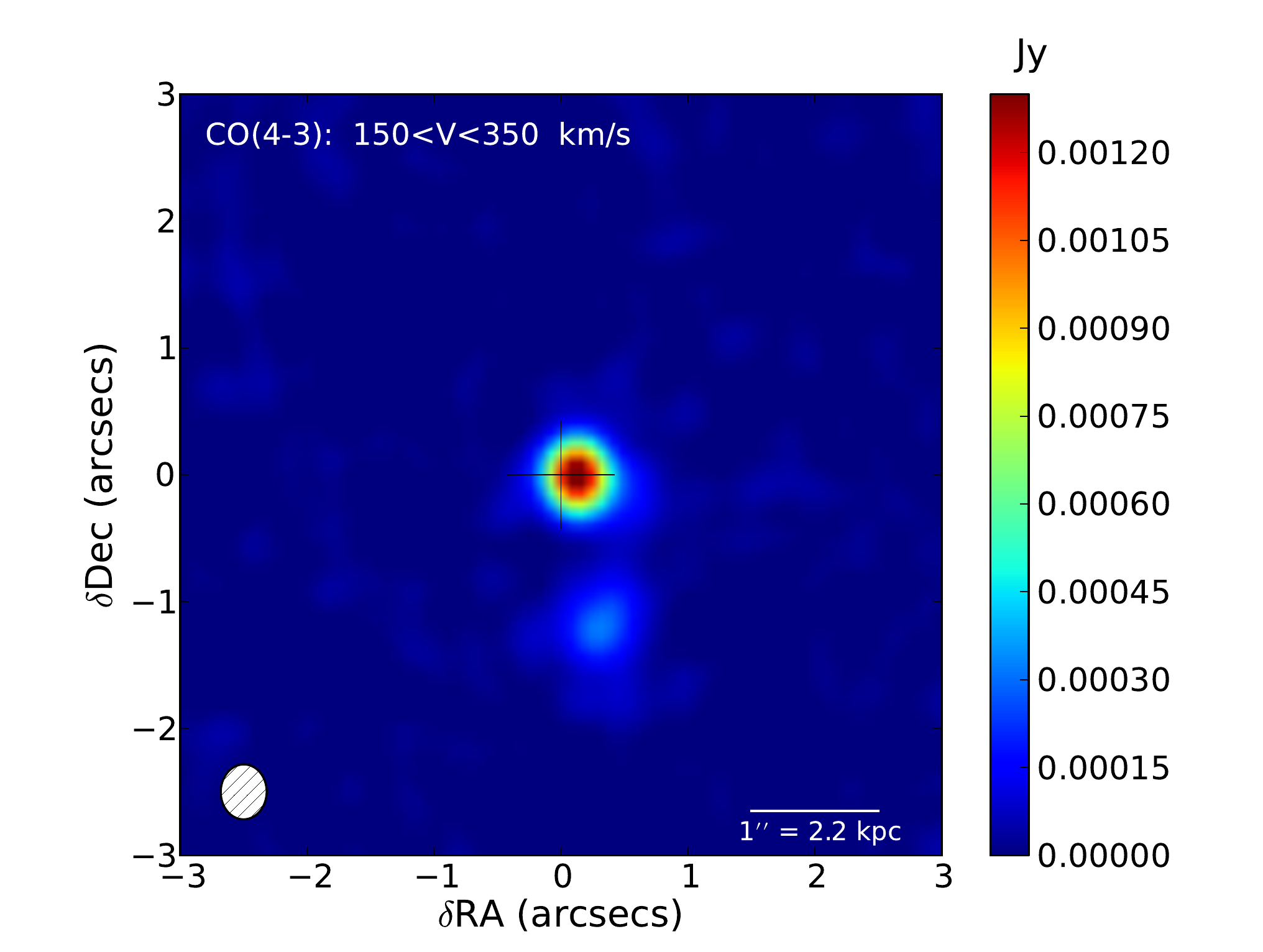}
                \includegraphics[width=6cm]{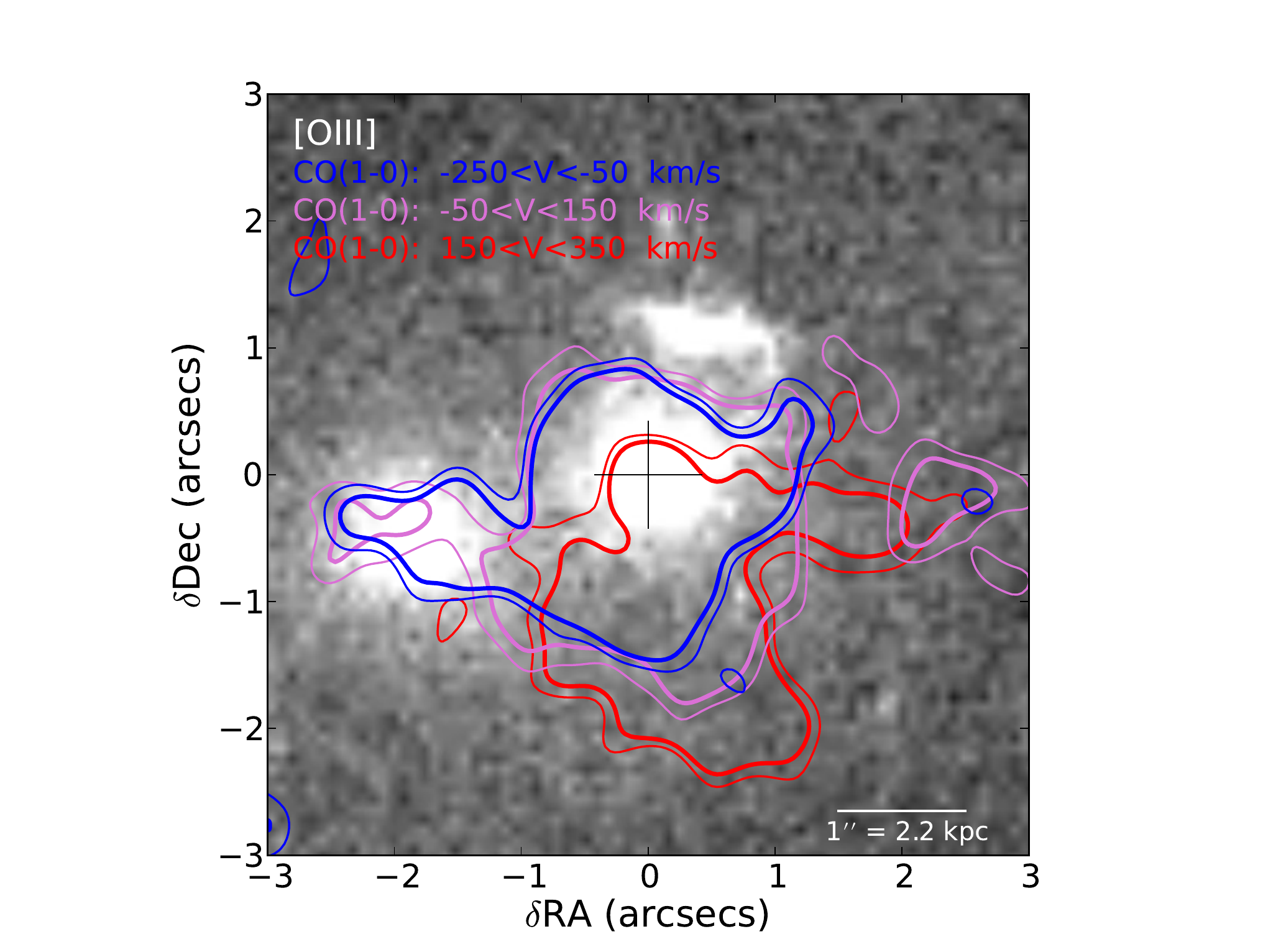}
                \includegraphics[width=6cm]{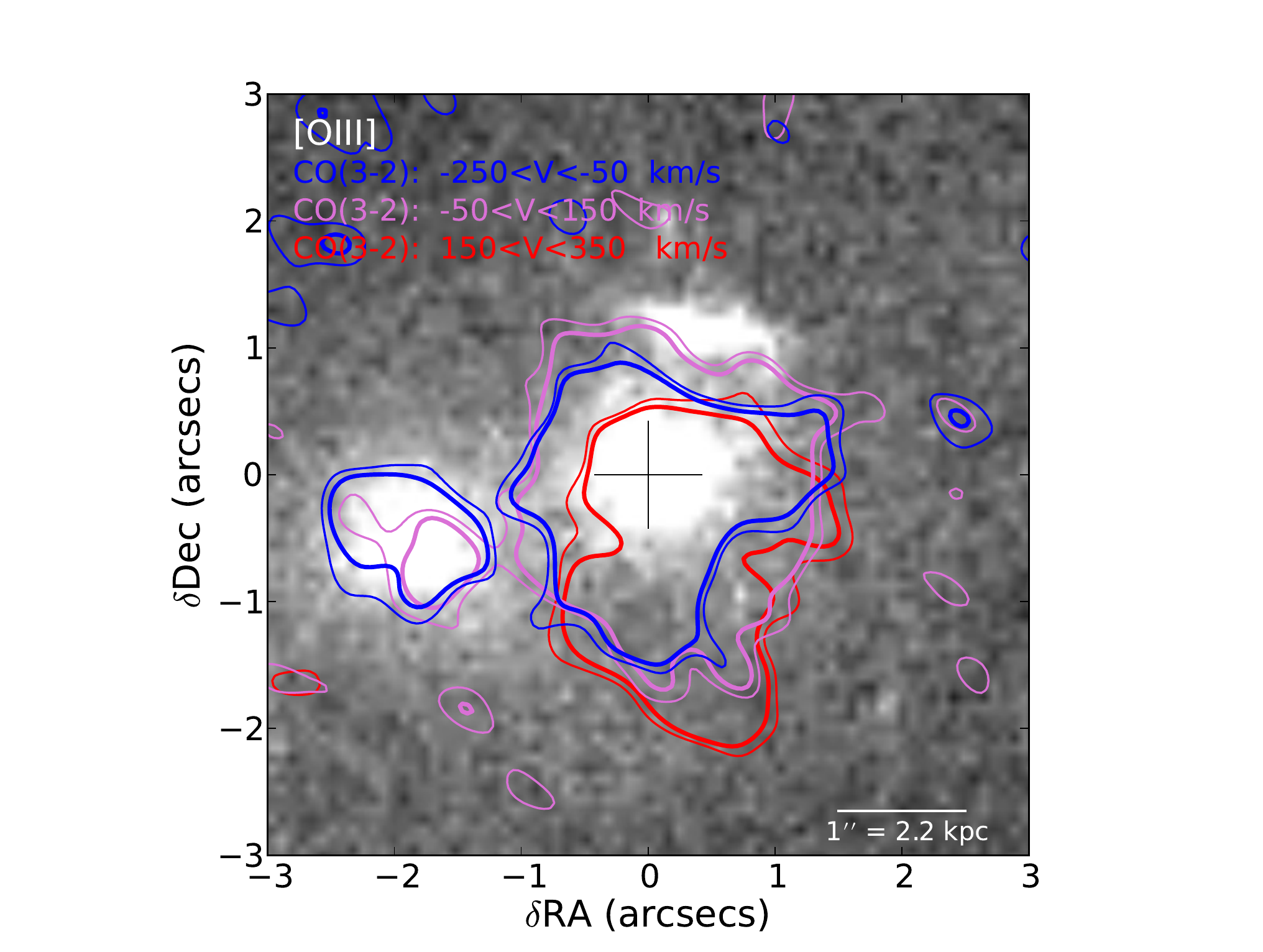}
                \includegraphics[width=6cm]{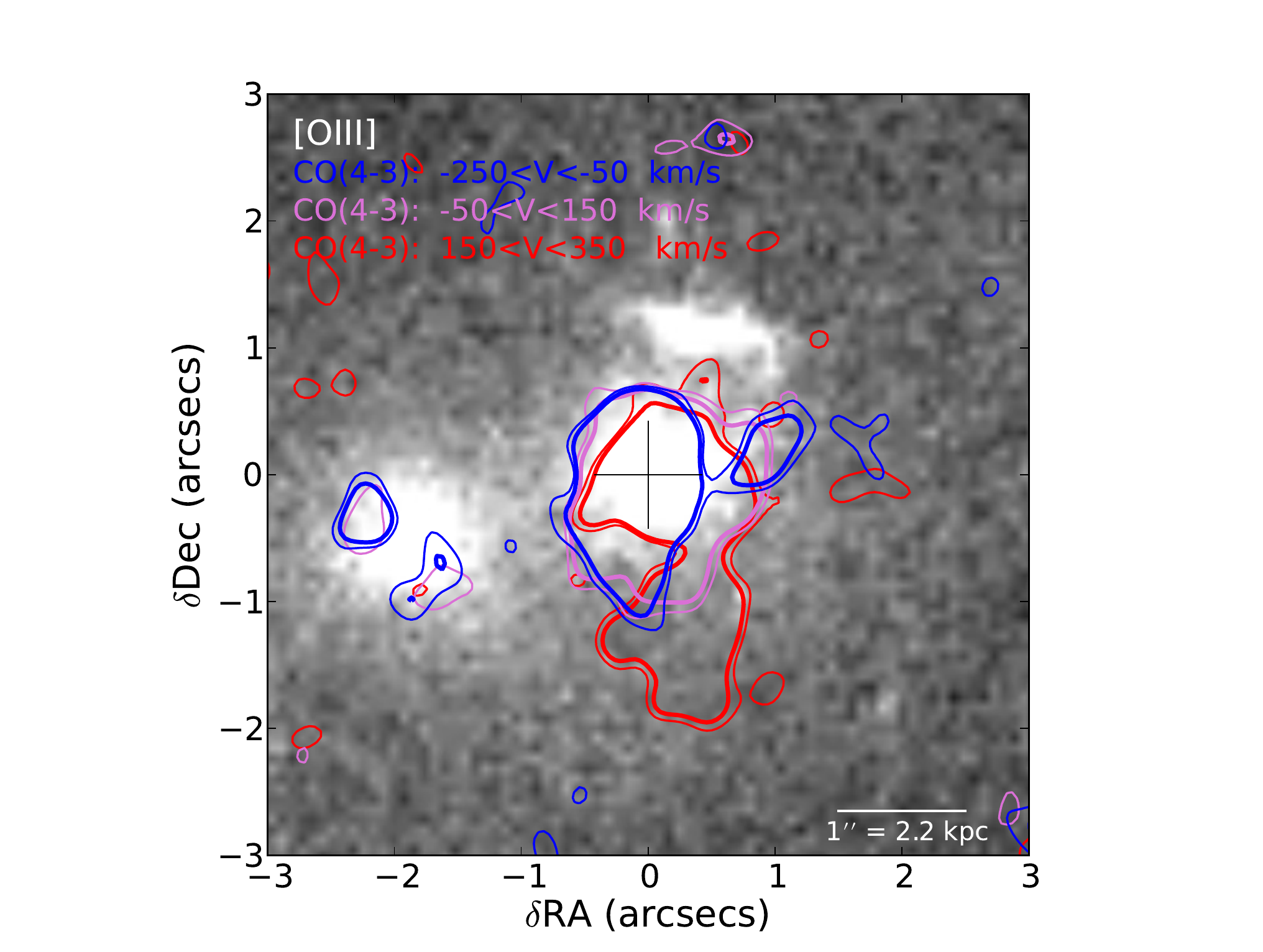}
                \caption[]{{\it Upper three panels:} \cooz, ~\cott ~and \coft\ emission in different velocity ranges. Filled ellipses show the beam size at each frequency. The same applies to all of the following figures. {\it Lower panel:} CO contours with levels 2$\sigma$ and 3$\sigma$ over the \oiii\ image in the optical showing the extent of the molecular gas near systemic velocity.}
                \label{fig:disk_detections}
        \end{center}
\end{figure*}

The CO data indicate an overall gas extent that exceeds  $\sim$12($\pm$2) kpc along the east--west axis (Fig.~\ref{fig:disk_detections}). Its shape is determined by the superposition of spatially resolved structures and its total mass in \htwo , as inferred from the high-resolution \cooz\ data, is 7.7($\pm$0.8)$\times$10$^9$ \msun (Fig. \ref{fig:galaxy_spectra}). The \htwo\ mass is computed as in \citet{solomon97}.
\begin{equation}\label{eq:mass}
M_{H_2}=3.25\times10^7 \alpha S _{CO(1-0)} \Delta V \nu_{obs}^{-2}D_L^2/(1+z) \msun ,
\end{equation}
where $\alpha$ is the CO intensity to \htwo\ mass conversion factor, 
$S _{CO(1-0)} \Delta V$ is the integrated line flux in Jy\kms , $\nu_{obs}$ is the observed \cooz\ frequency in GHz, and $D_L$ is the source luminosity distance in megaparsec.  Throughout our work we adopted an $\alpha$ value of 0.8\msun/(K\kms pc$^2$) for consistency with the literature assuming a lower conversion factor for ULIRGs than for the Milky Way \citep{downes98}.

\begin{figure*}
        \begin{center}
                \includegraphics[width=6cm]{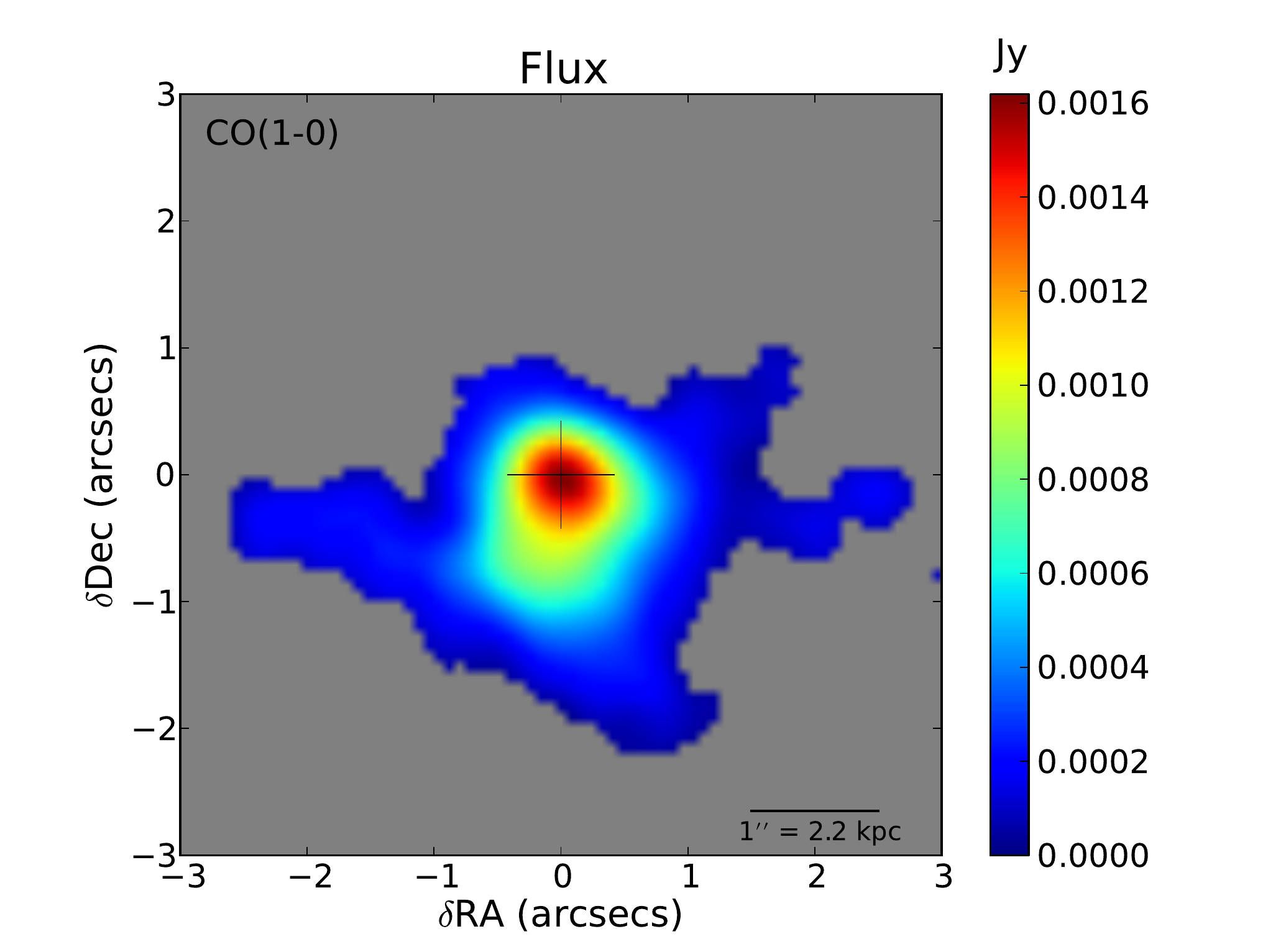}
                \includegraphics[width=6cm]{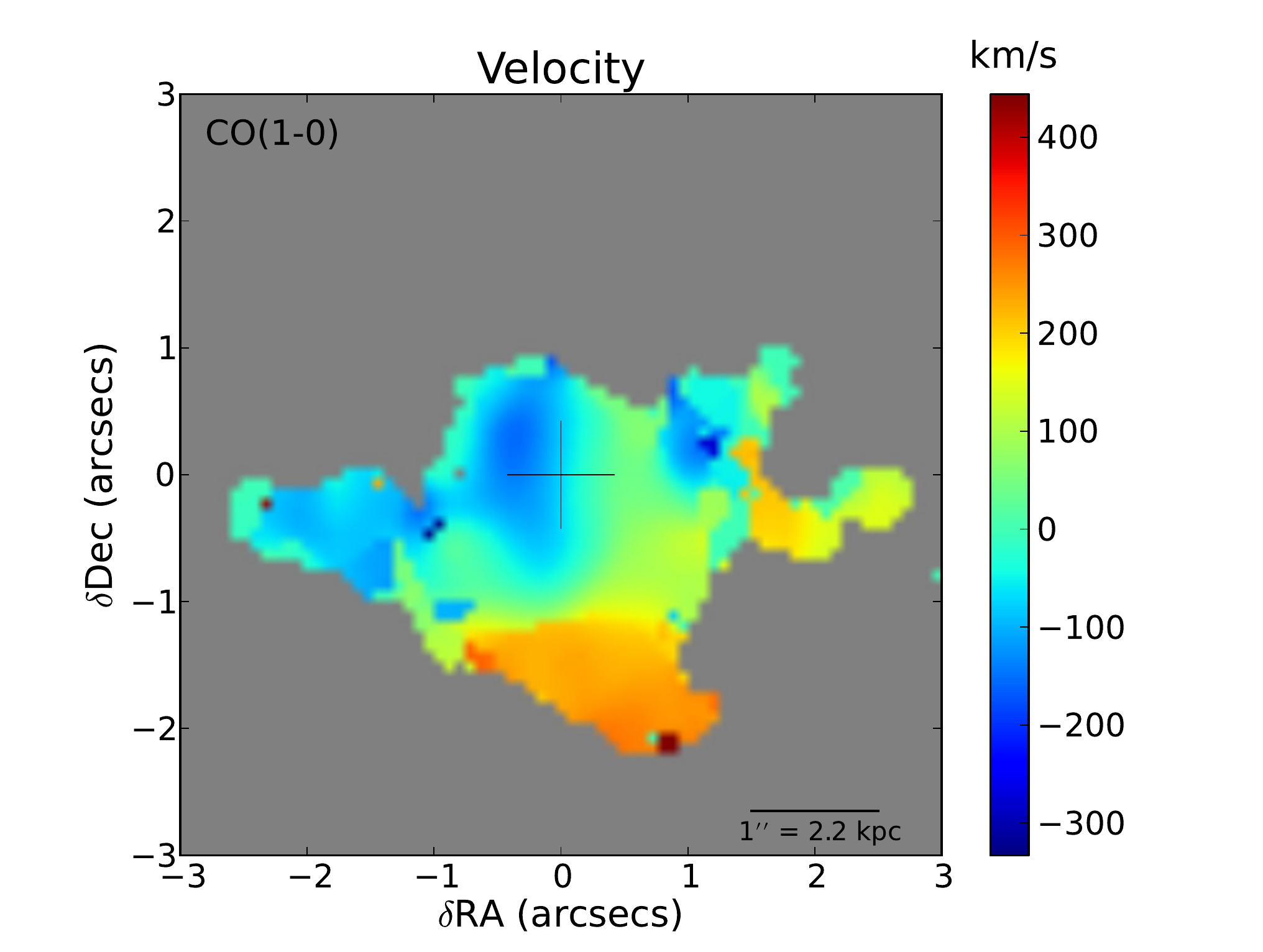}
                \includegraphics[width=6cm]{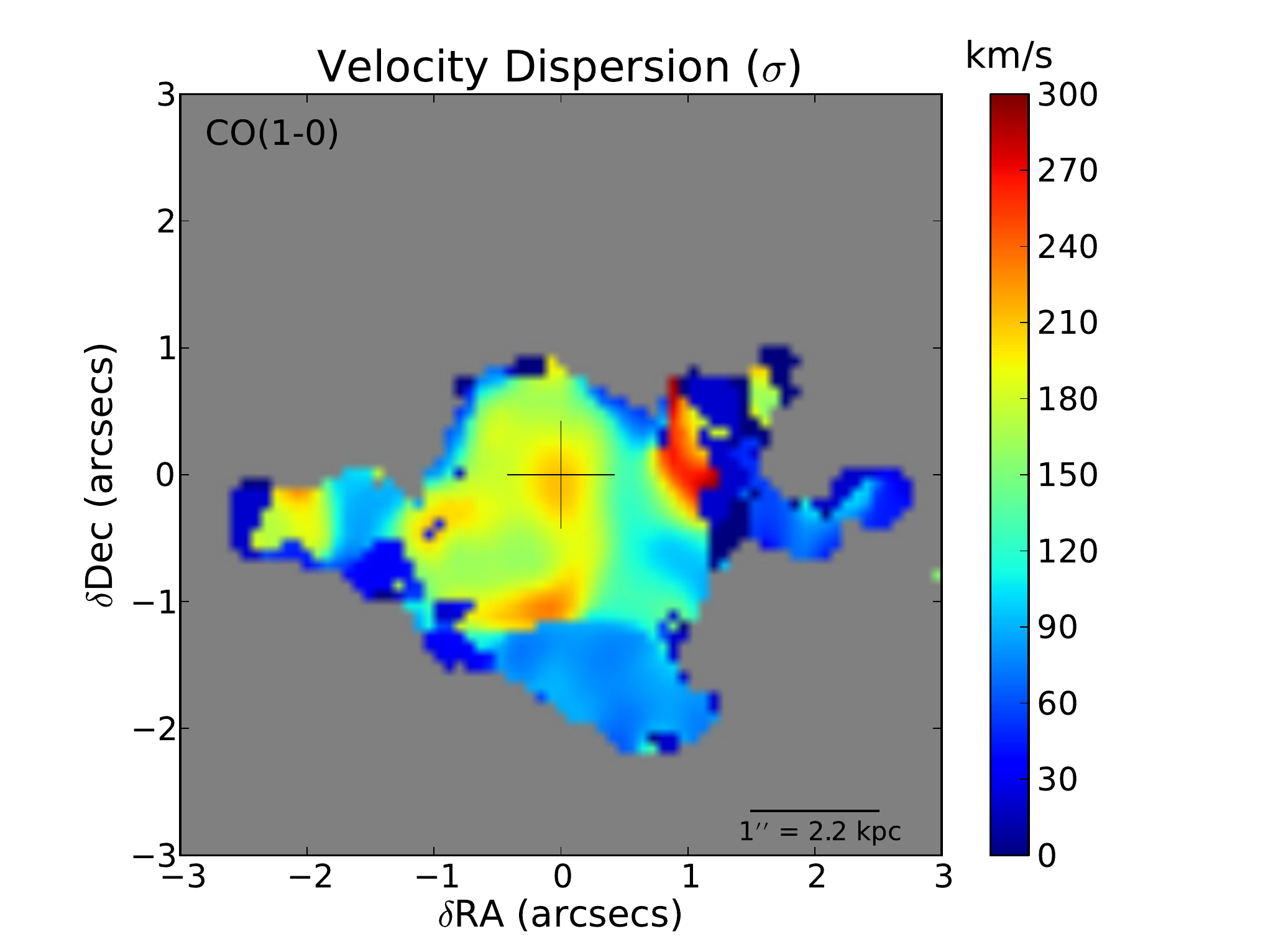}

                \includegraphics[width=6cm]{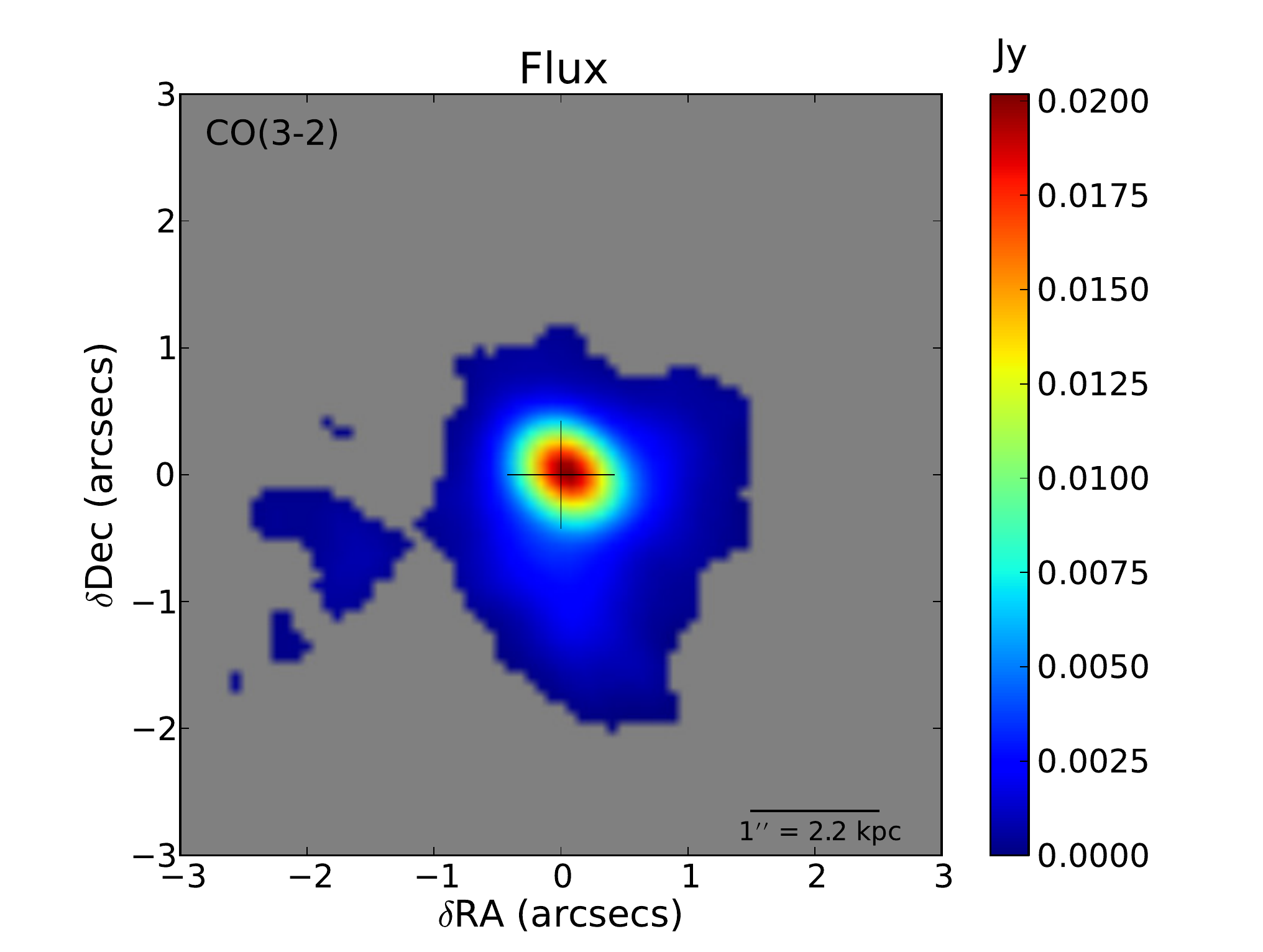}
                \includegraphics[width=6cm]{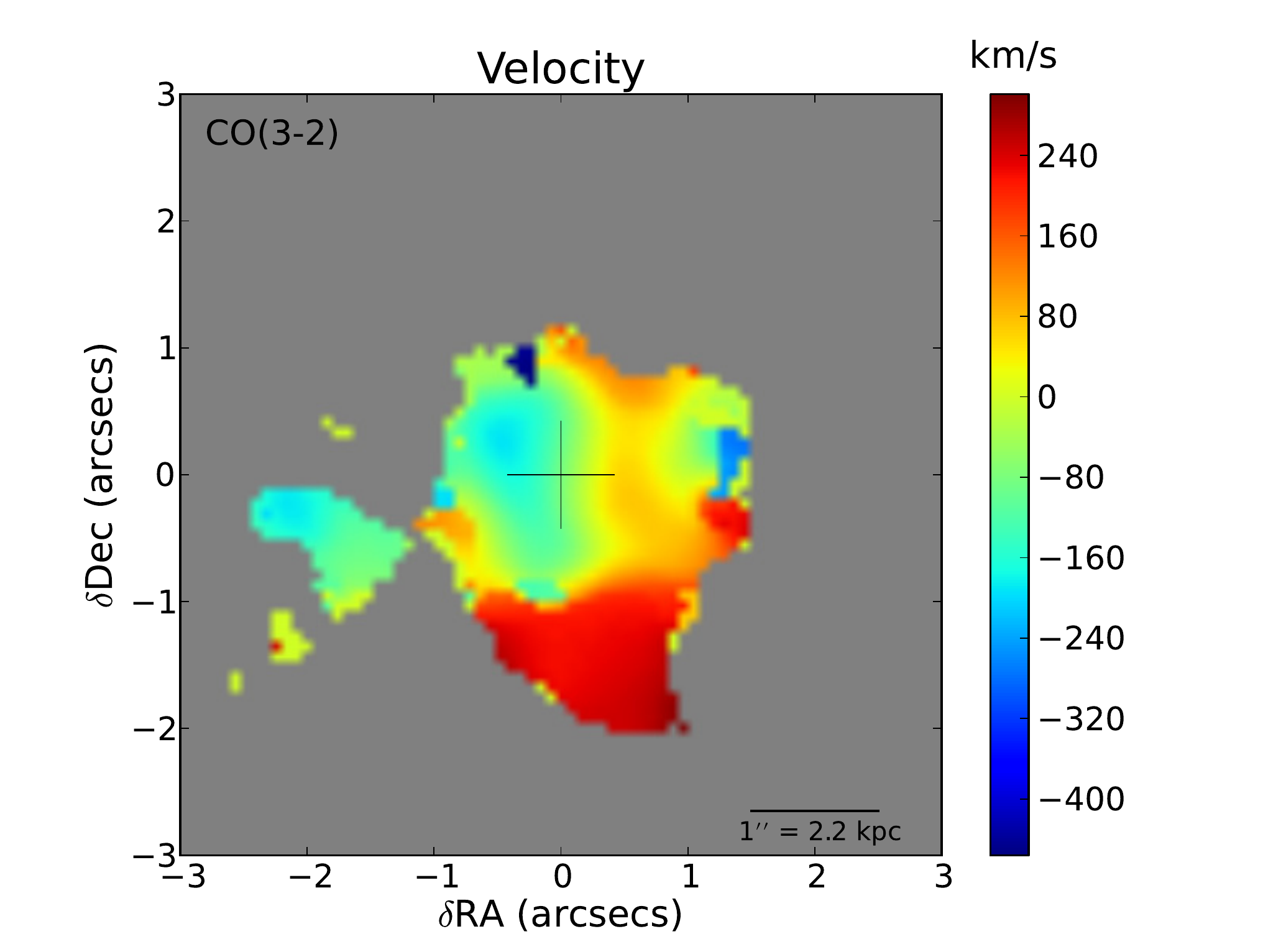}
                \includegraphics[width=6cm]{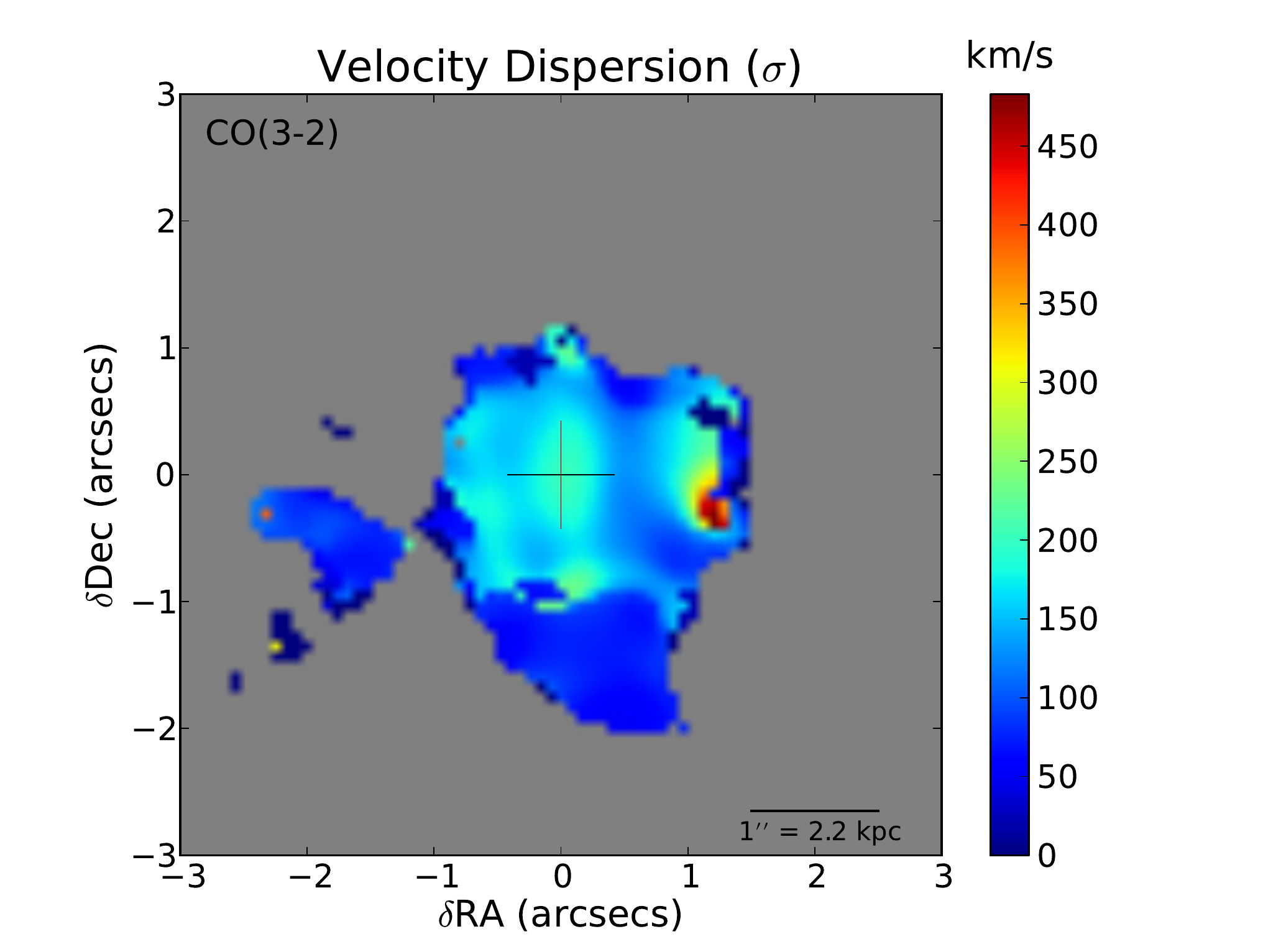}

                \includegraphics[width=6cm]{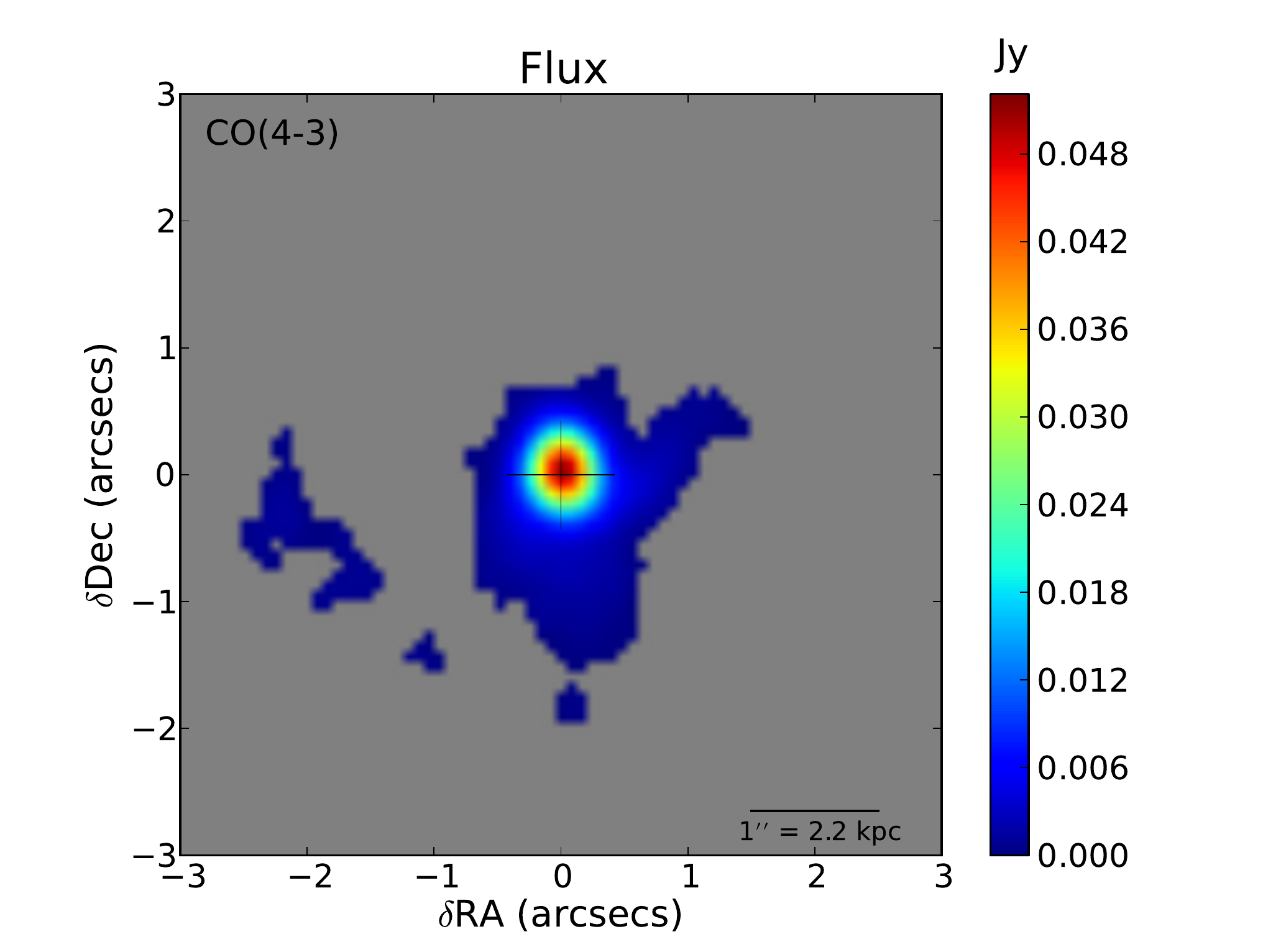}
                \includegraphics[width=6cm]{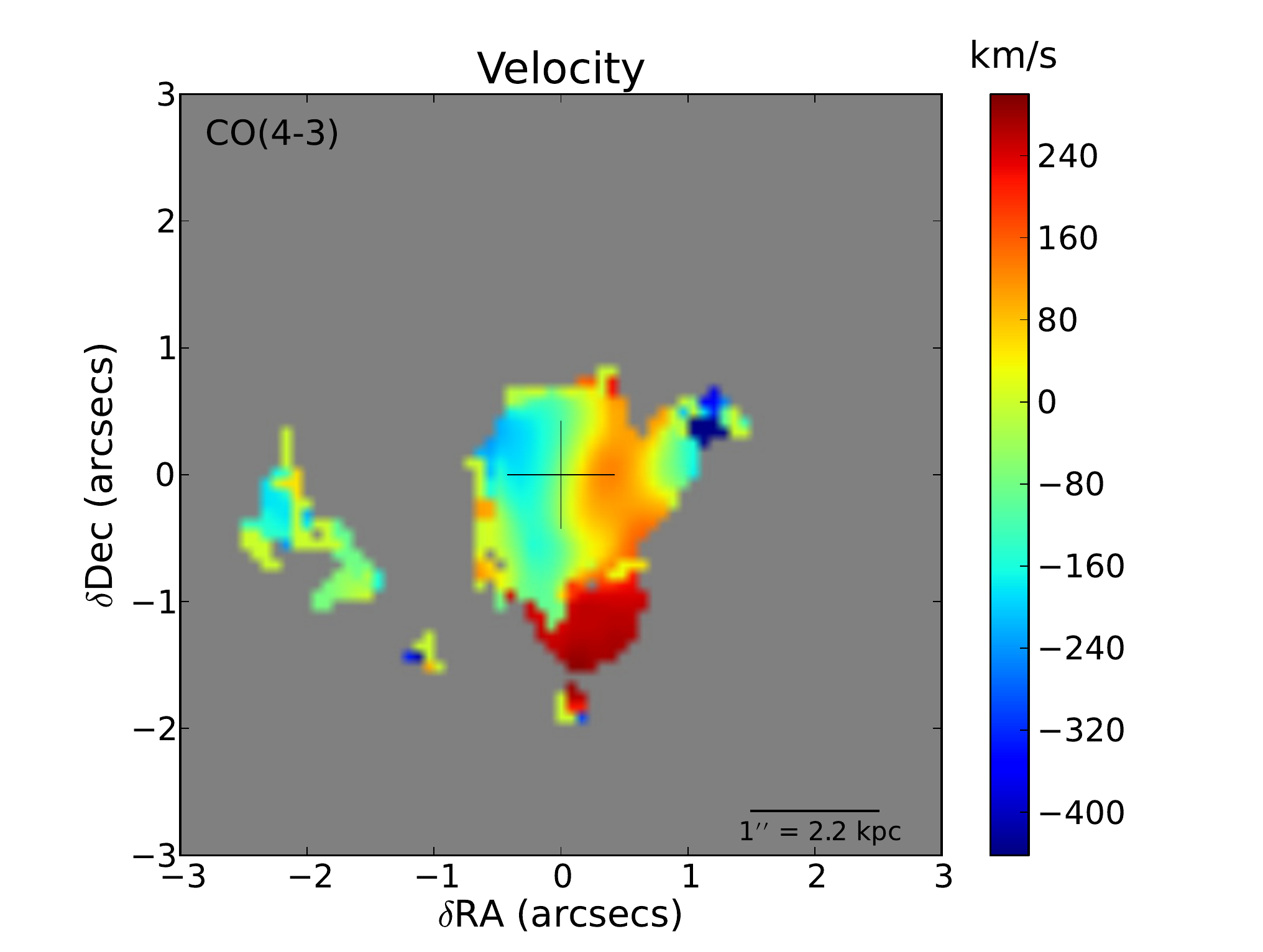}
                \includegraphics[width=6cm]{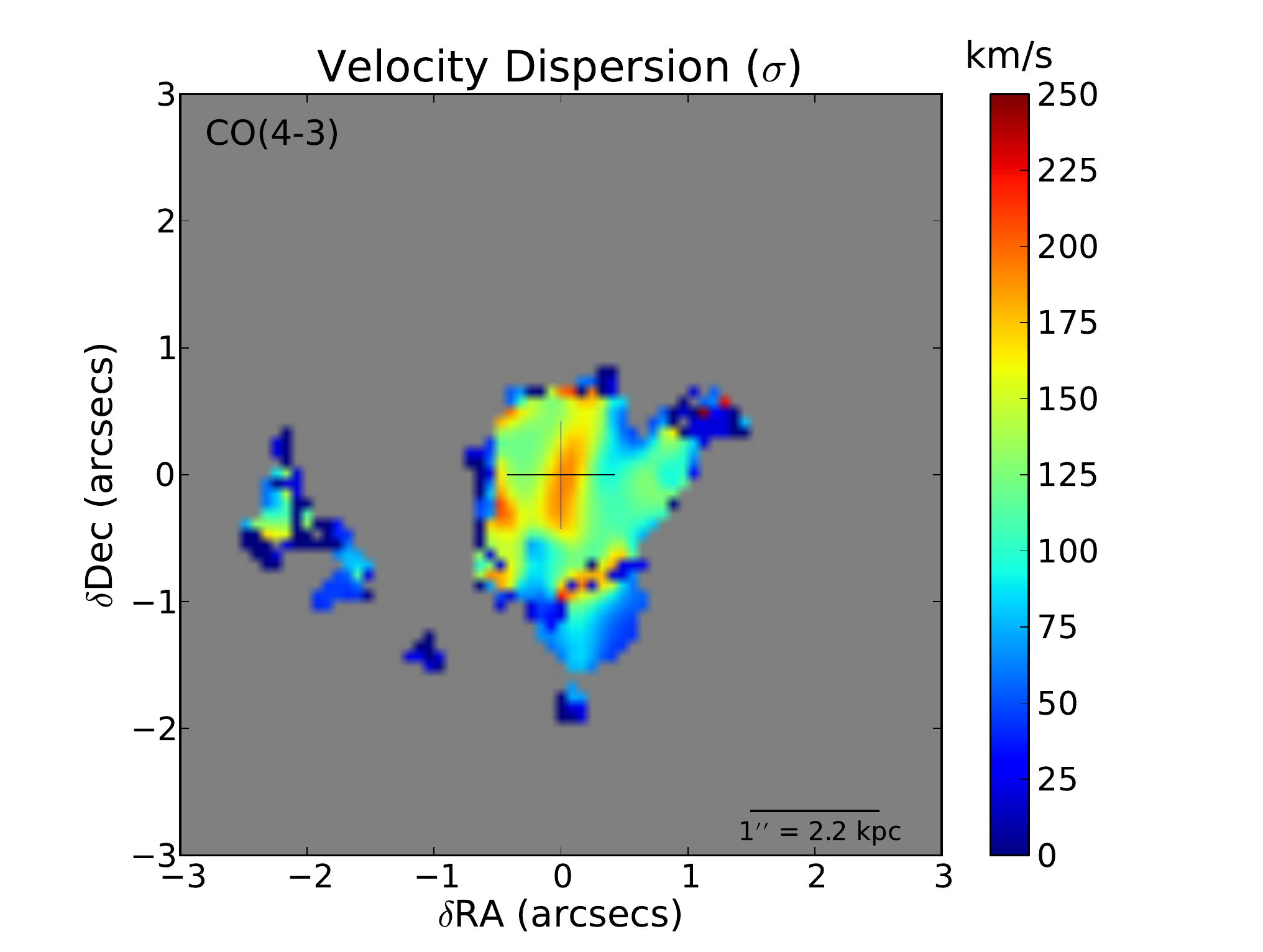}           
                \caption[]{\cooz, ~\cott ~and \coft ~momenta maps for pixels with flux exceeding three times the noise root mean square. {\it Left:} Flux. {\it Middle:} Velocity. {\it Right:} Velocity dispersion.}
                \label{fig:momenta}
        \end{center}
\end{figure*}

A dynamically settled disk is detected around the main nucleus, from which the radio jet is launched. Its \htwo\ mass within 0.8\arcsec\ (radius 1.8kpc) is 3.8($\pm$0.4)$\times$10$^9$\msun. Compared to the \cooz\ emission, the \cott\ or the \coft\ emission in the disk is more nucleated (Fig.~\ref{fig:disk_detections}). In all CO lines, the disk is centered at the main nucleus, and it follows a progression from the northeast to the southwest with increasing velocity (from negative to positive). Depending on the emission line and the galaxy side examined for the kinematics, the  projected disk circular velocity is 150-200\kms . The disk rotation pattern is best revealed by the fitting of a Gaussian line profile in every spatial pixel of the ALMA data (see Fig.~\ref{fig:momenta} for the momenta maps). The rotation is clearly seen for the inner 2 kpc, and it agrees with the results of \citet{imanishi16,imanishi18} for comparable or smaller scales. Counter-rotating blobs or high-velocity dispersion regions are also seen in the momenta maps, for example to the west of the nucleus. 

A blueshifted stream from the primary nucleus reaches, in the plane of the sky, the secondary nucleus (see Fig. ~\ref{fig:disk_detections}; -300$<$V$<$-100\kms\ panel). This stream could either be a true bridge or a tail of the main nucleus in front of the secondary nucleus,  similarly to the tail in projection between, for example, M51 and NGC5195 \citep{toomre72}. In this scenario, the system originates from the major merger of a gas-rich galaxy with an elliptical galaxy that has a (potentially small) disk. The companion elliptical has gone through a series of turn-arounds and pericenters, creating loops and spiral extensions at each pericenter. Another scenario is that the system originates from the major merger of two gas-rich galaxies with bulges, and that gas stripping led to a ratio of secondary-disk flux over primary-disk flux that is low for the \oiii\ and even lower for the CO. 

\subsection{Galactic components (out of dynamical equilibrium) south of the primary nucleus}
\label{sec:southern_structure}
\begin{figure*}
        \begin{center}
                \includegraphics[width=6cm]{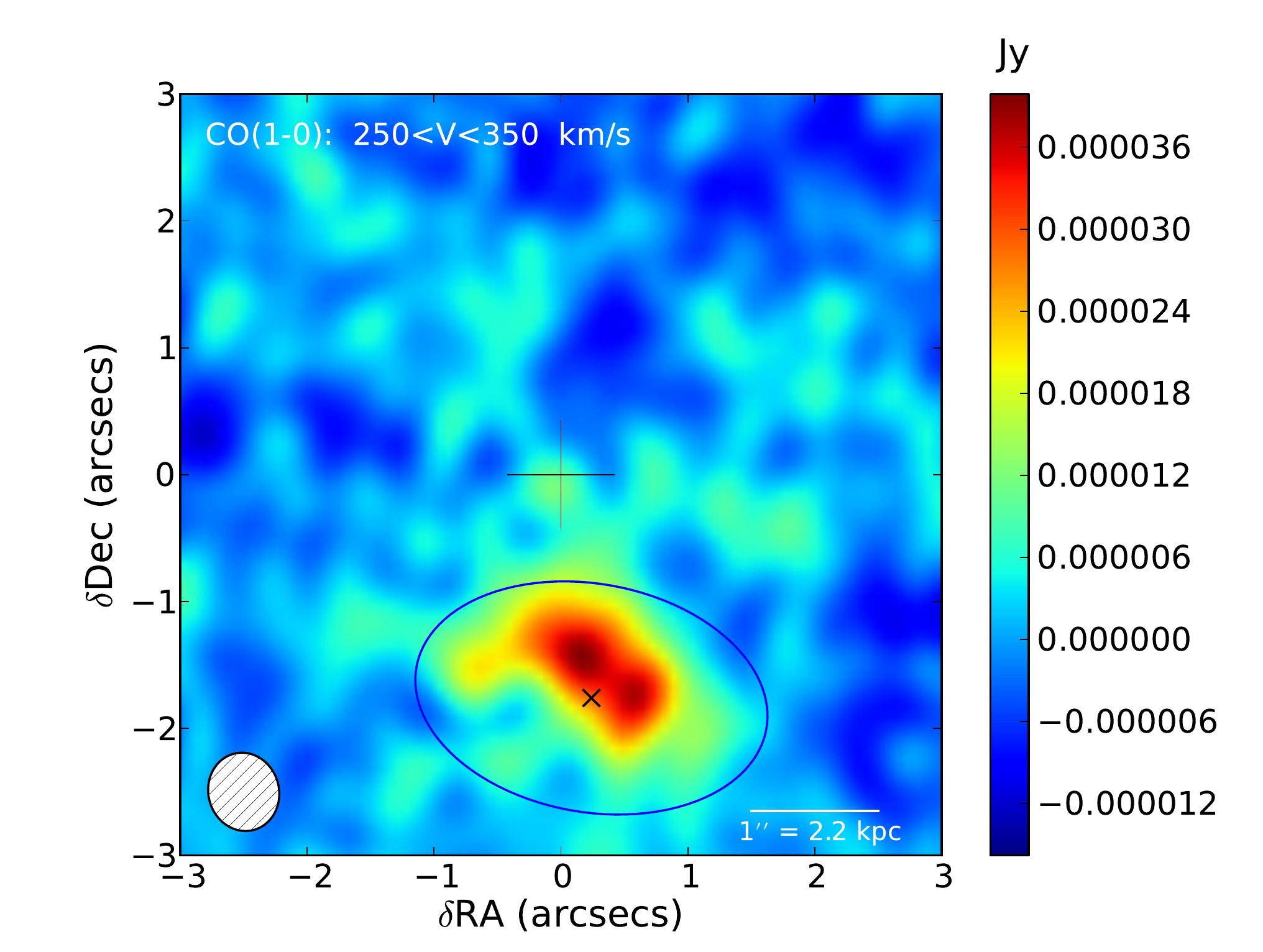}             
                \includegraphics[width=6cm,height=4.5cm]{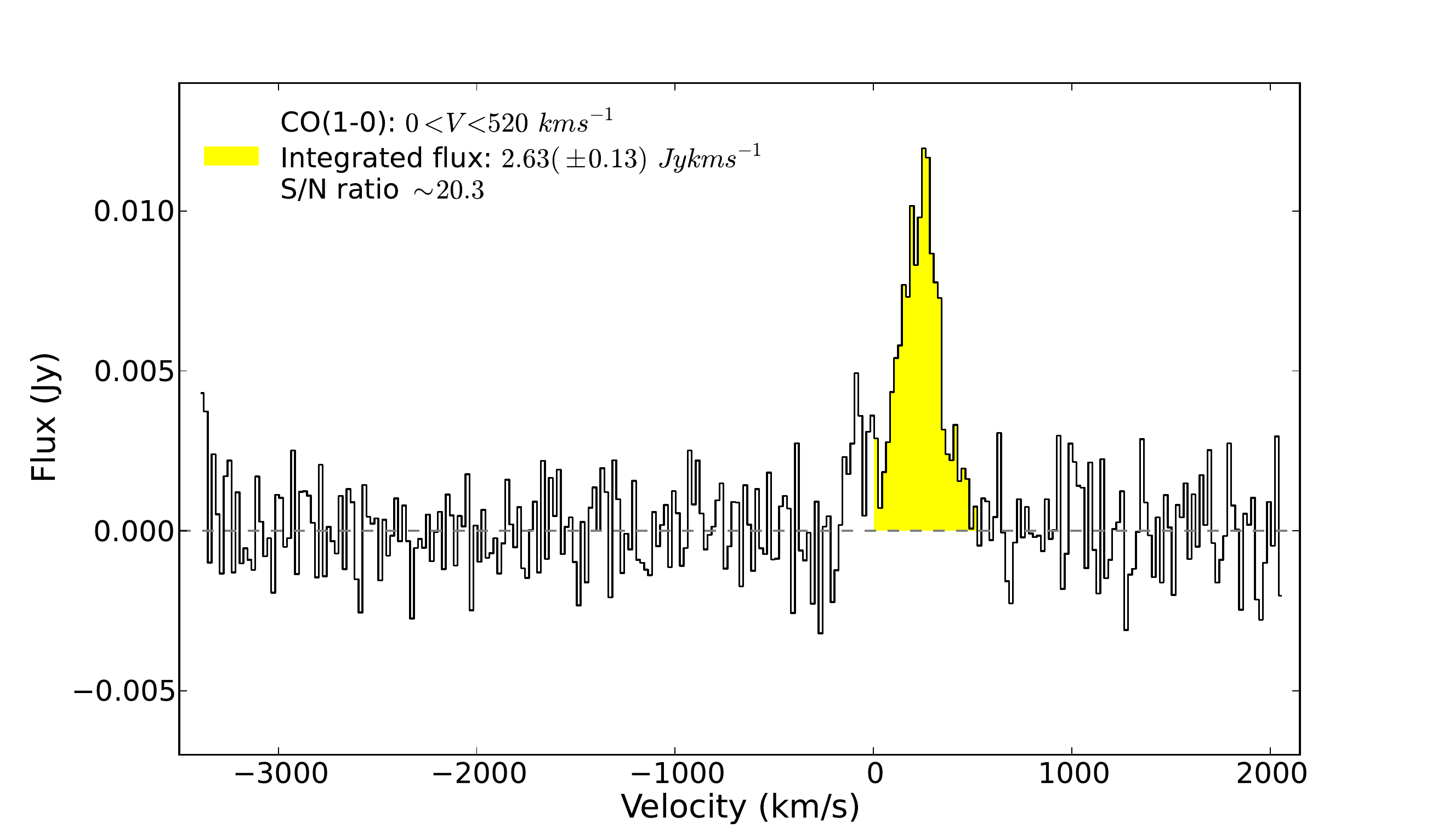}
                \includegraphics[width=6cm]{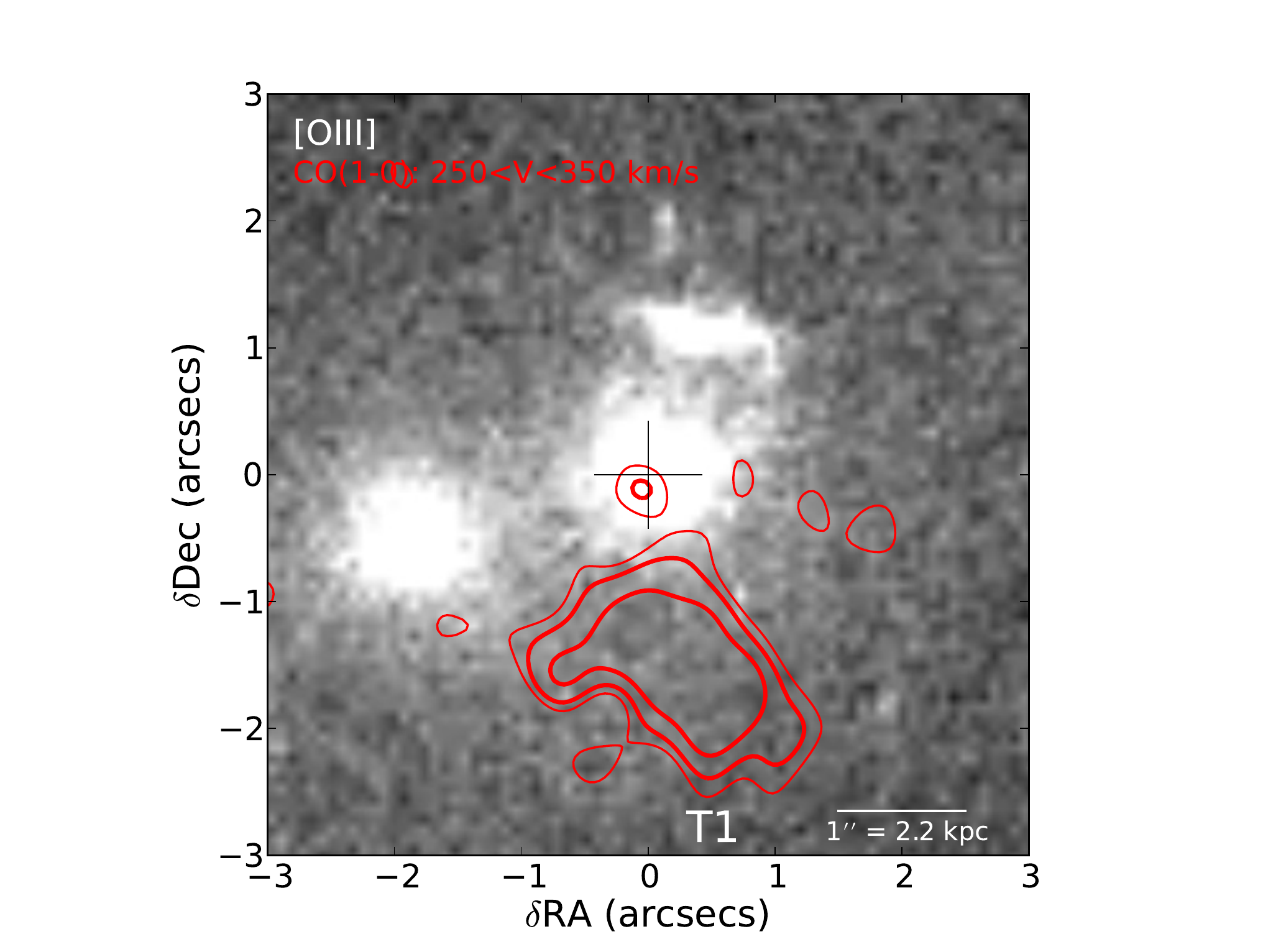}          
                \includegraphics[width=6cm]{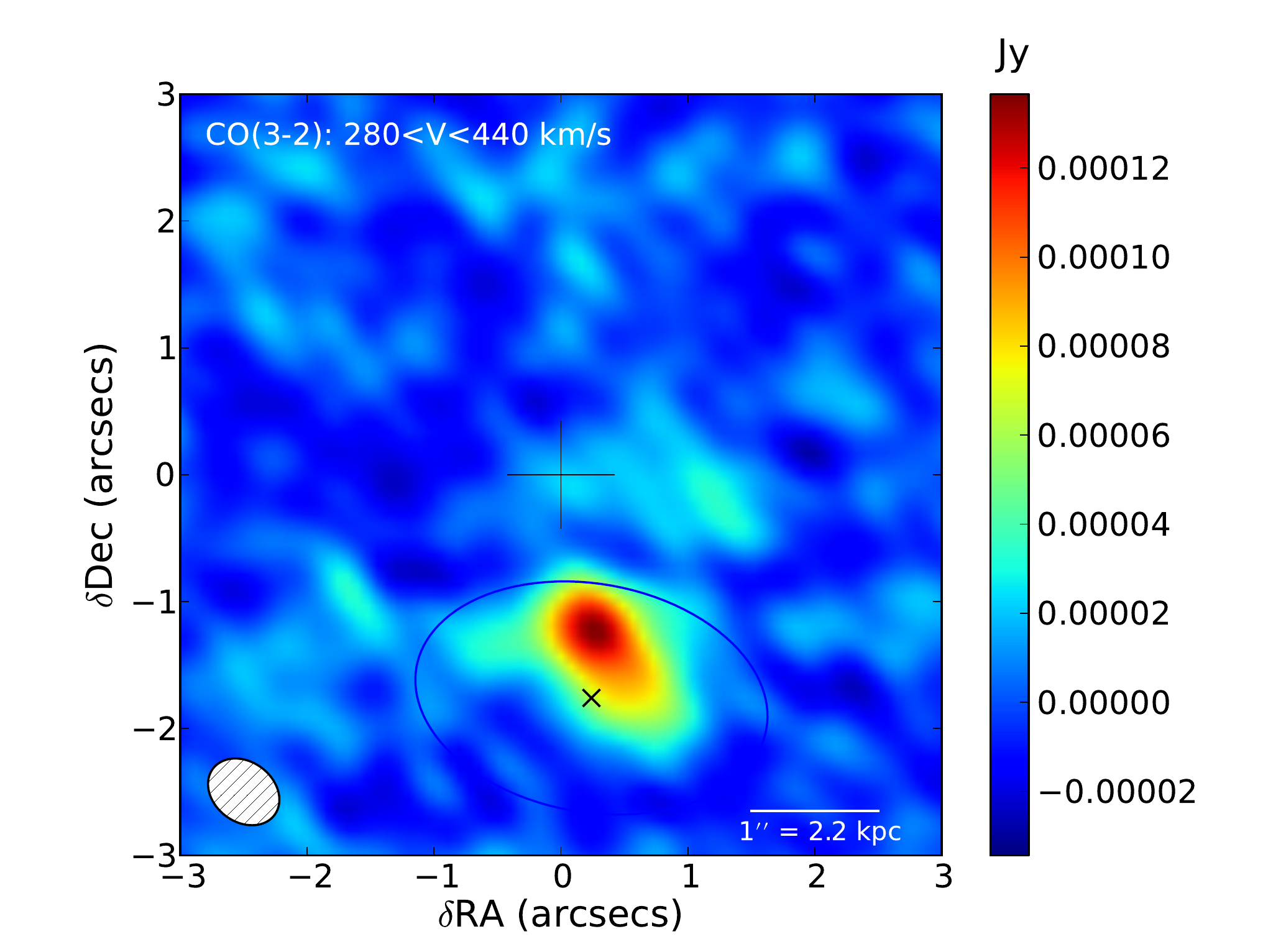}
                \includegraphics[width=6cm,height=4.5cm]{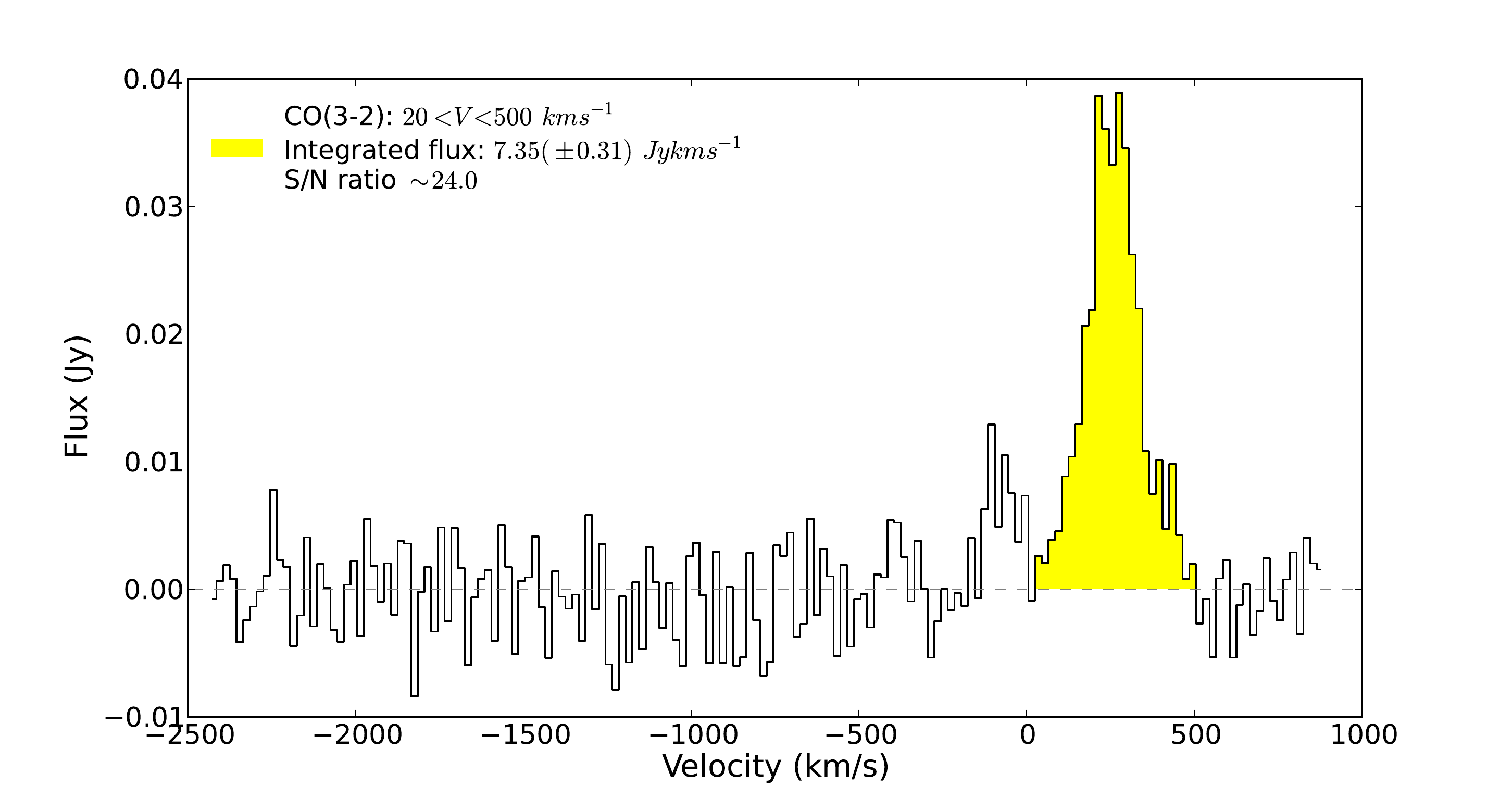}
                \includegraphics[width=6cm]{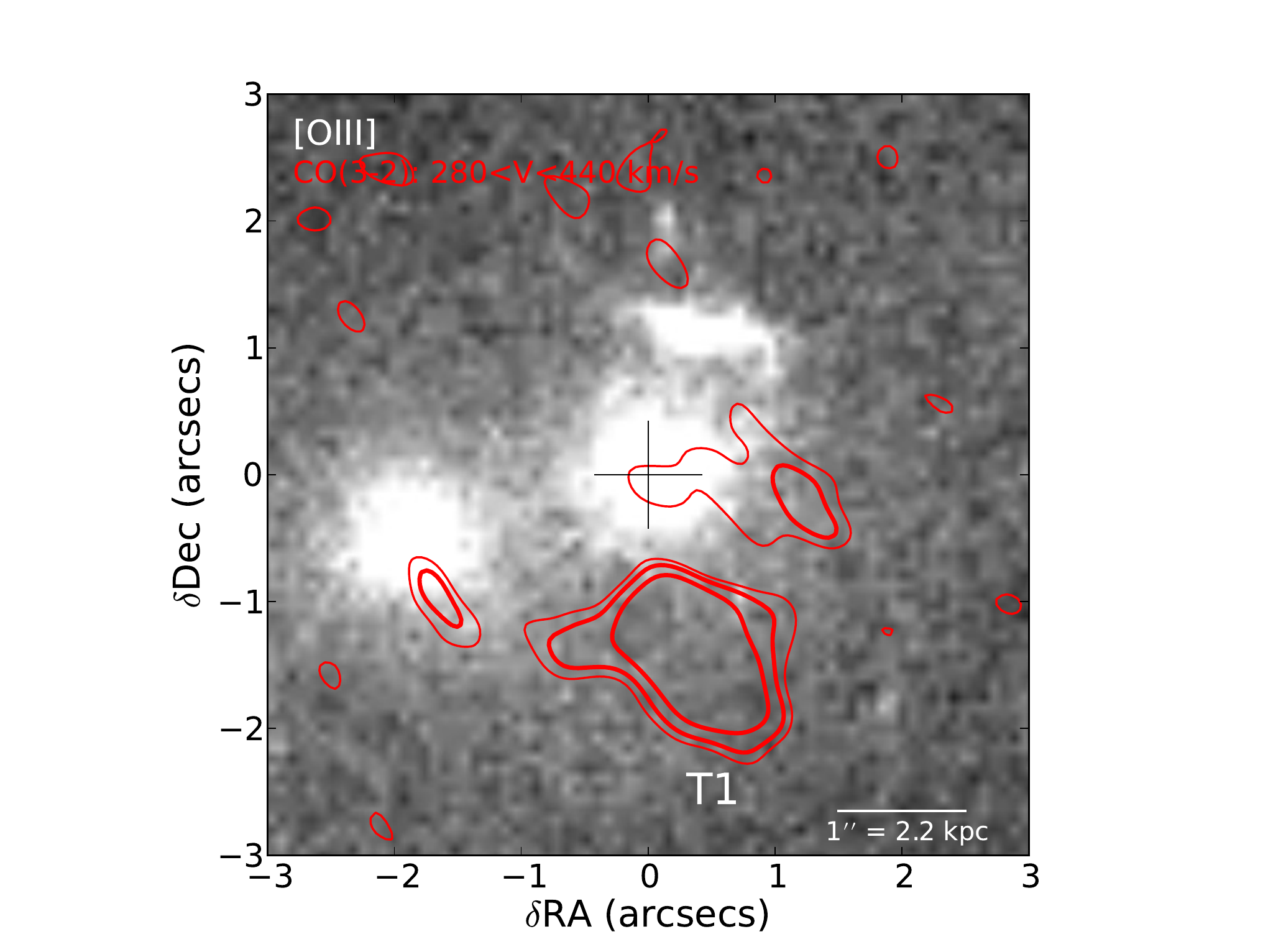}             
                \includegraphics[width=6cm]{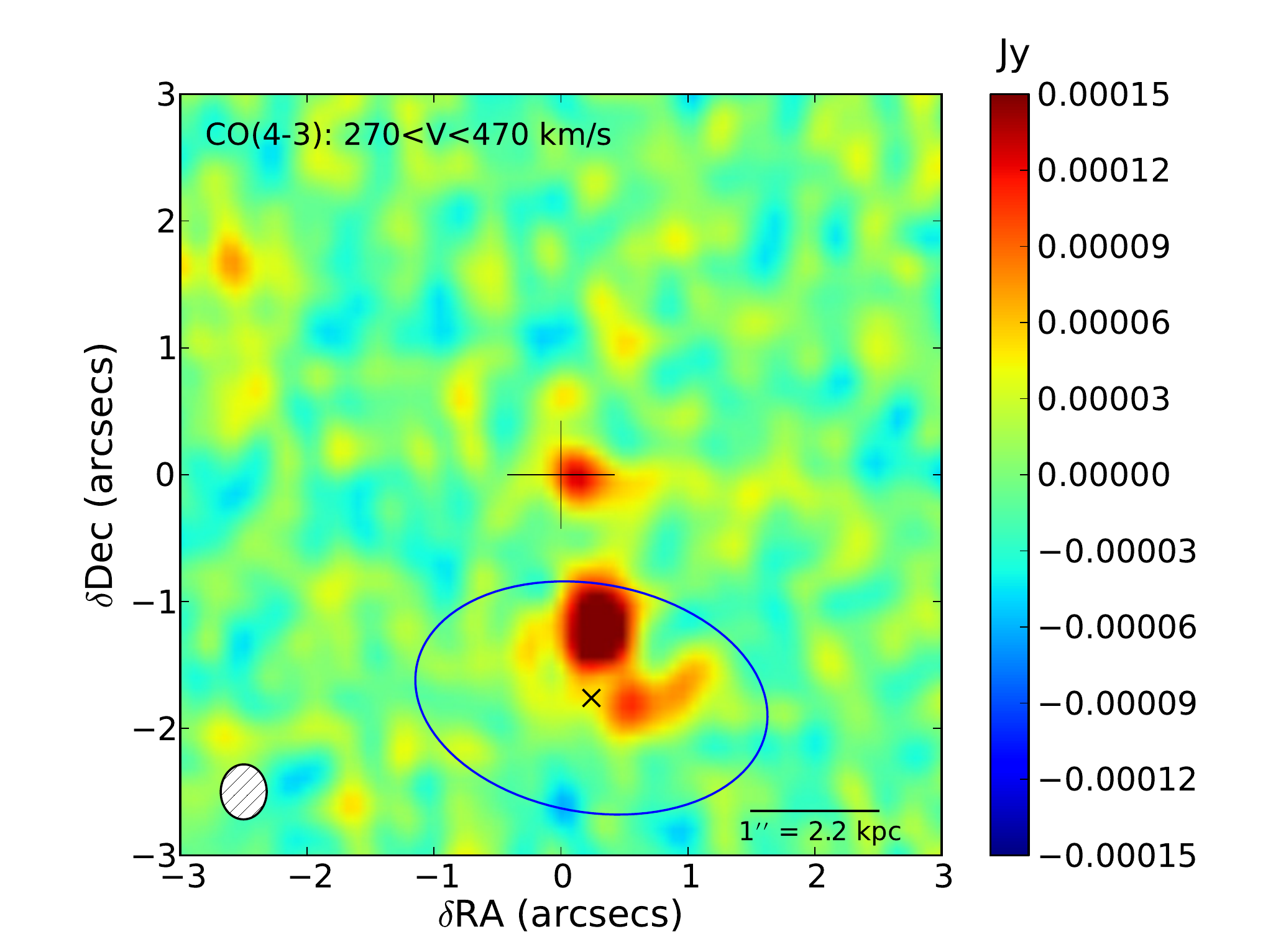}                     
                \includegraphics[width=6cm,height=4.5cm]{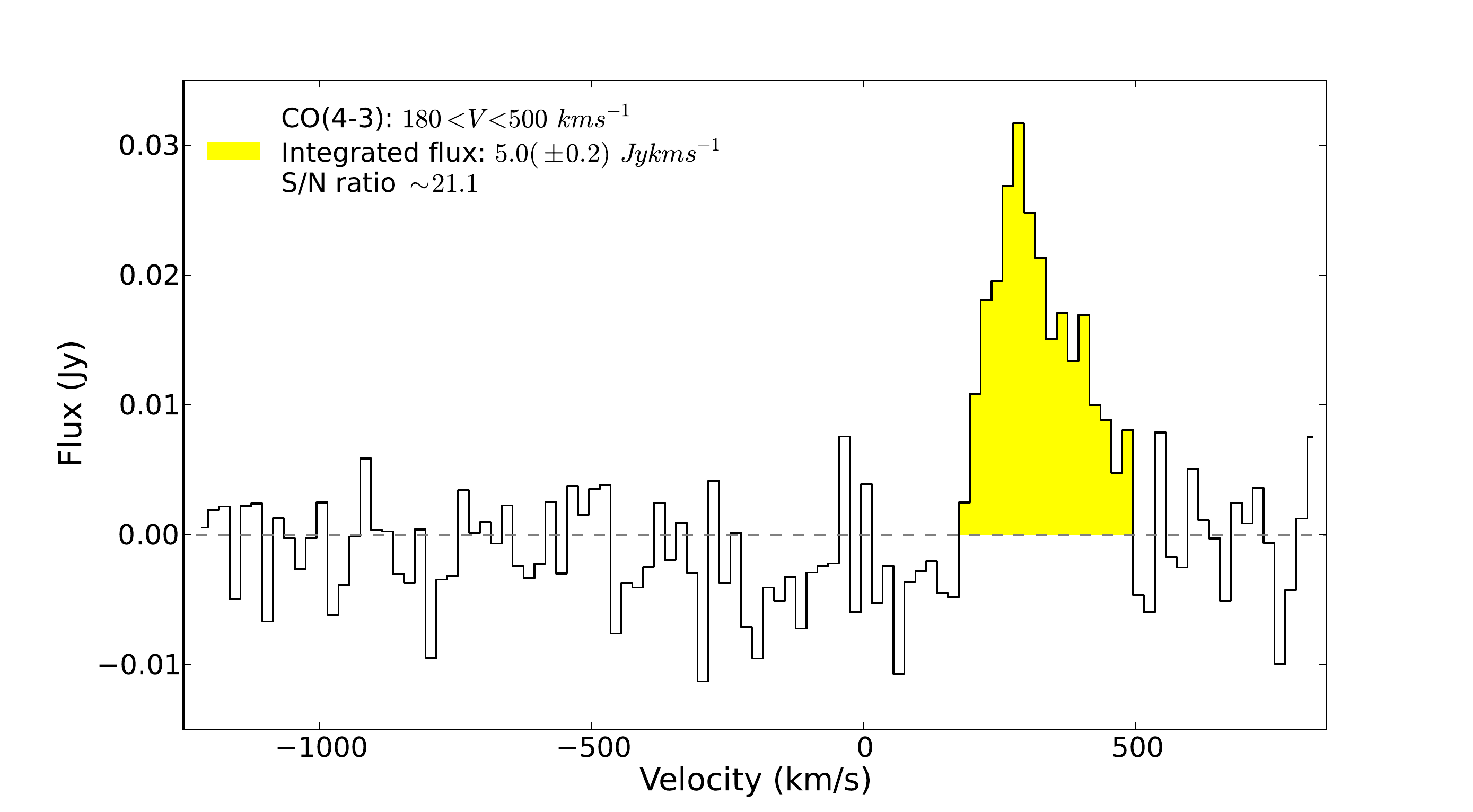}
                \includegraphics[width=6cm]{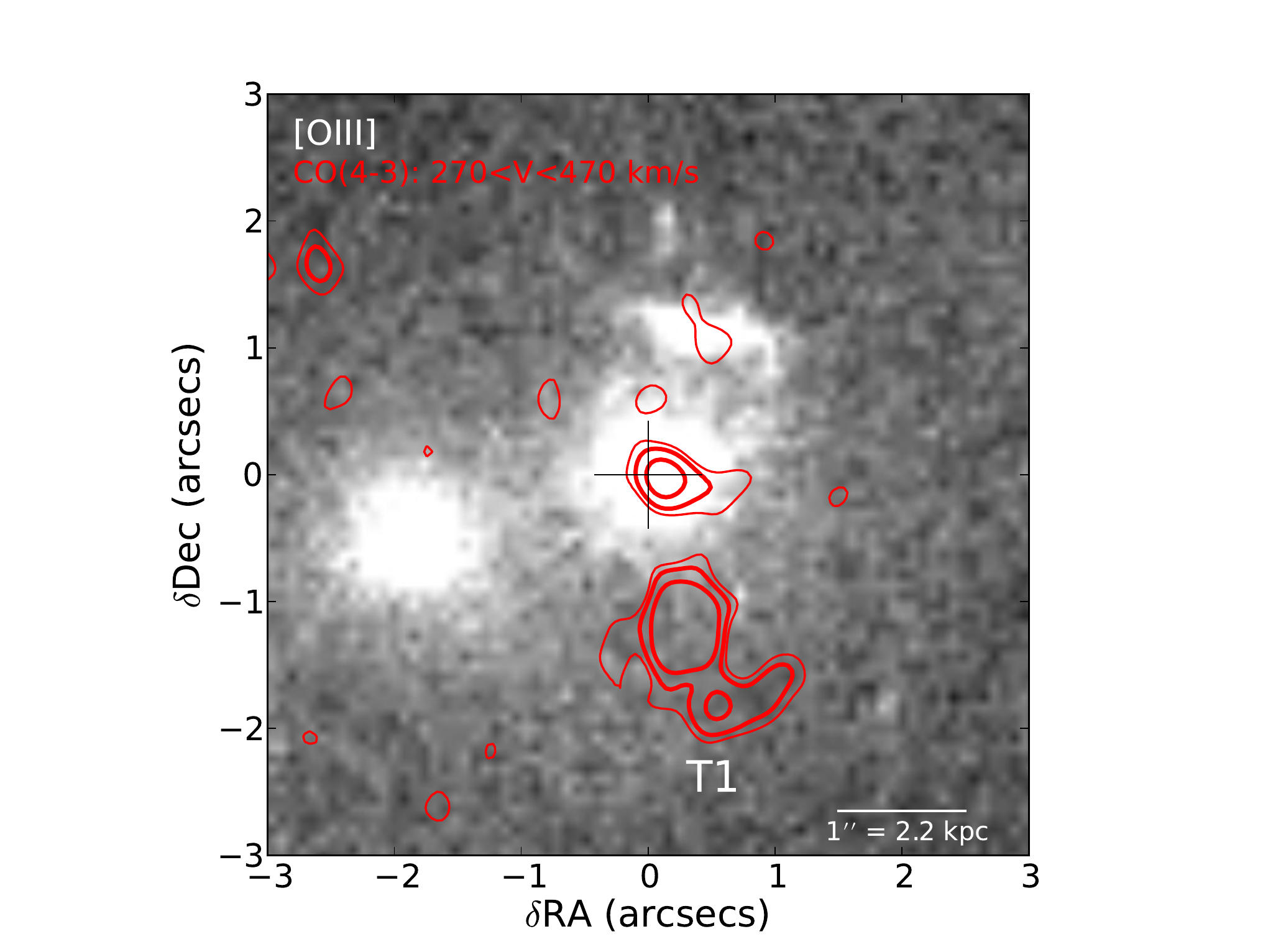}
                \includegraphics[width=6cm]{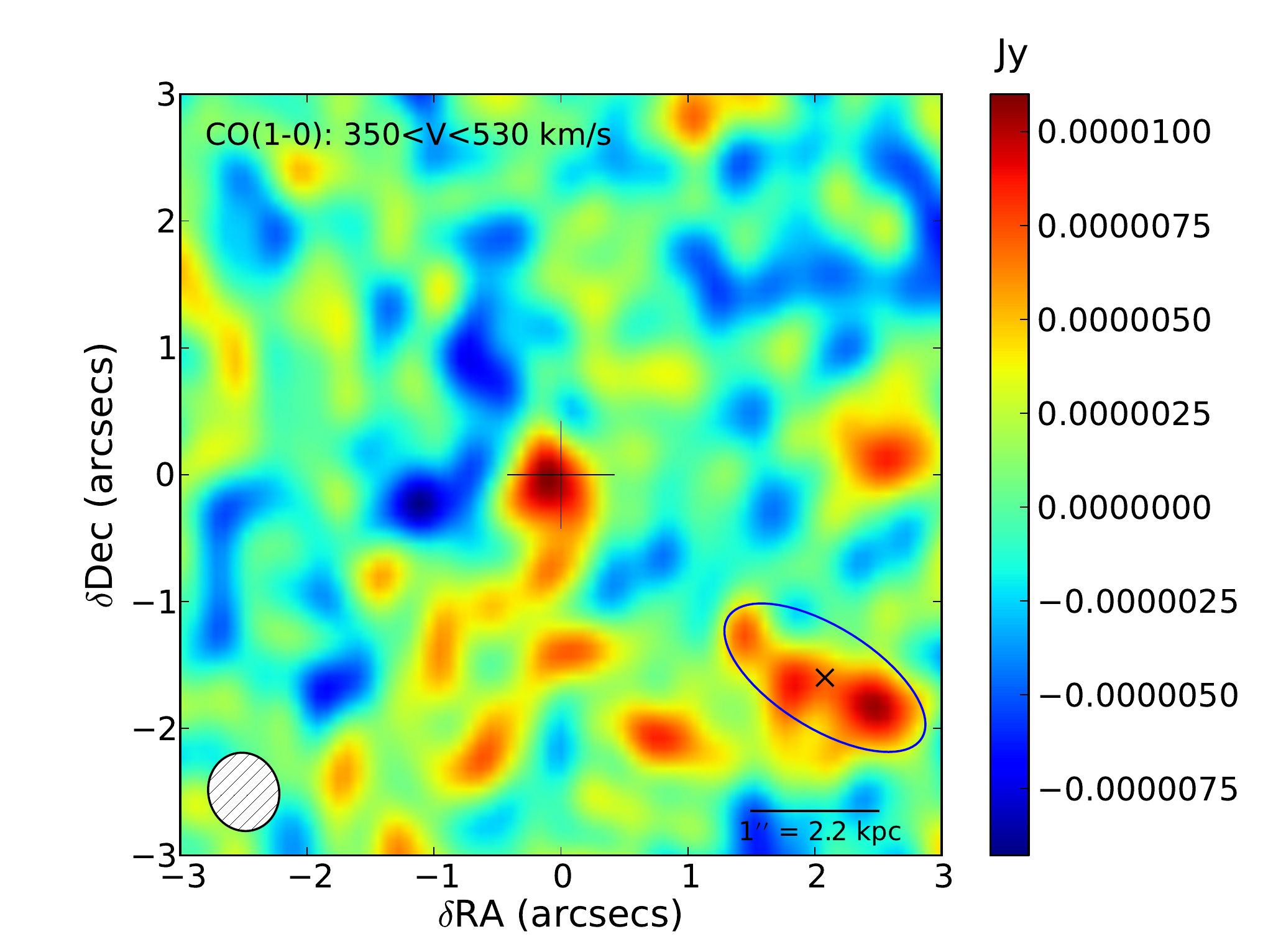}
                \includegraphics[width=6cm,height=4.5cm]{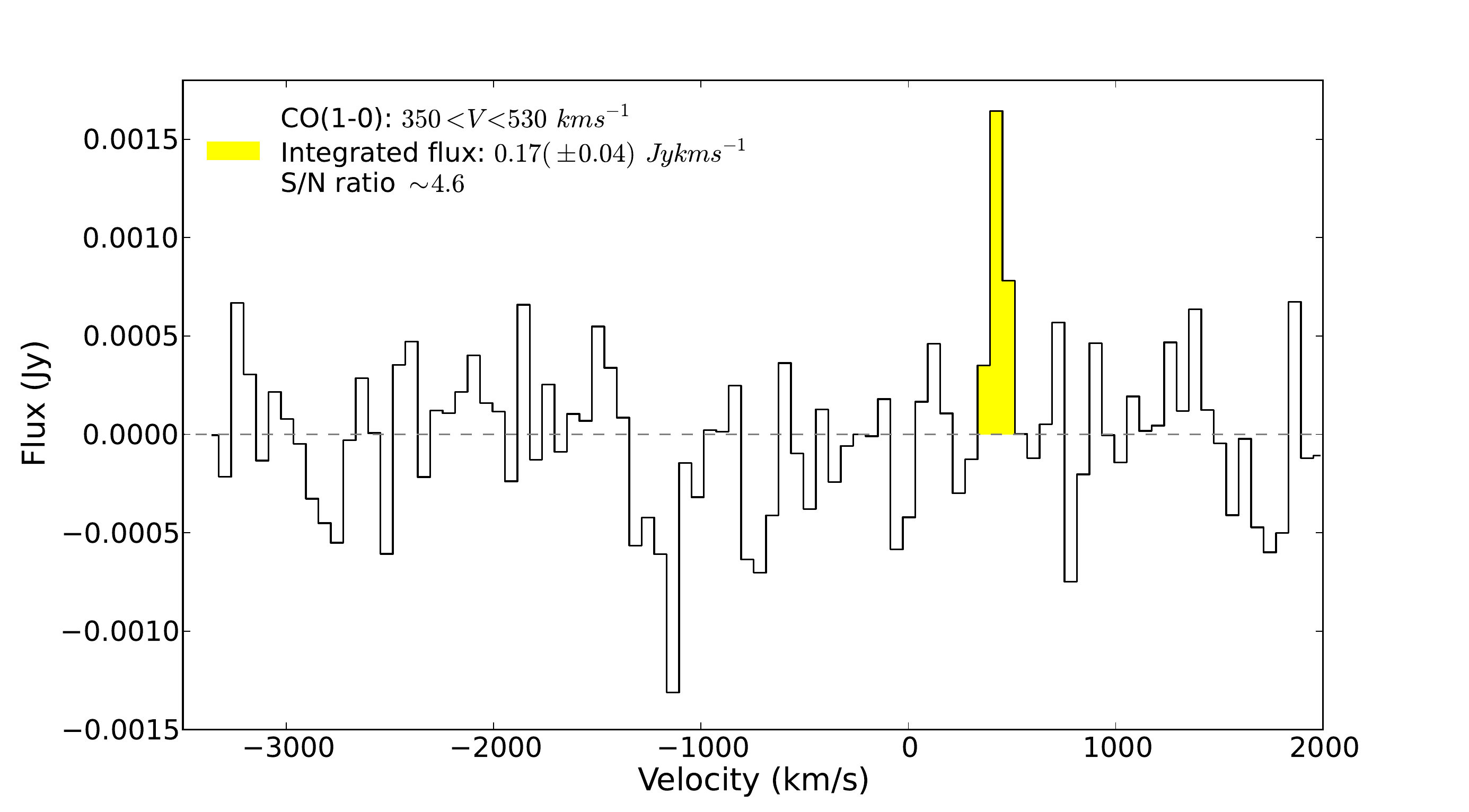}
                \includegraphics[width=6cm]{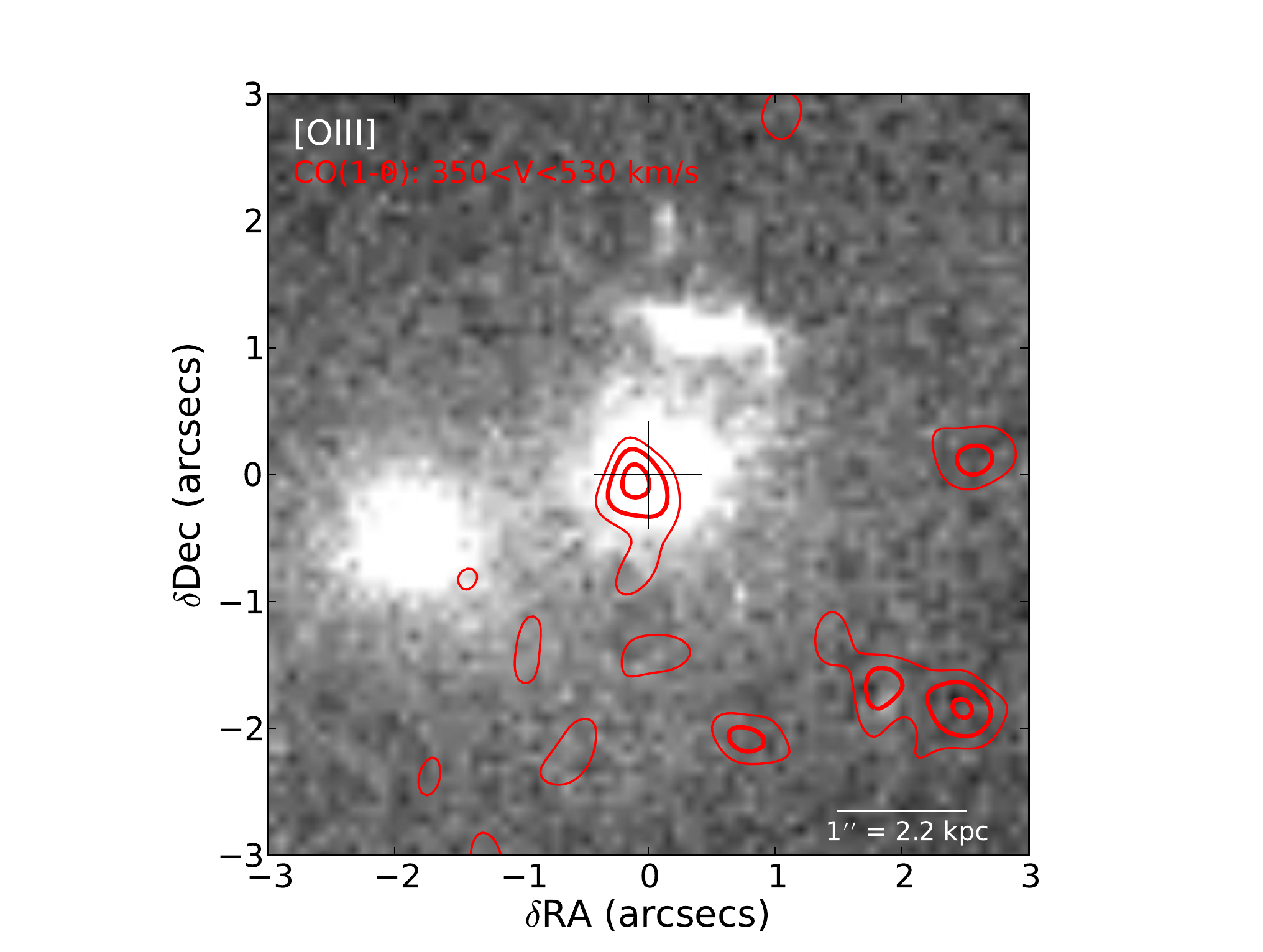}
                \caption[]{Redshifted part of the emission south of the main nucleus. The contour levels in the right panels are: 2$\sigma$,3$\sigma$,5$\sigma$ for the first three rows and  2$\sigma$,3$\sigma$,4$\sigma$ for the last row.}
                \label{fig:south_pos}
        \end{center}
\end{figure*}

\begin{figure*}
        \begin{center}
                \includegraphics[width=6cm]{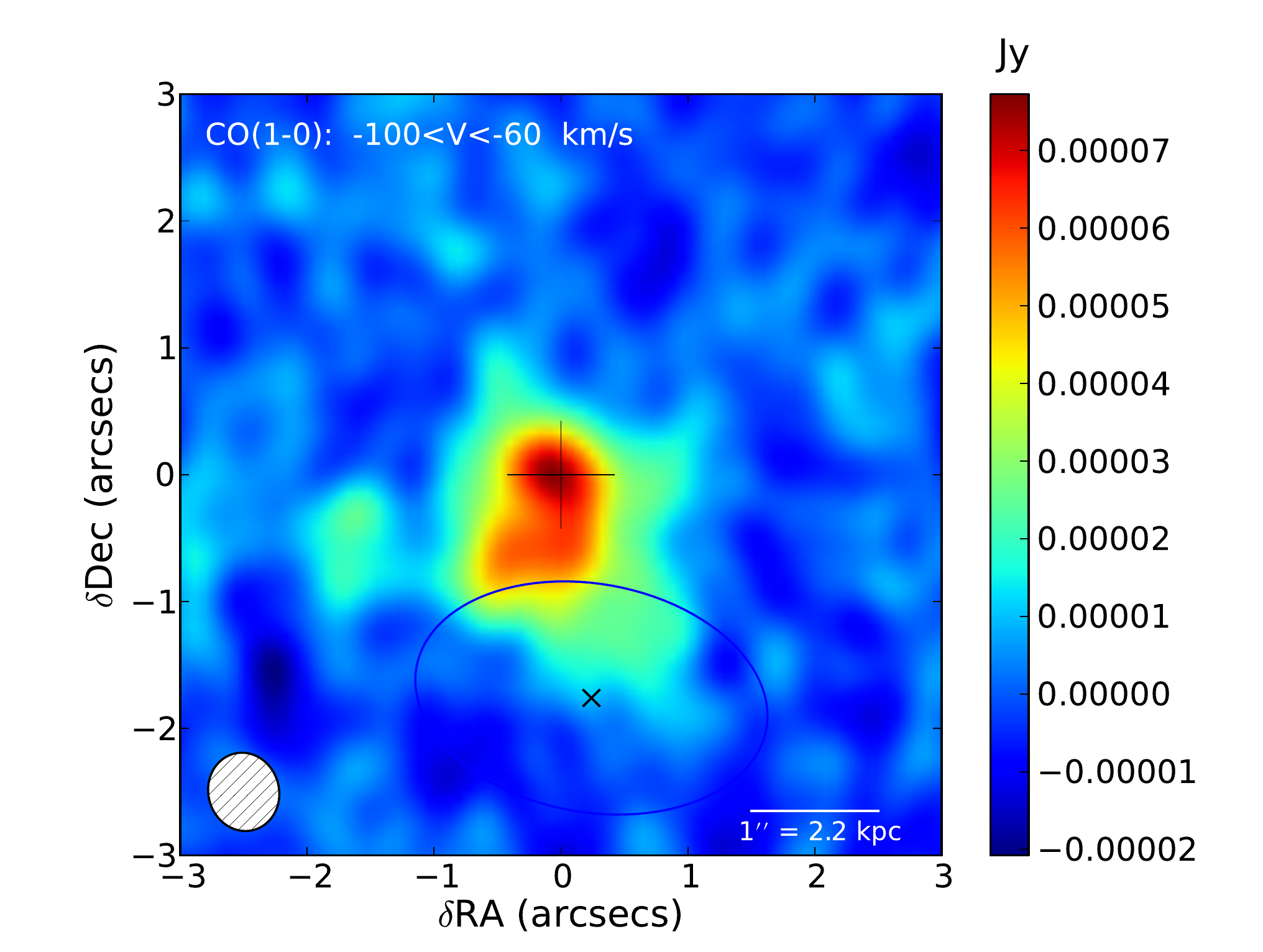}                            
                \includegraphics[width=6cm,height=4.5cm]{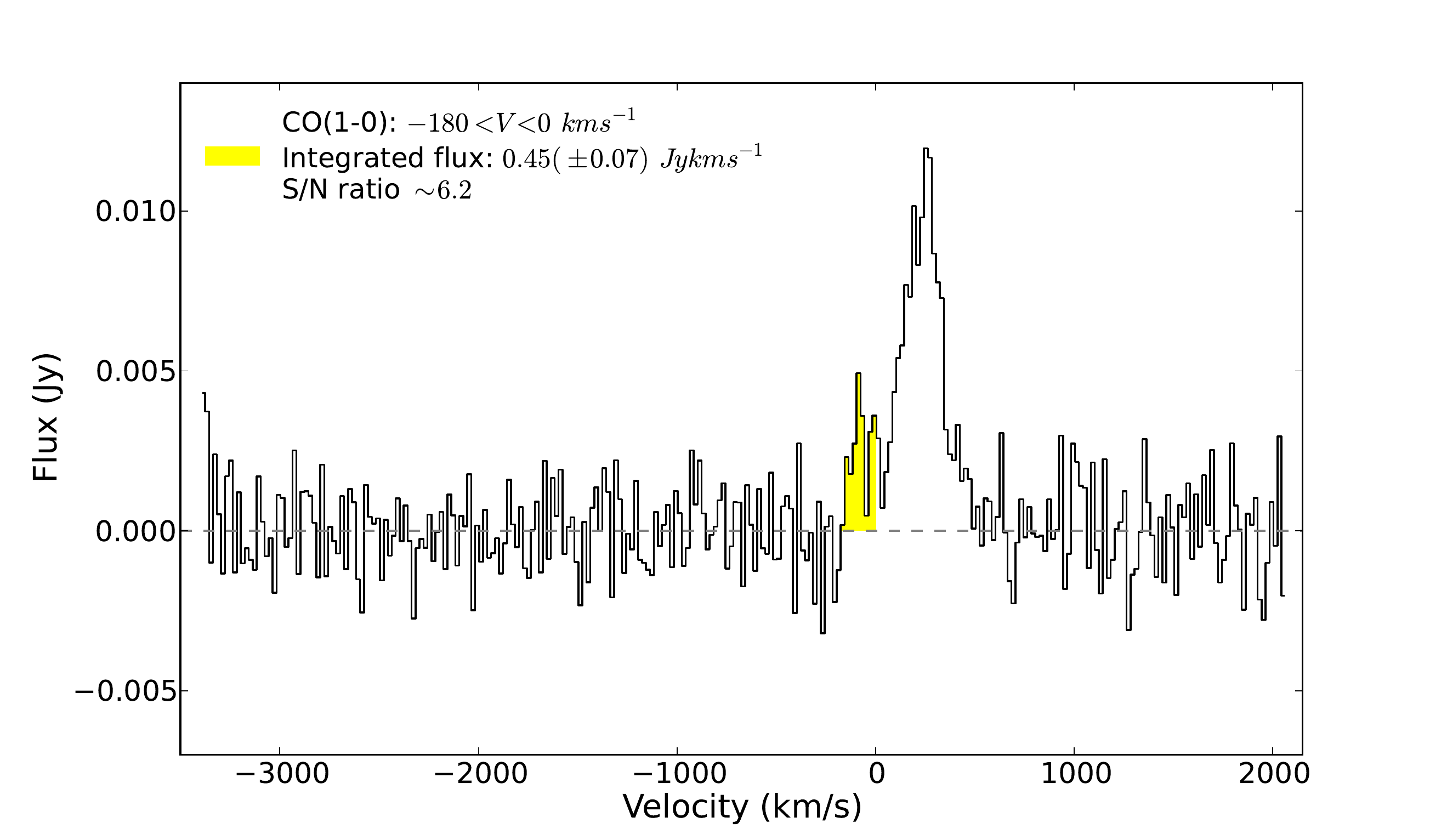}
                \includegraphics[width=6cm]{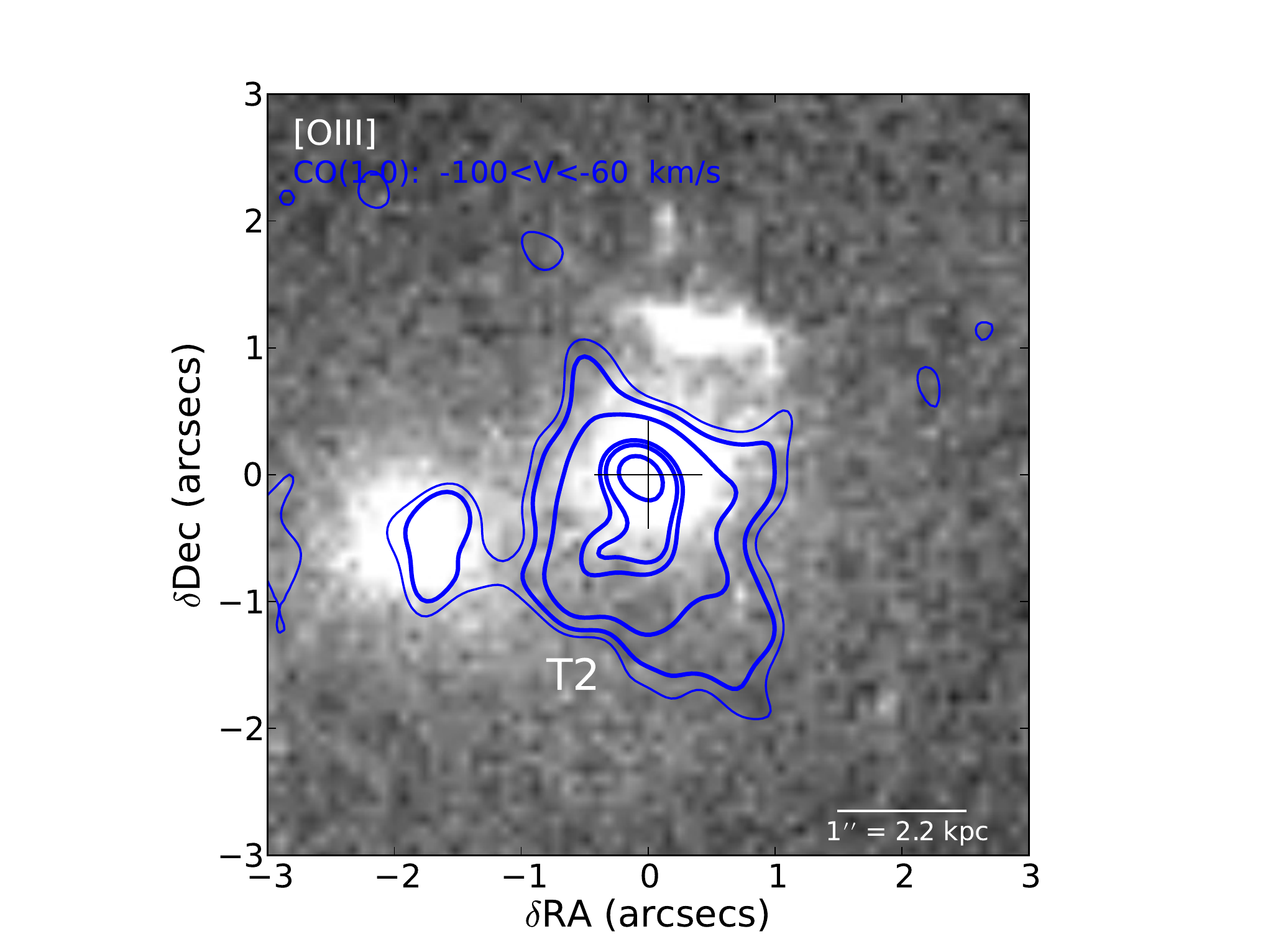}            
                \includegraphics[width=6cm]{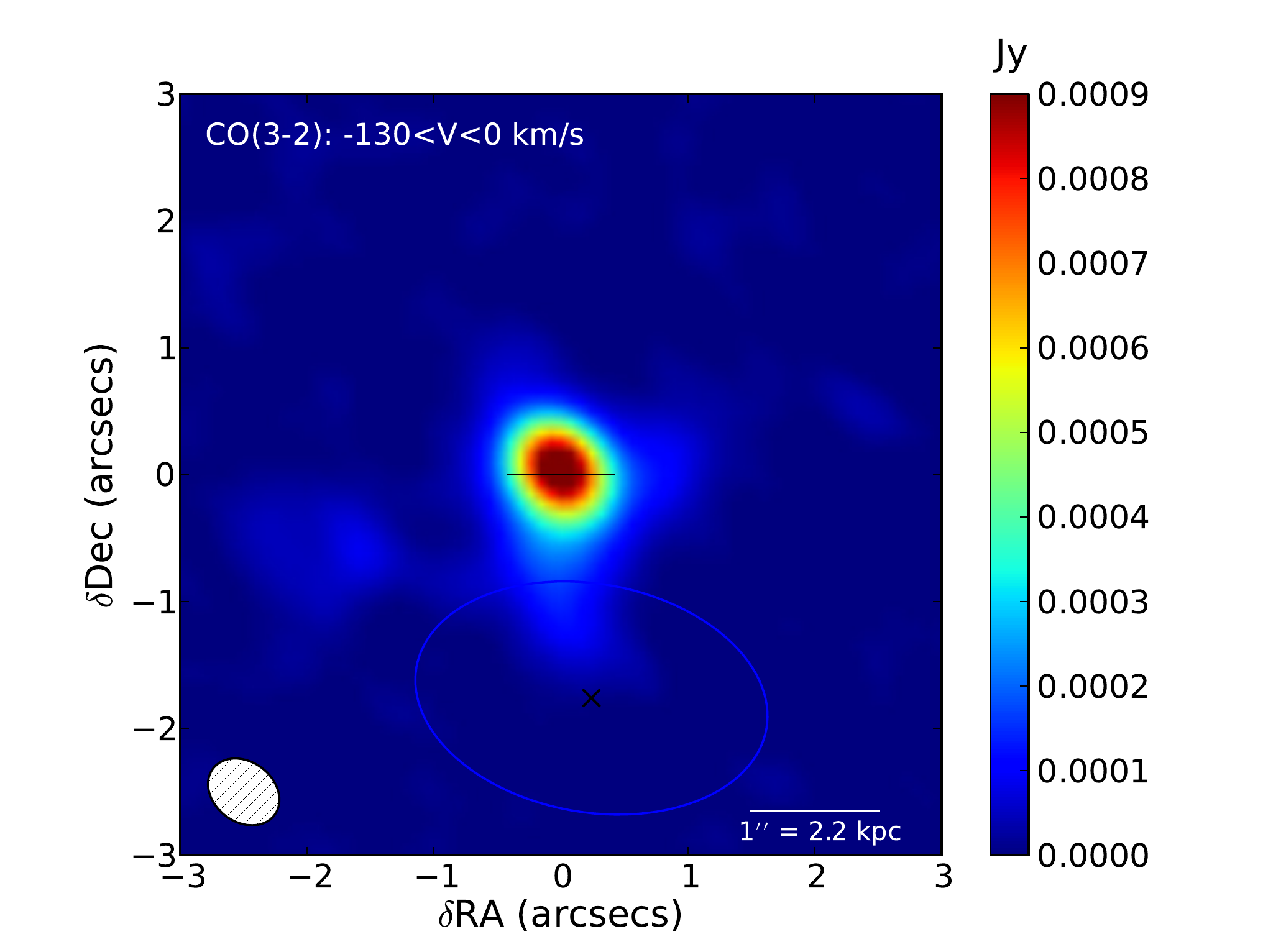}            
                \includegraphics[width=6cm,height=4.5cm]{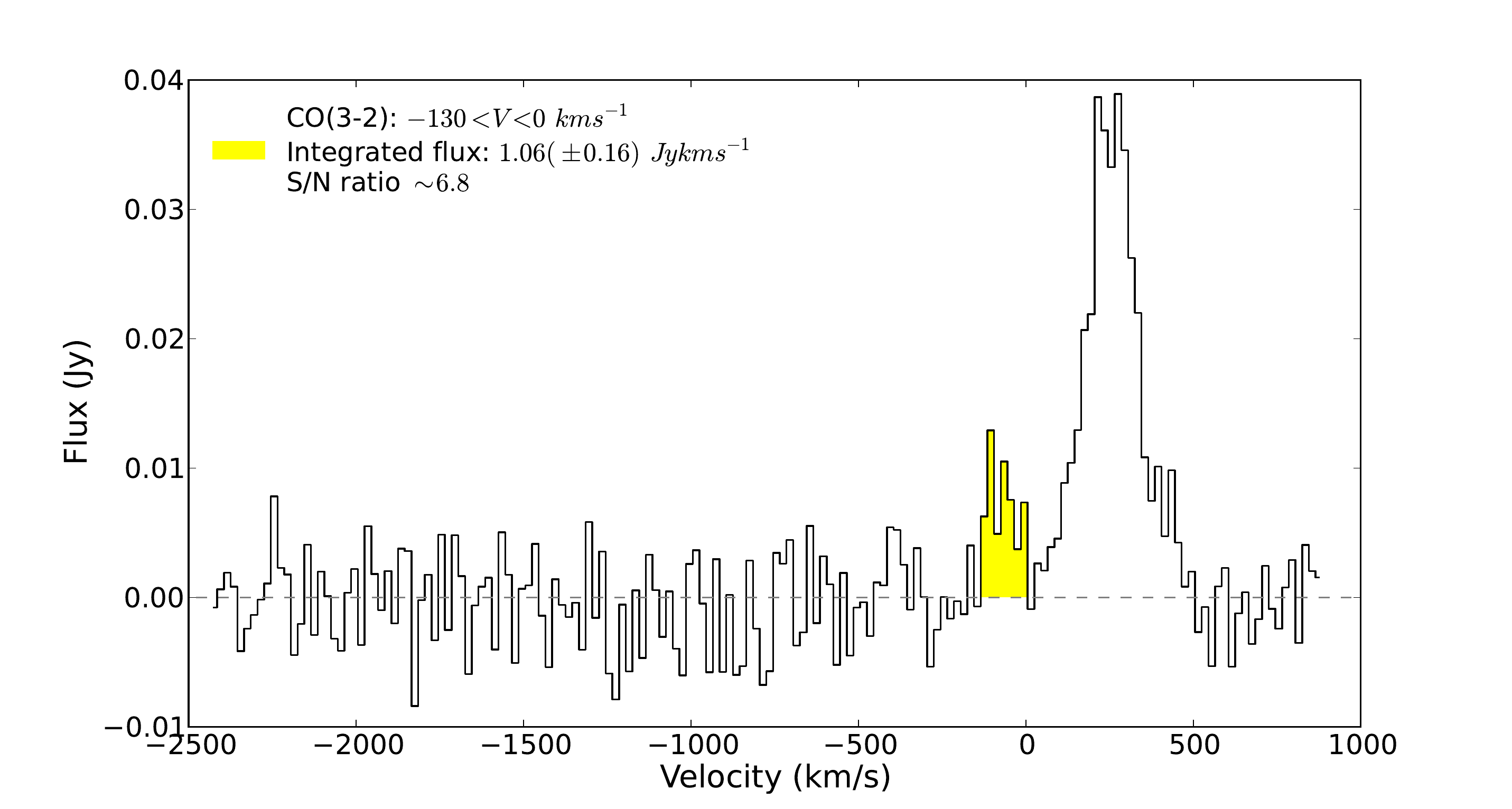}
                \includegraphics[width=6cm]{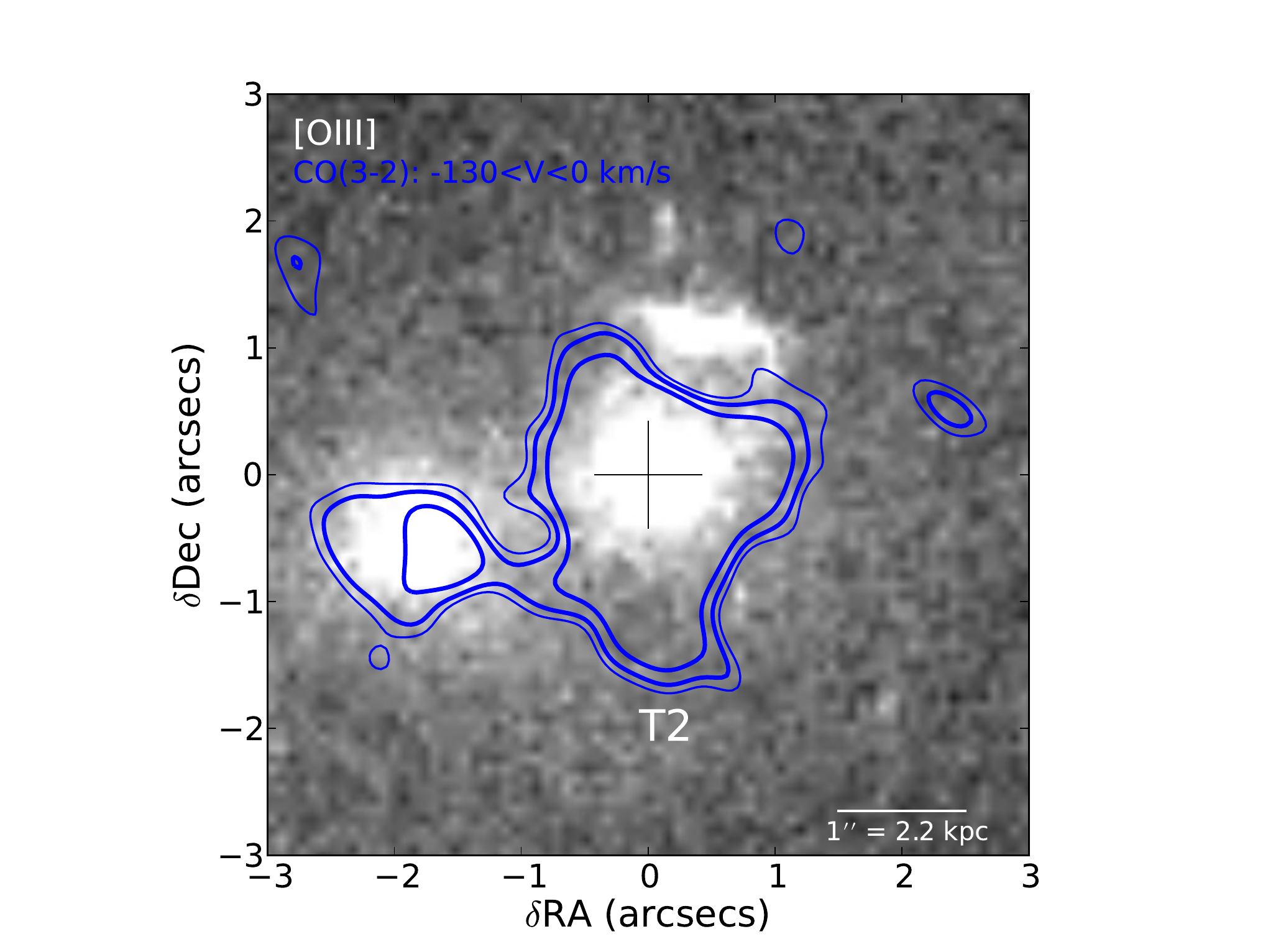}            
                \caption[]{Blueshifted part of the emission south of the main nucleus. The contour levels in the right panels are: 2$\sigma$,3$\sigma$,5$\sigma$,10$\sigma$,11$\sigma$,13$\sigma$ for \cooz\ and  2$\sigma$,3$\sigma$,5$\sigma$ for \cott\ .
                } 
                \label{fig:south_neg}
        \end{center}
\end{figure*}

A tail-like structure south of the main nucleus is the brightest extra-nuclear region detected in our data (Figs. \ref{fig:south_pos}, ~\ref{fig:south_neg}). Its total flux, as deduced from the \cooz\  high-resolution cube is 3.1($\pm$0.6) Jy\kms , which corresponds to a mass of  1.7($\pm$0.3)$\times$10$^9$\msun. 

This structure is detected in all CO(1$-$0), (3$-$2), and (4$-$3) lines and is brightest in  \cott . The detected emission is so bright that it dominates the kinematics of the region: its velocity dispersion is low, typical of that of a spiral arm (60-80\kms ), while the velocity dispersion of the disk exceeds 120\kms\ (Fig.~\ref{fig:momenta}). Its mean velocity is high (200-300\kms ) compared to that of the disk,  meaning that its redshifted component dominates.

The redshifted emission primarily originates from a structure that spirals out from the nucleus to the southwest. In its inner part, the gas reaches velocities of $\sim$300\kms. Further away, towards the southwest, the gas velocity exceeds 500 \kms (Fig.~\ref{fig:south_pos}). The structure is either a spiral arm that has conserved its initial angular momentum, or a tidal tail, or material that was accreted in a polar ring geometry during the merger. In any case, it can be related to the large-scale structure seen by \citet{dasyra14} at scales $>$4 kpc. 
Adding its mass to the above calculated mass of the tail-like structure, the mass of the south structure reaches 1.8($\pm$0.4)$\times$10$^9$\msun.
A secondary, fainter structure with different orientation and angular momentum vector, which spirals out from the nucleus to the southeast is also seen in the inner part of the emission. The same applies to the inner part of the blueshifted emission partly coinciding with the redshifted emission (Fig.~\ref{fig:south_neg}). It might be yet another spiral arm or twisted tidal tail, or even part of a second bridge to the secondary nucleus.  Due to the spatial overlap of the blueshifted and the redshifted components, we cannot reliably identify wind candidates or ascribe masses to individual components.

\subsection{Other structures as wind components or candidates}
\label{sec:outflow_emission}

\subsubsection{Nuclear wind in CO}\label{sec:res_nuclear}
\begin{figure*}
        \begin{center}
                \includegraphics[width=6cm]{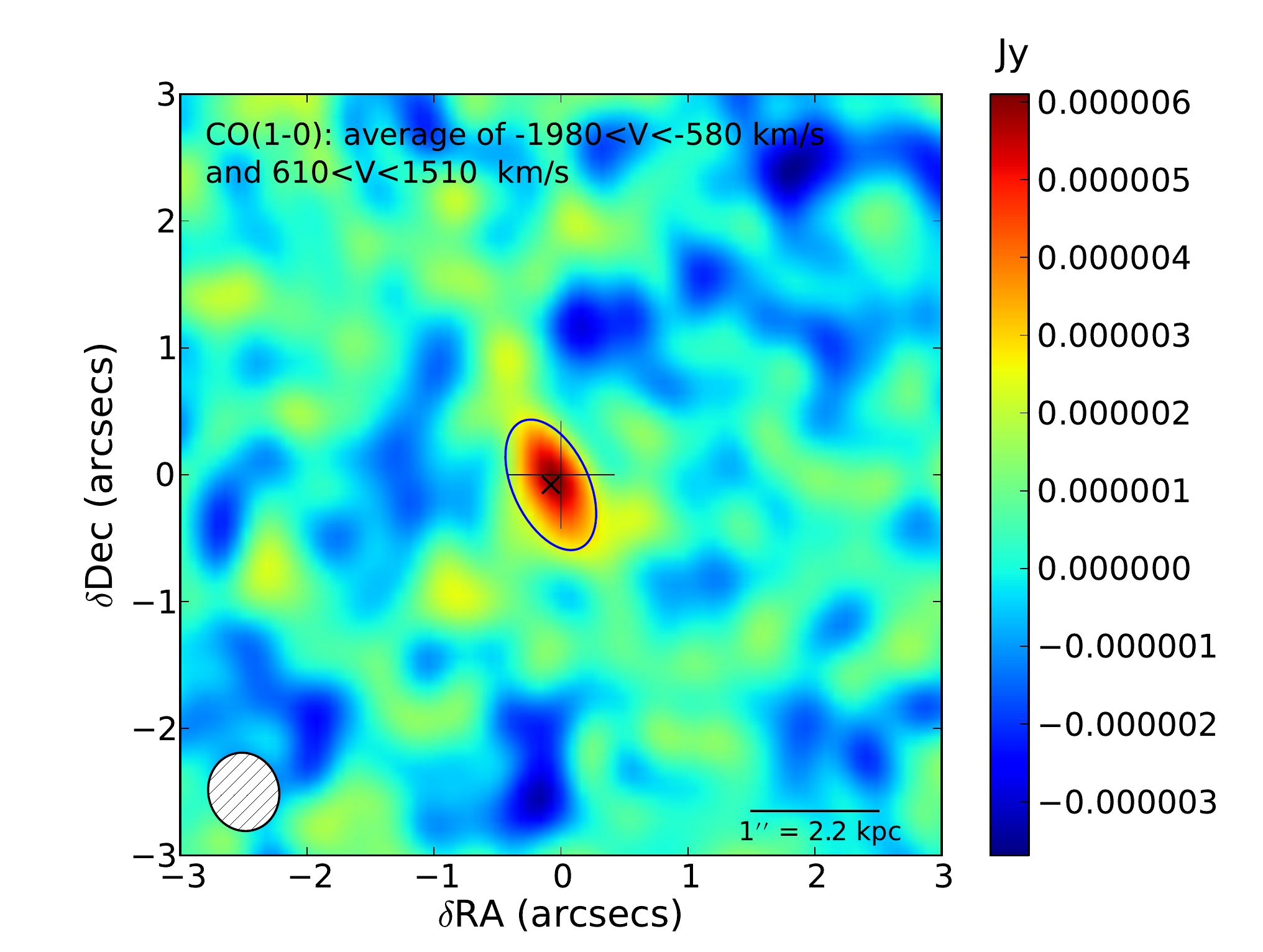}
                \includegraphics[width=6cm,height=4.5cm]{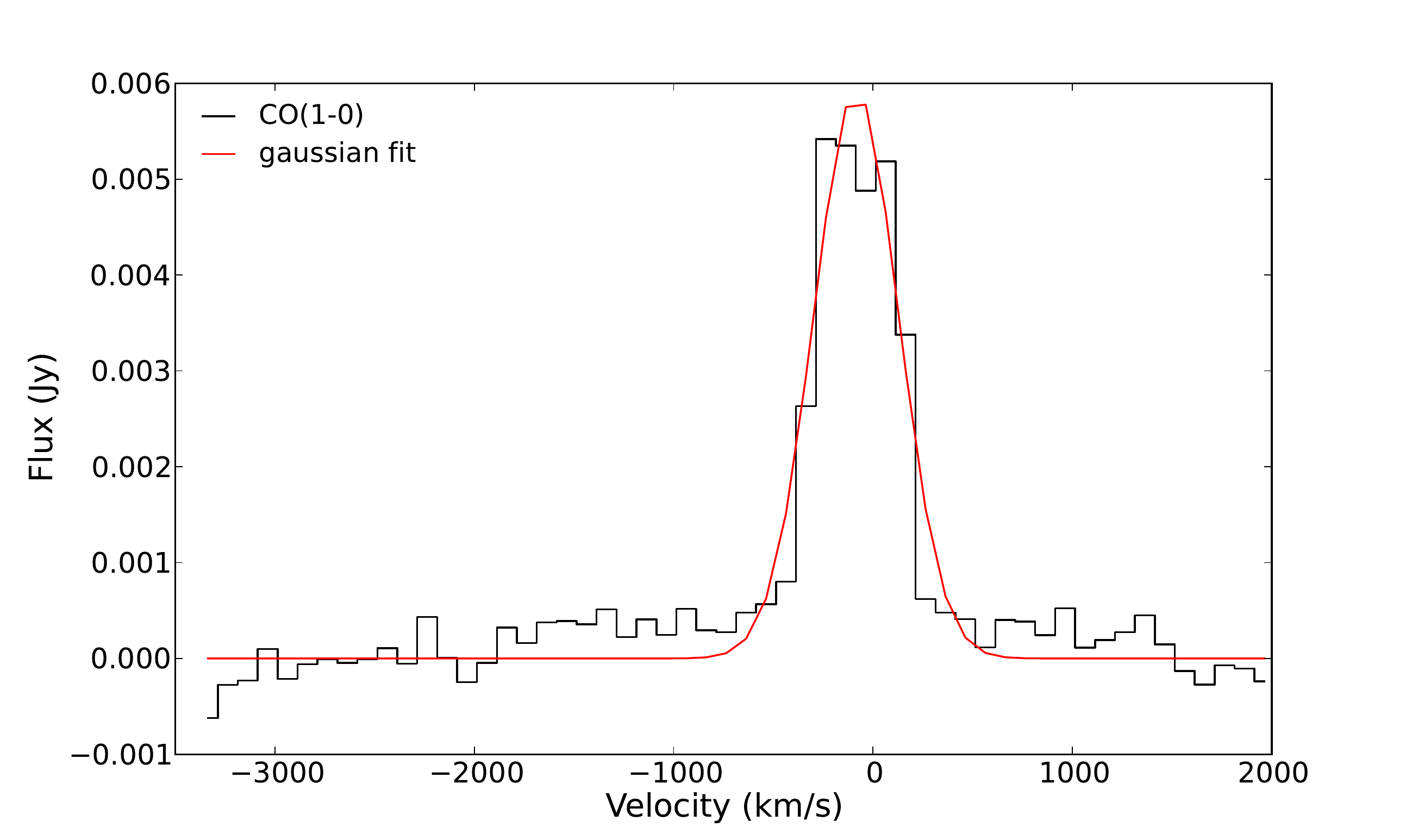}
                \includegraphics[width=6cm,height=4.5cm]{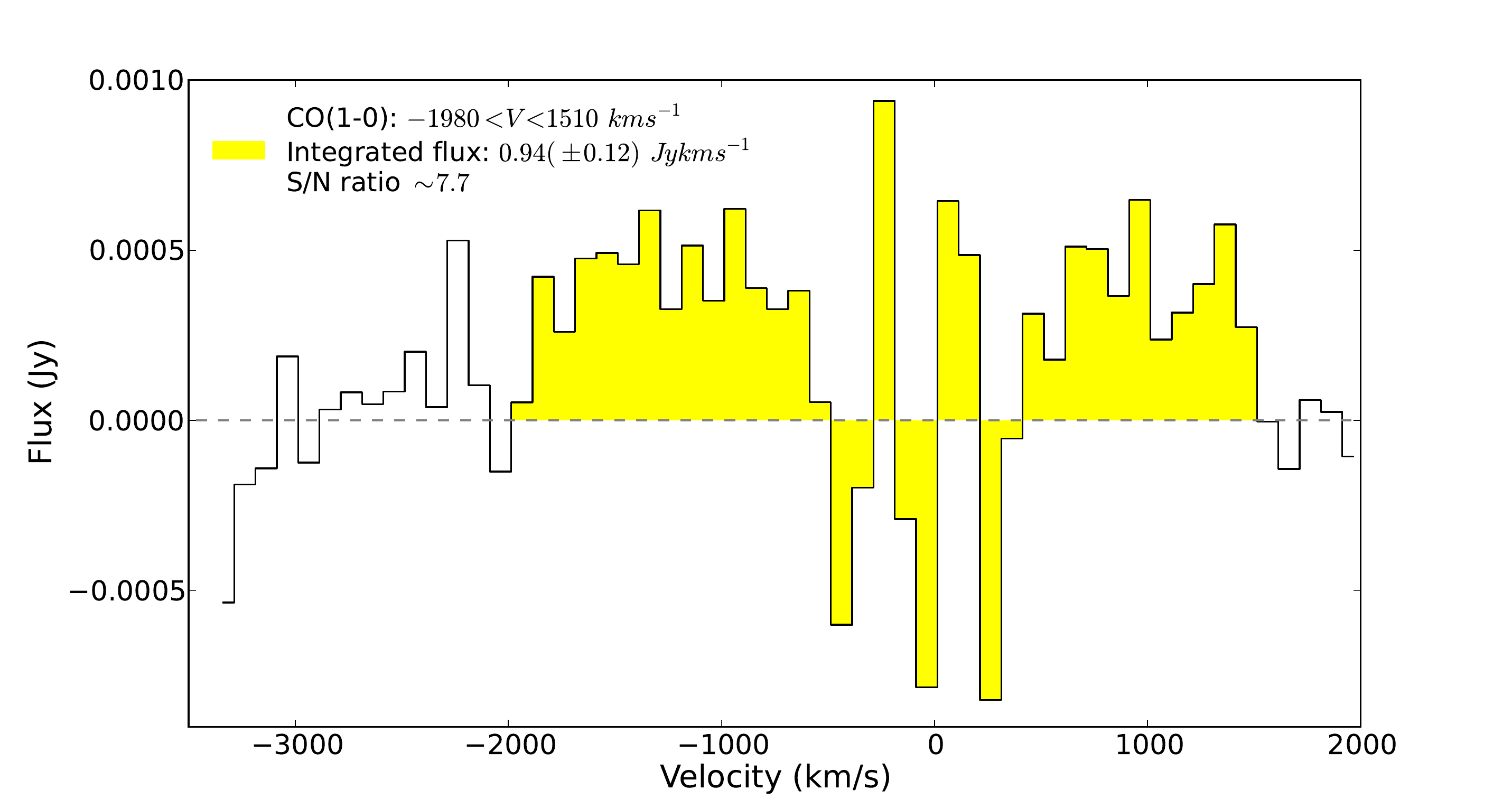}
                \caption[]{Nuclear wind candidate. Average image of the blueshifted and the redshifted emission of the nuclear wind (left). A Gaussian fit to the nuclear spectrum is performed and fit residuals are presented in the middle and right panels, respectively. 
                }
                \label{fig:cen_wind_detection}
        \end{center}
\end{figure*}

A tentative wind detection in the nuclear region is seen in the \cooz\ data (Fig. \ref{fig:cen_wind_detection}). Components at multiple velocities are seen due to line-of-sight effects and/or to an intrinsic distribution of velocities (potentially caused by the different efficiency of acceleration of the molecular gas at different densities). The molecular wind  terminal velocity exceeds -1800\kms\ for the blueshifted wing and 1400\kms for the redshifted wing. \citet{holt03}, in their optical spectroscopic study of 4C12.50, also found a very broad wind in the nucleus of the system. The \oiii\ line profile required three individual Gaussian components to be fit; one of them was centered at -1980\kms , with a width of 1944\kms . The terminal velocity of the \oiii\  was thus -2950\kms .  A Gaussian fit to the data provides the residual emission underlying the disk emission (Fig. \ref{fig:cen_wind_detection}), indicating a residual flux of 0.94($\pm$0.12) Jy\kms. This corresponds to a nuclear wind mass of  5.3($\pm$0.7)$\times$10$^8$ \msun. 

Despite the strong continuum and the high dynamic range (of order 1000; Fig. \ref{fig:spectra}), the wind detection survives the test of altering pipeline parameter values (e.g., first- and second-order continuum fit, different spectral window ranges for the continuum fit) in both datasets, that is, in our high-resolution and in the deep cubes. In previous work carried out by the team using Plateau de Bure (PbD) data \citep{dasyra14, dasyra_combes12} this nuclear emission was not detected because the spectral window employed was not sufficient (the PdB data reached -1500\kms\ , whereas the detection in the current ALMA data exceeds -1800\kms\ ; Fig. \ref{fig:cen_wind_detection}) and the PdB sensitivity was not adequate to detect it: the PdB data were sensitive down to 1.1mJy/beam.

\subsubsection{Extended wind: high-velocity CO in radially extending filaments}
\label{sec:res_ridge}

{\it Herschel} observations indicated the possible  existence of an extended molecular wind in 4C12.50: unlike in most other ULIRGs, the outflowing OH molecules in 4C12.50 are primarily seen in emission \citep{spoon13}. This result indicated that many OH molecules are located in lines of sight free of background emission or that the background emission is low compared to the OH emission. A region in which an extended outflow is likely to be detected in 4C12.50 is north of the main nucleus \citep{zaurin07}. There, a very distinct \oiii\ ridge is seen in optical \hst\ data, referred to as an "arc" by \citet{tadhunter18}. Our ALMA data indicate that the molecular gas disk is rather abruptly cut along this ridge at low velocities (see CO line contours near systemic velocity; Fig.~\ref{fig:disk_detections}). This result could be indicative of a turnover of a tidal tail or stream or of an interface where a gas phase transition from molecular to ionized occurs. The transition could be due to the deposition of energy by some mechanism, for example, stellar or AGN radiation, collision-related shocks, or jet. Filaments that radially extend from the nucleus towards the ridge are additionally seen in the \oiii\ image. Two of them are also distinct in the stellar continuum image. \citet{holt03} found kinematically distorted \oii\ emission along a 160\degr -position-angle slit that contained such filaments. 

\begin{figure*}
        \begin{center}                                                  
                \includegraphics[width=6cm]{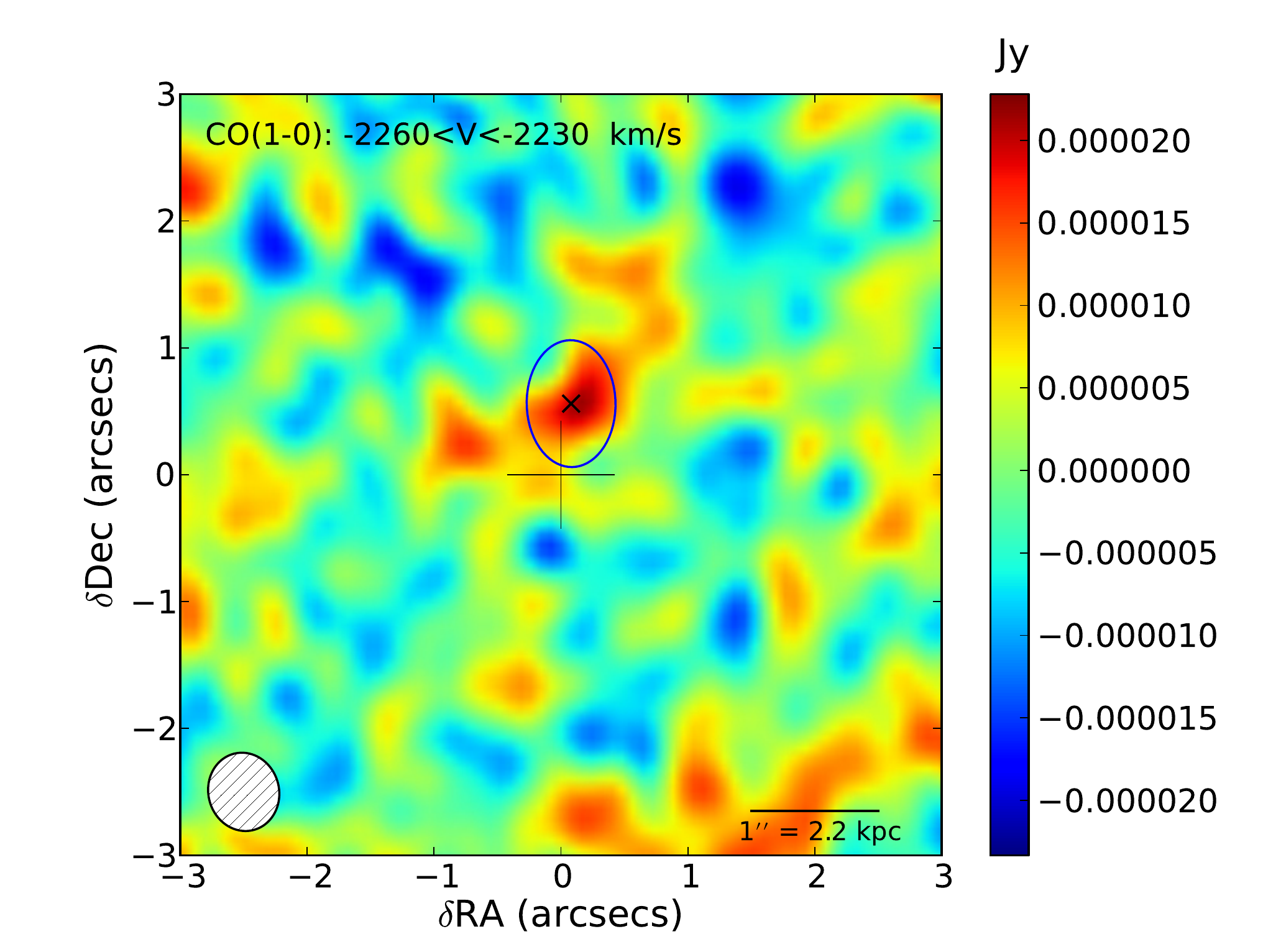}
                \includegraphics[width=6cm,height=4.5cm]{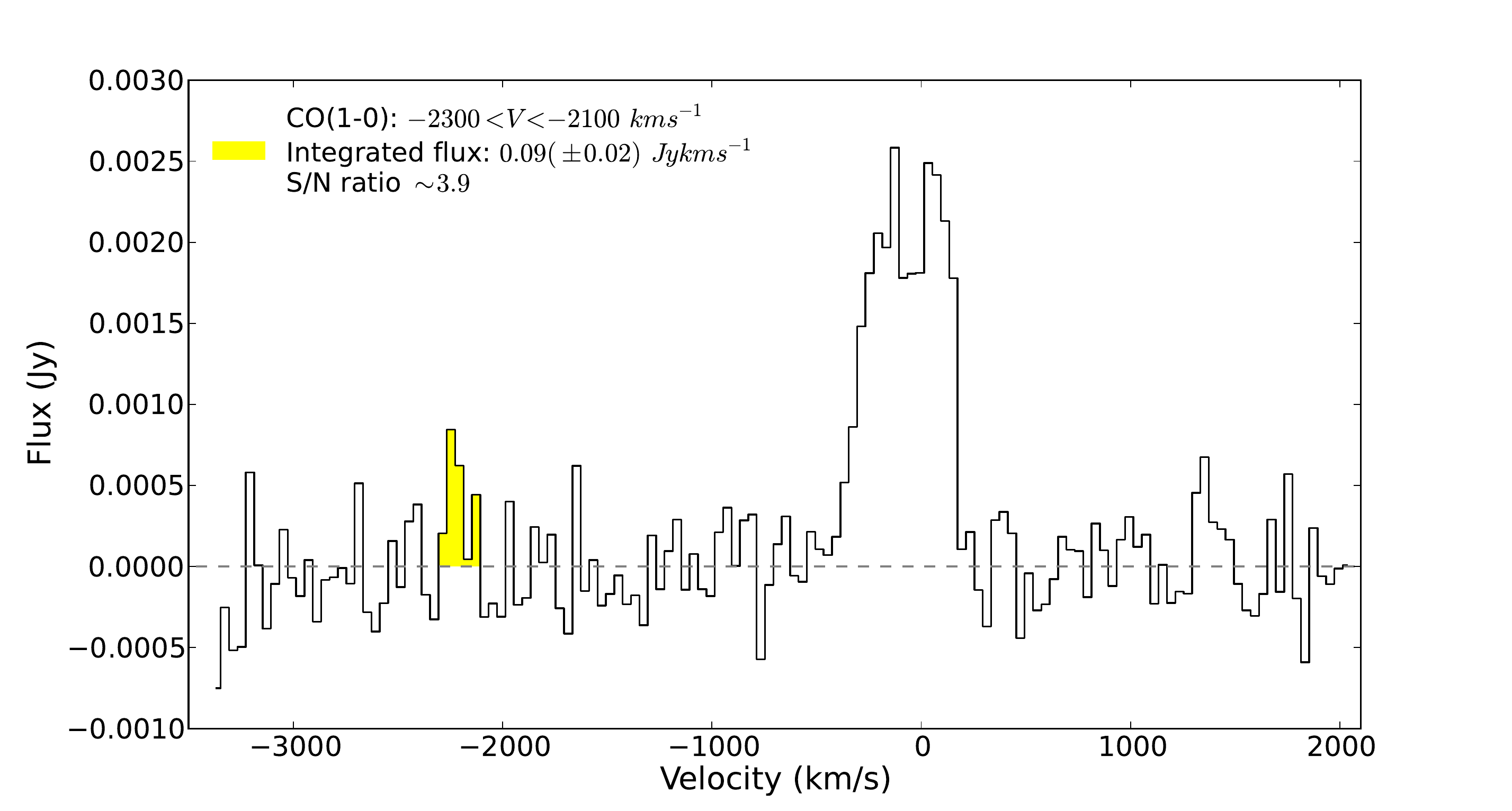}
                \includegraphics[width=6cm]{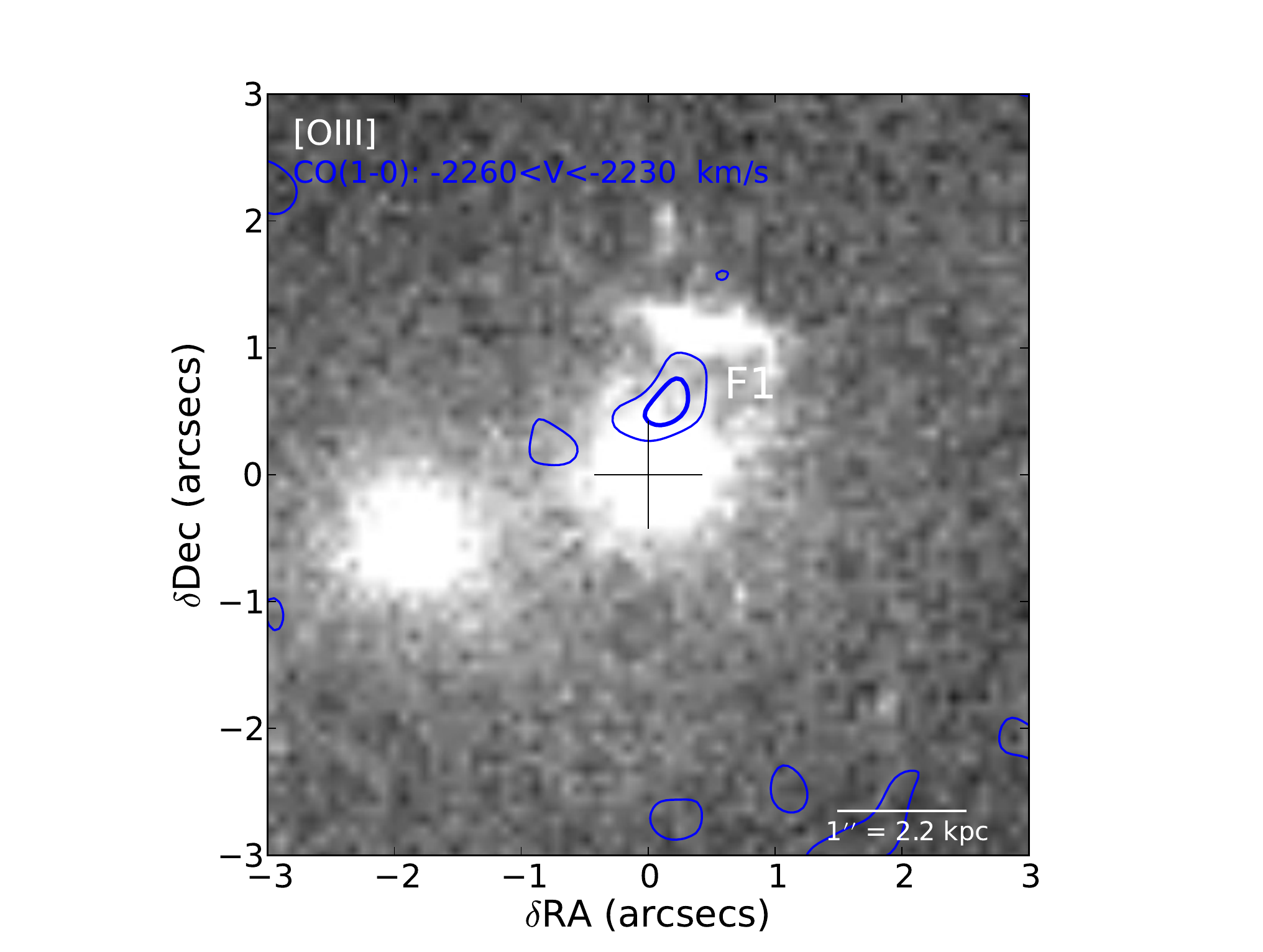}         
                \includegraphics[width=6cm]{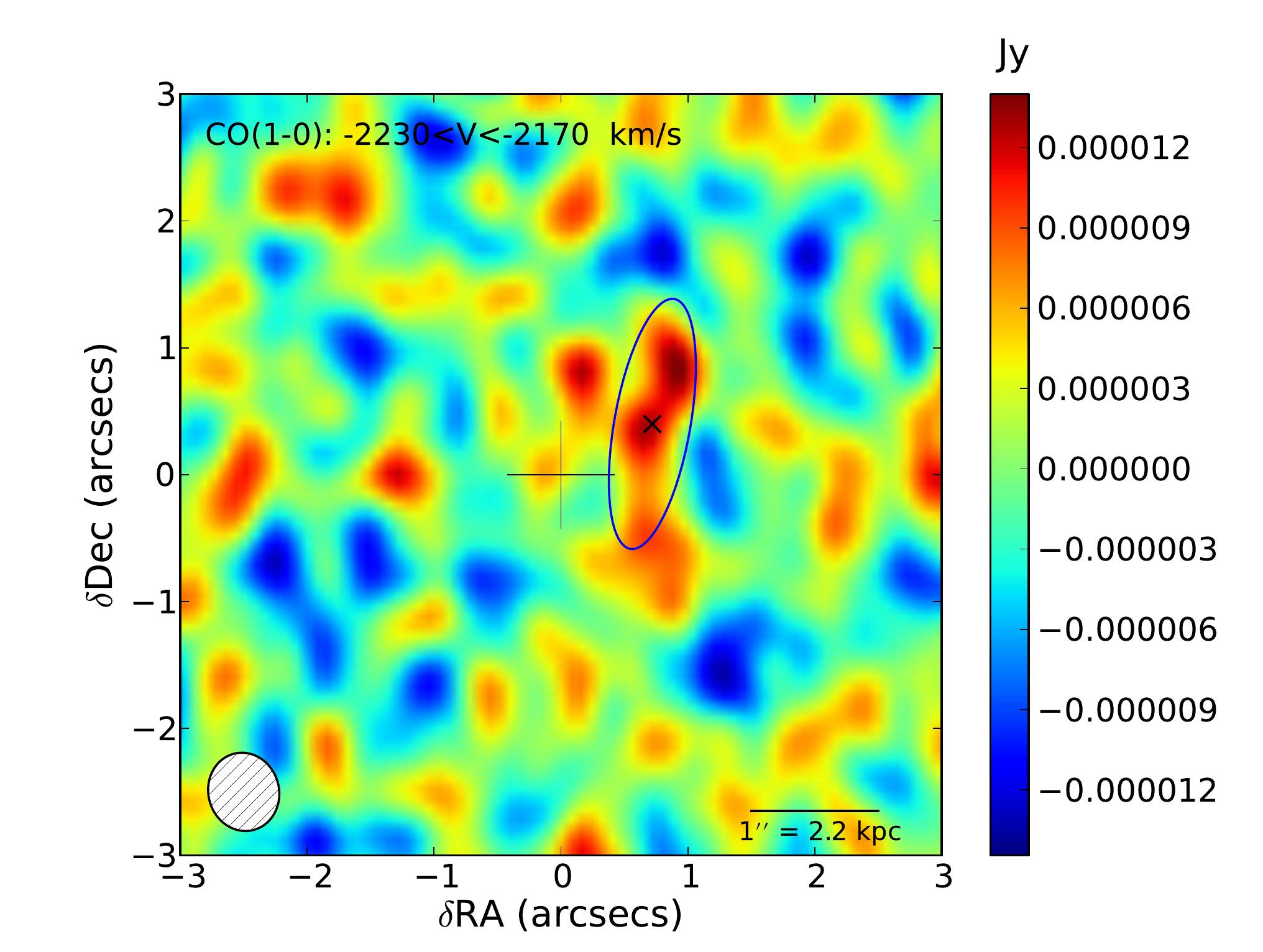}
                \includegraphics[width=6cm,height=4.5cm]{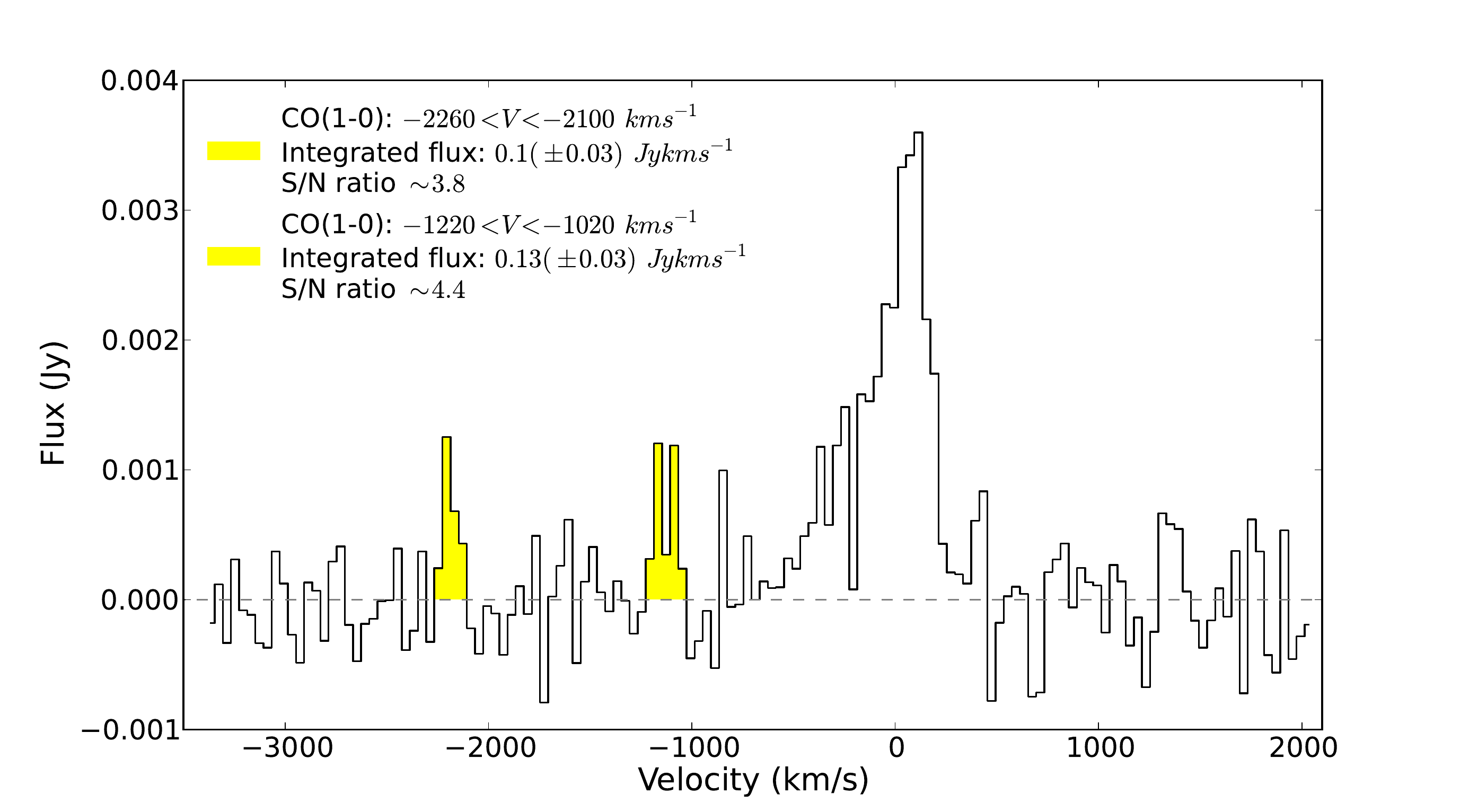}
                \includegraphics[width=6cm]{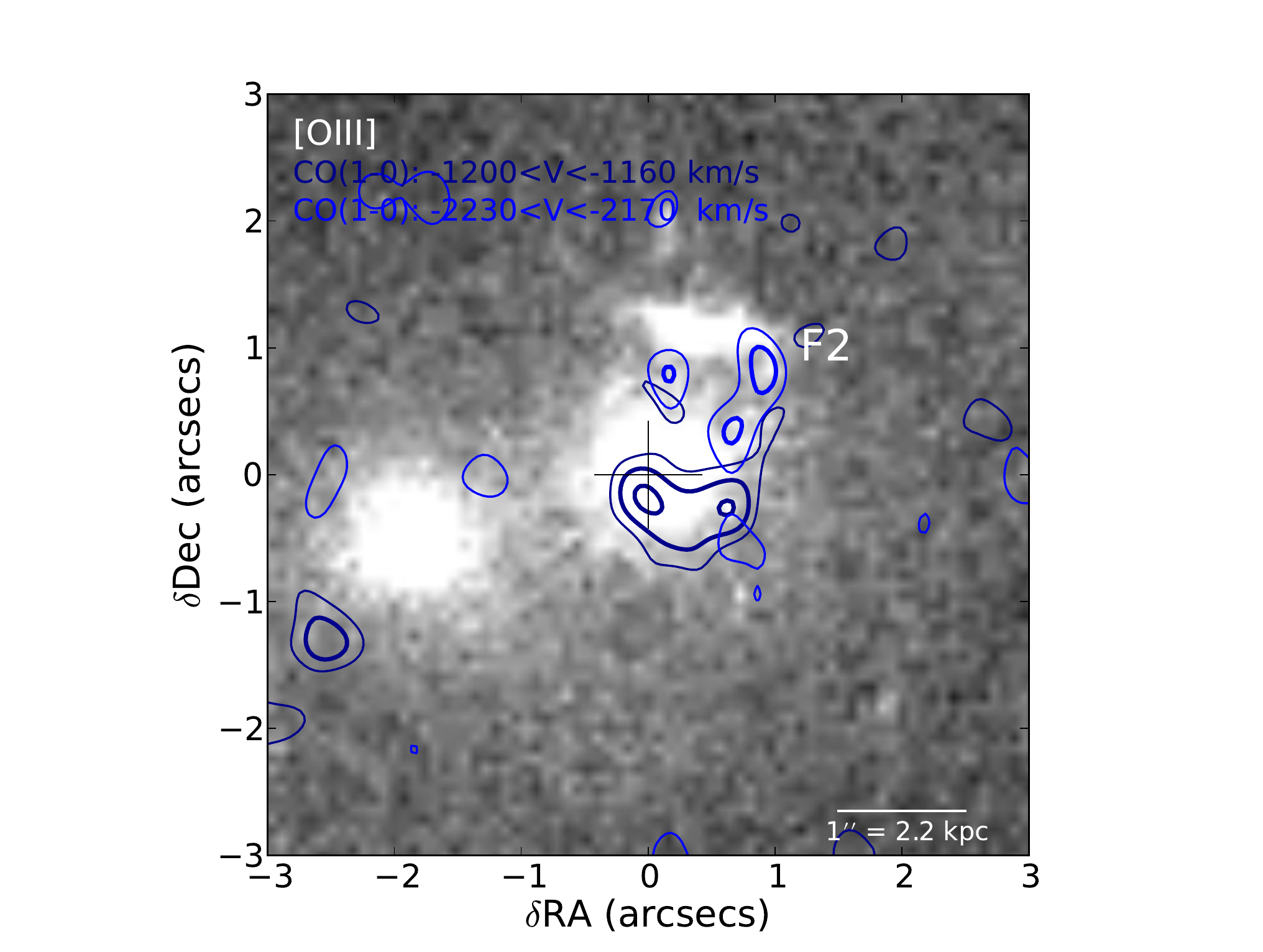}                          
                \includegraphics[width=6cm]{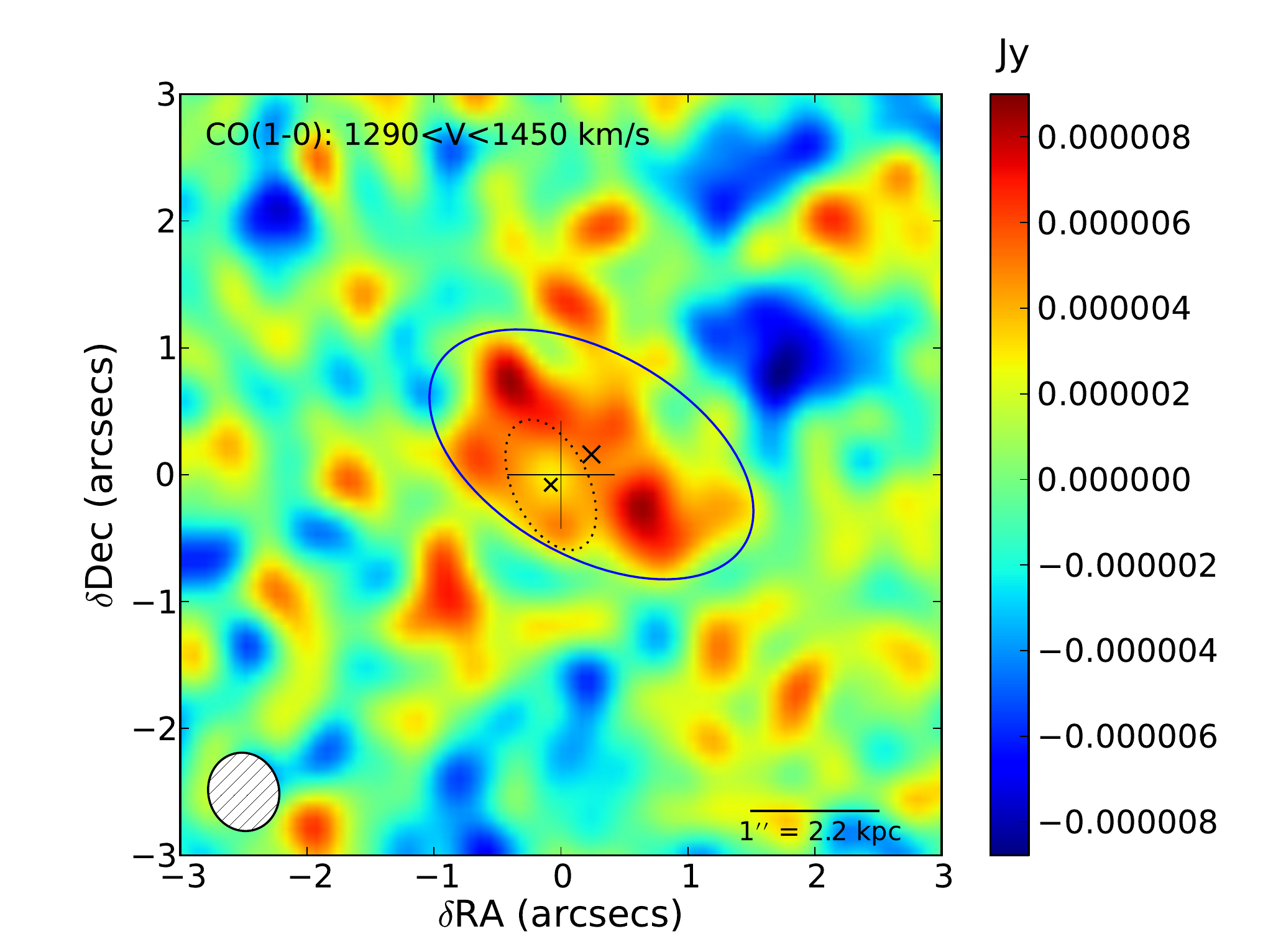}
                \includegraphics[width=6cm,height=4.5cm]{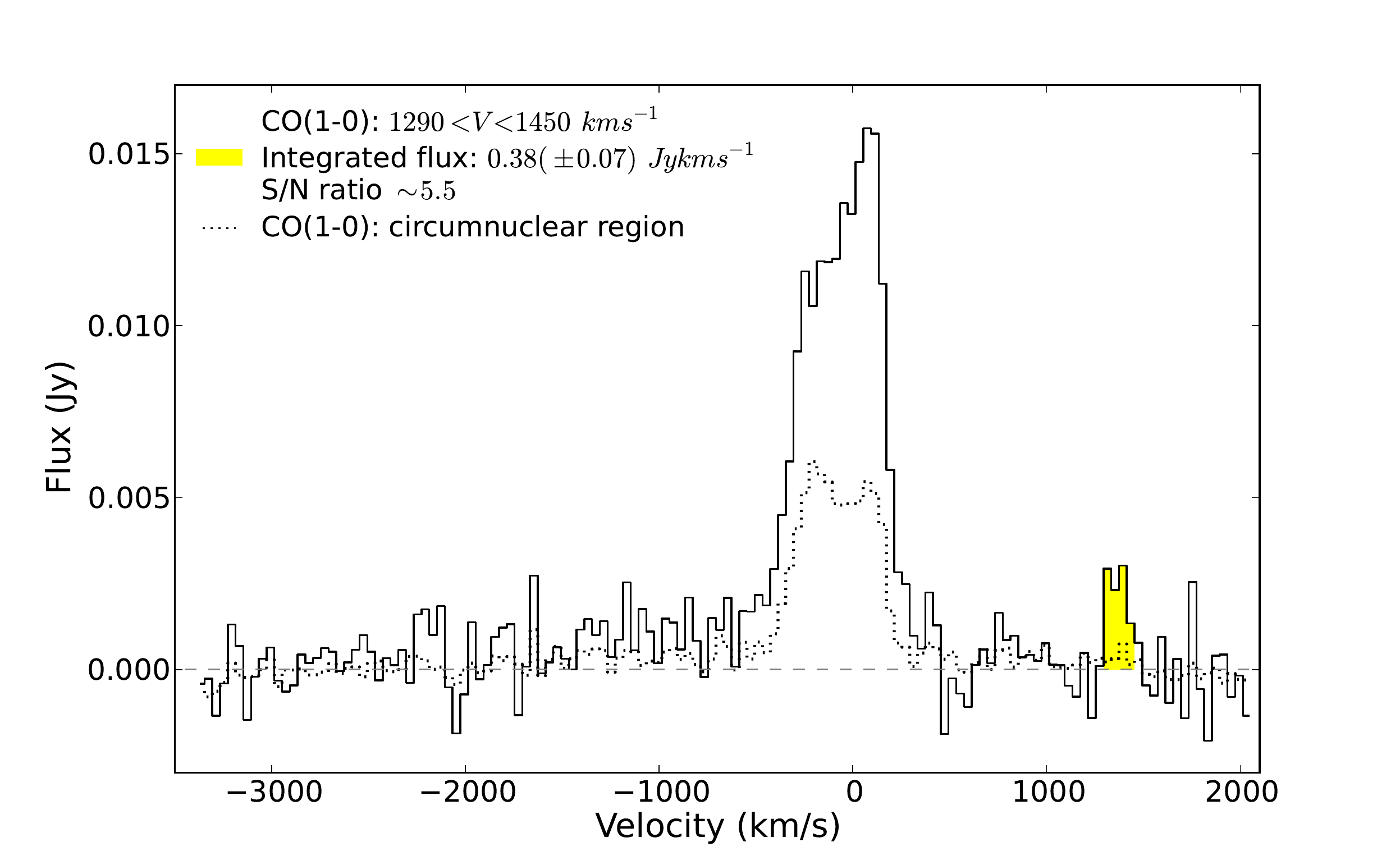}       
                \includegraphics[width=6cm]{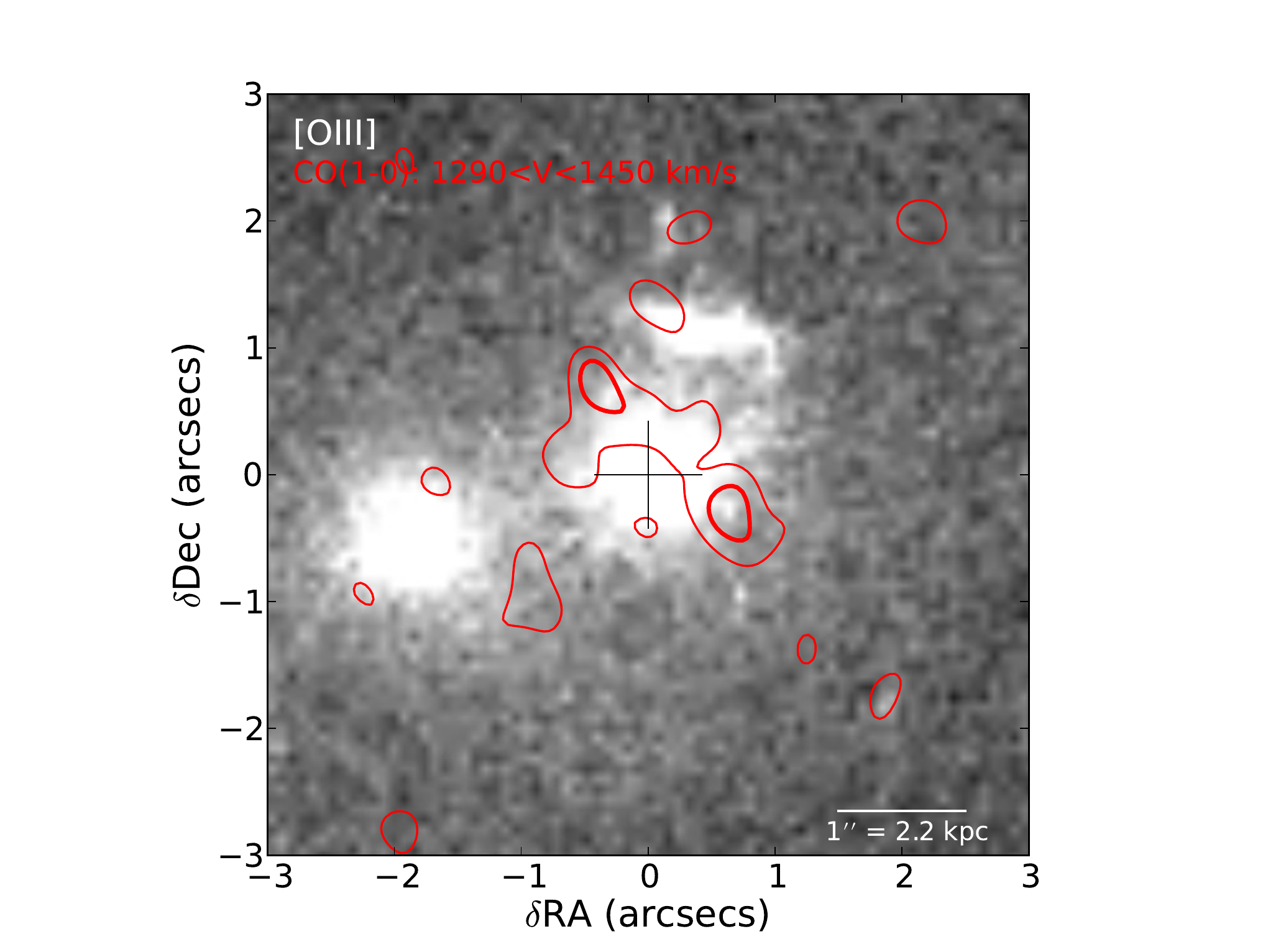}          
                \includegraphics[width=6cm]{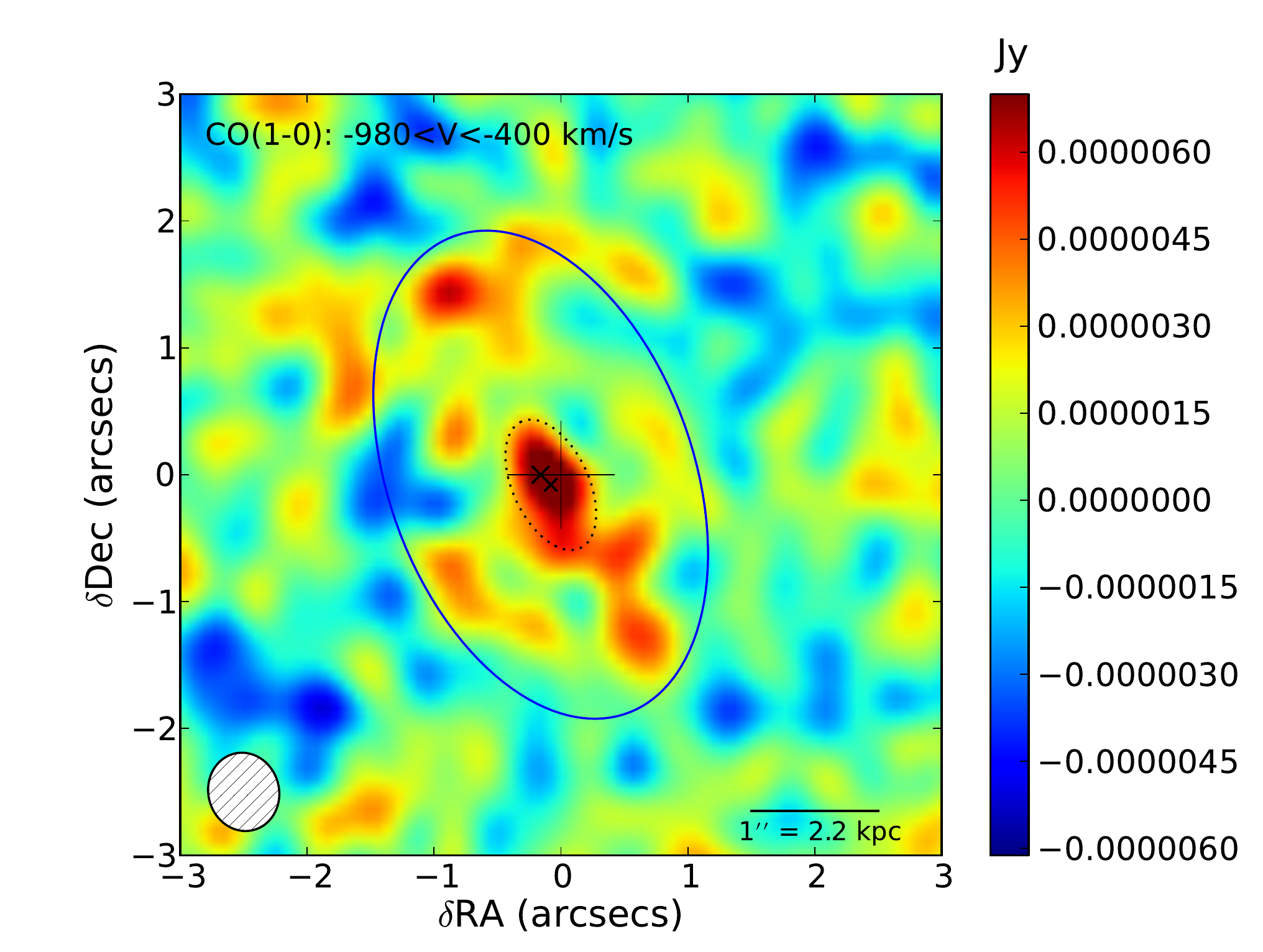}
                \includegraphics[width=6cm,height=4.5cm]{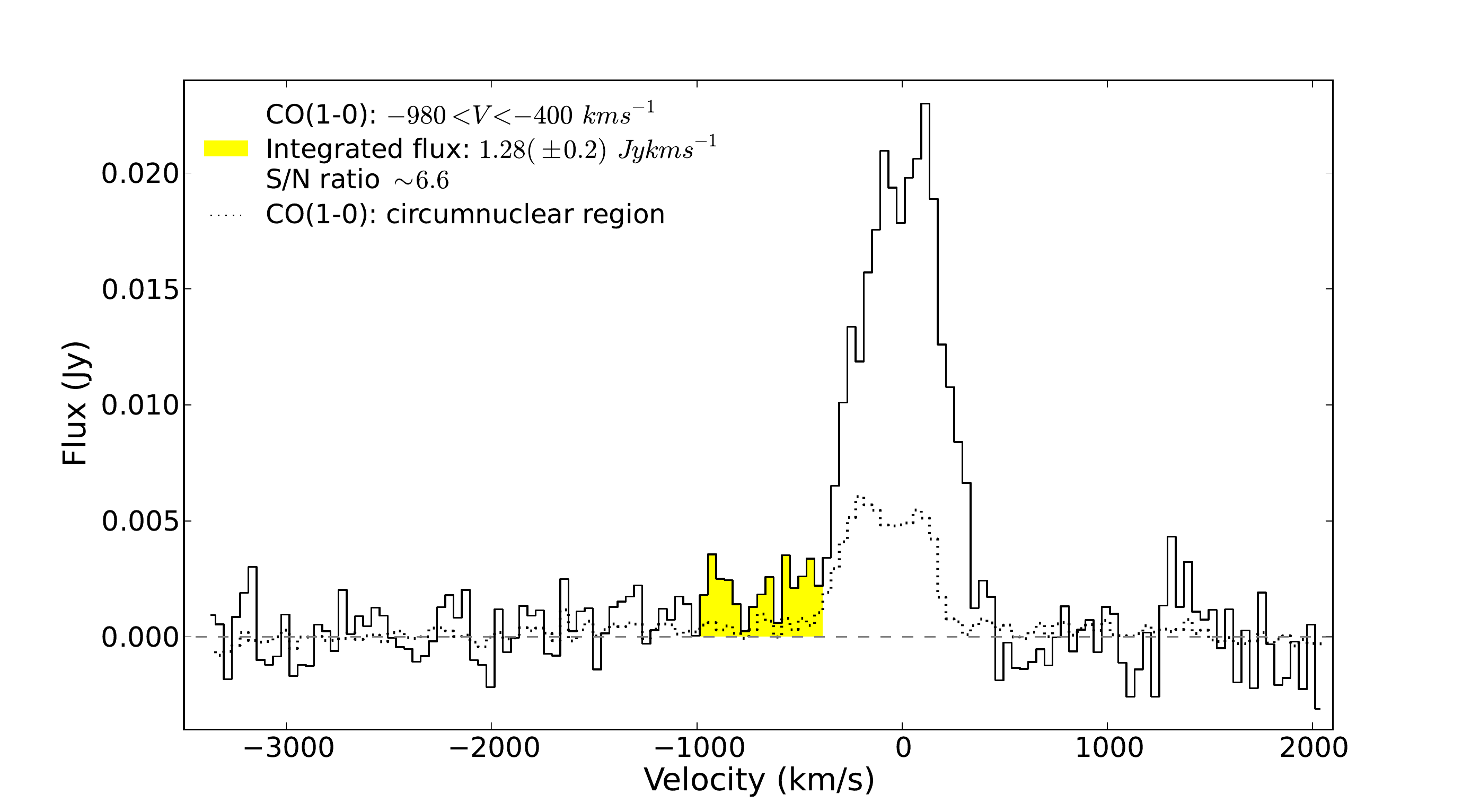}       
                \includegraphics[width=6cm]{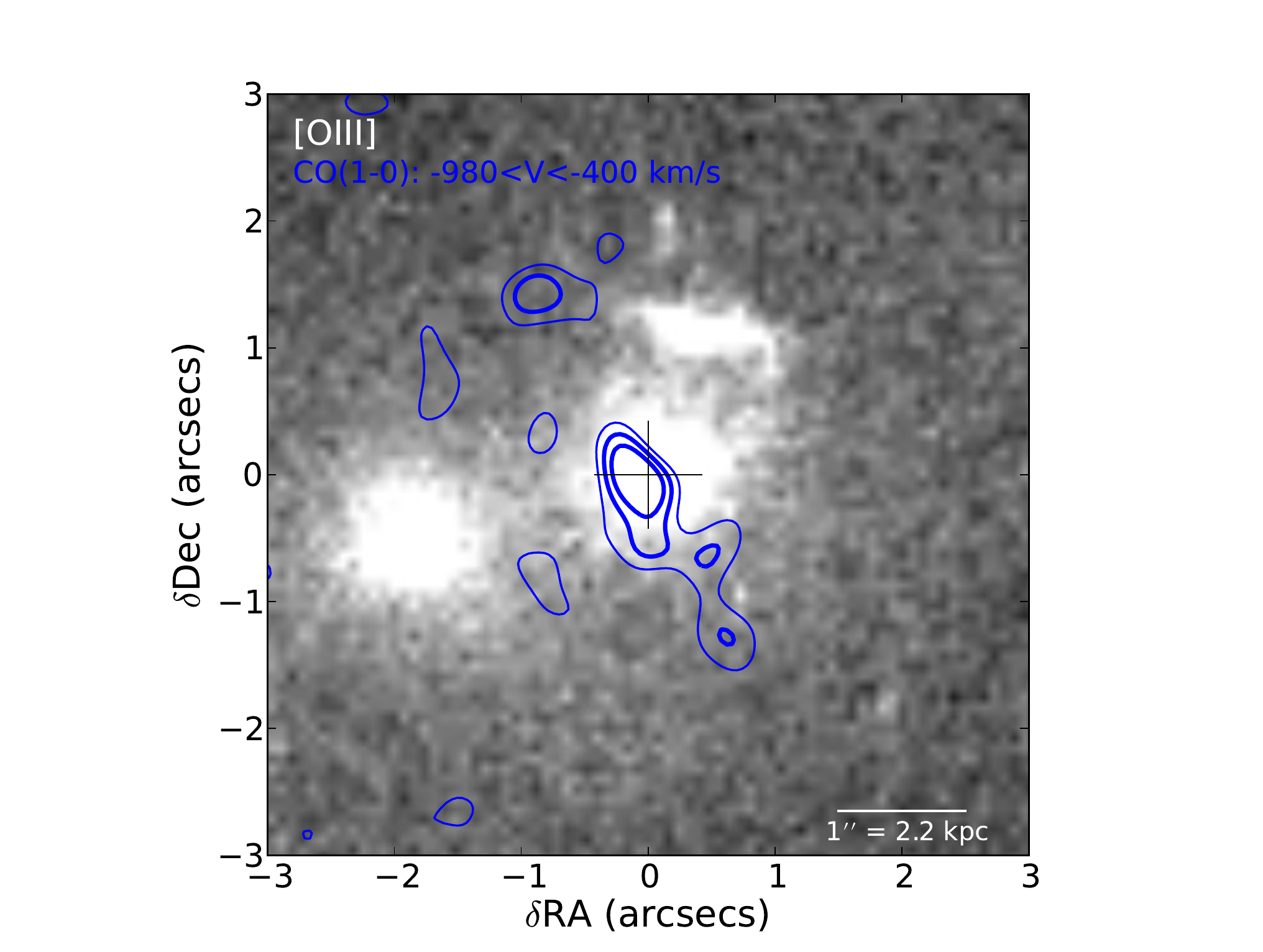}                                                                                                  
                \caption[]{ Tentative CO detections of high-velocity clouds in directions radially extending from the main nucleus to outer parts of the galaxy. The contour levels in the right panels are: 2$\sigma$,3$\sigma$ and 4$\sigma$. In the last two panels, the solid-line and the dotted spectra (middle) correspond to the emission of the solid-line and the dotted ellipses (left), respectively. The latter marks the extent of the nuclear wind which is shown in Fig. \ref{fig:cen_wind_detection}.
                }
                \label{fig:radial_outflows}
        \end{center}
\end{figure*}

High-velocity clouds are detected tentatively along these filaments in the ALMA data, with signal-to-noise ratios between 3 and 5 (Fig. \ref{fig:radial_outflows}). In a filament along a position angle of -10\degr , designated as F1 in Fig. \ref{fig:radial_outflows} and prominently seen in \oiii , we see \cooz\ at about -2200\kms .  
The total \htwo\ gas mass in region F1 is 0.5($\pm$0.1)$\times$10$^8$\msun. 

\begin{figure*}
        \begin{center}
                \includegraphics[width=10cm]{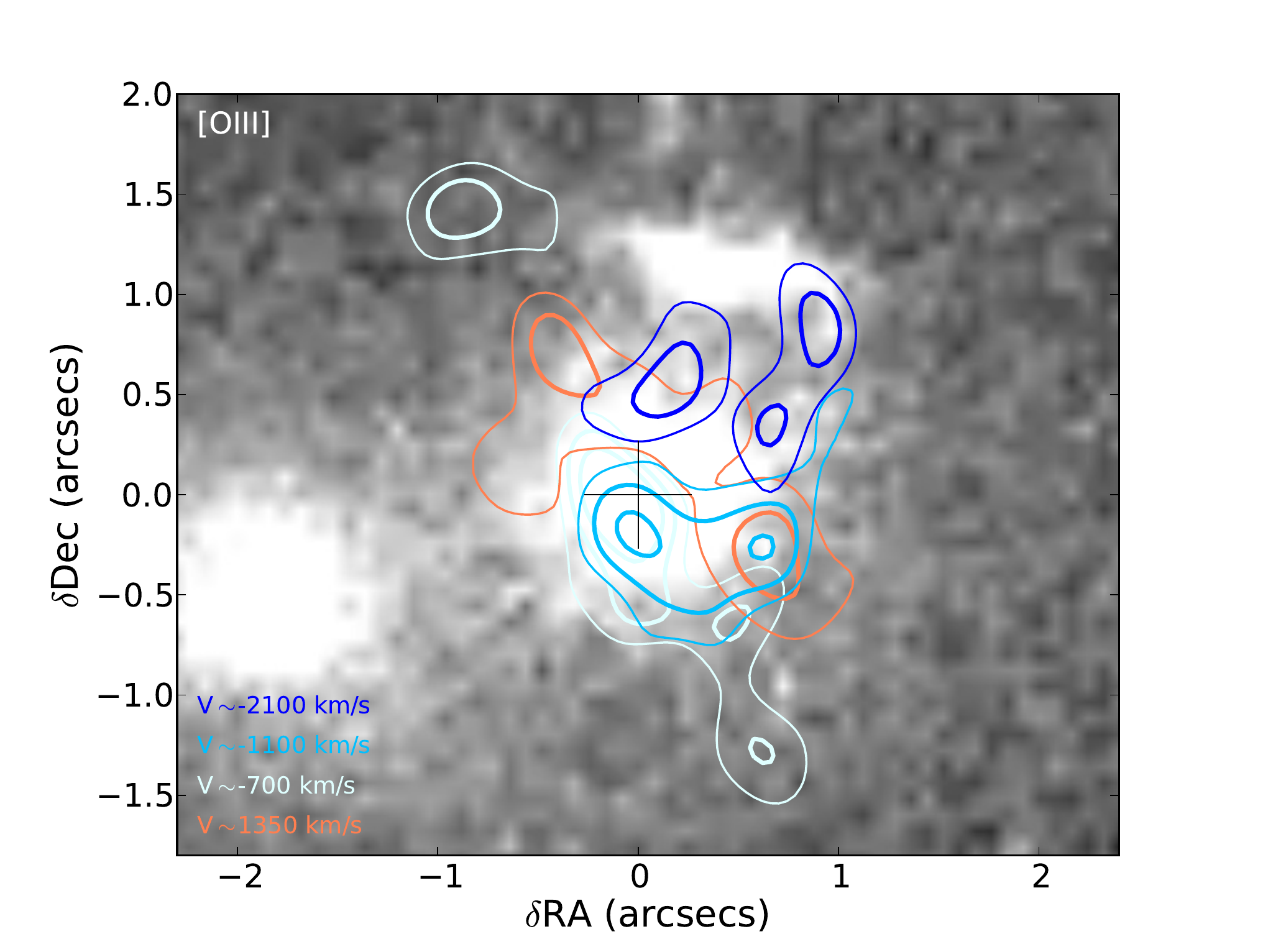}
                \includegraphics[width=10cm]{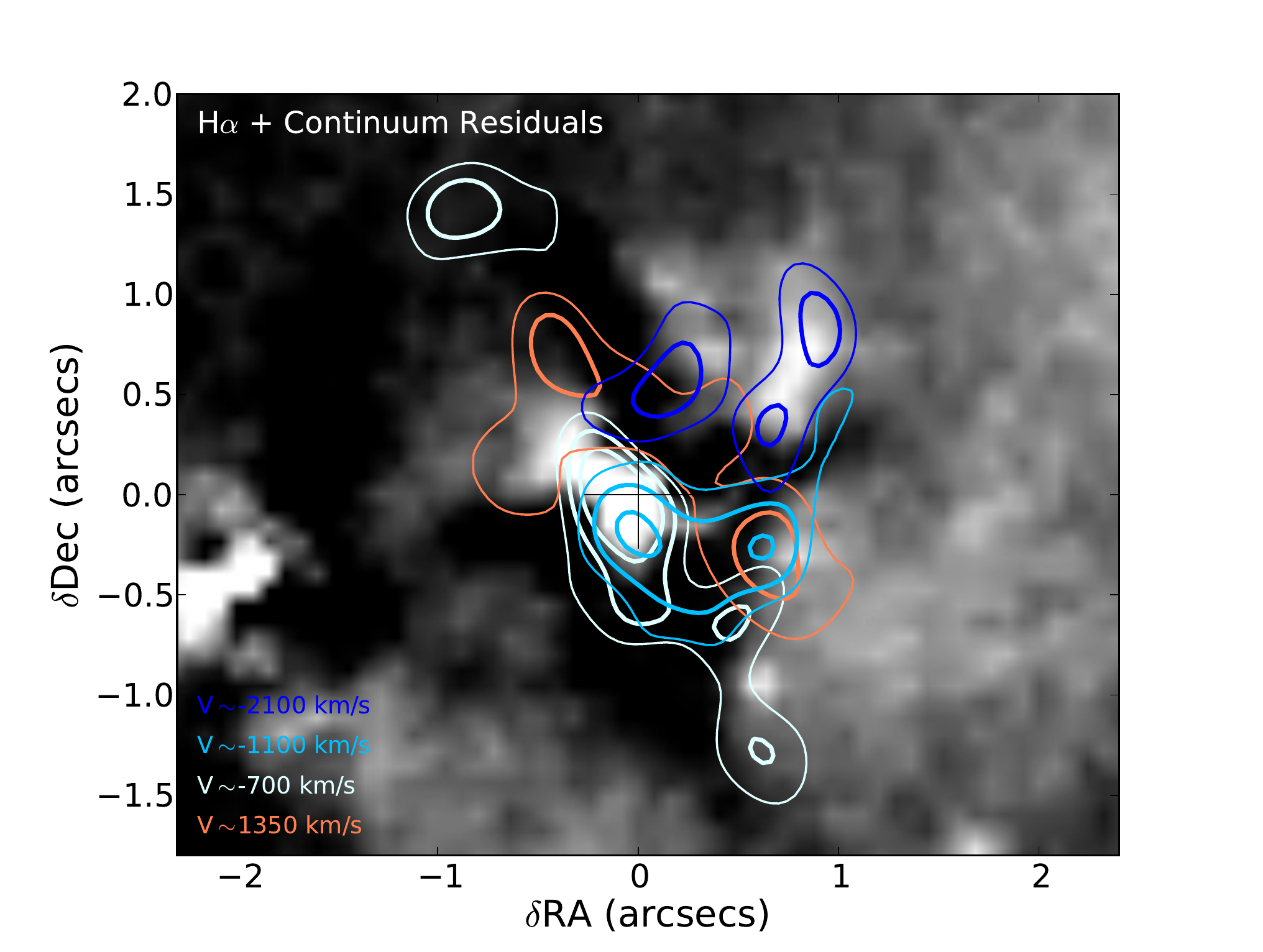}
                \includegraphics[width=10cm]{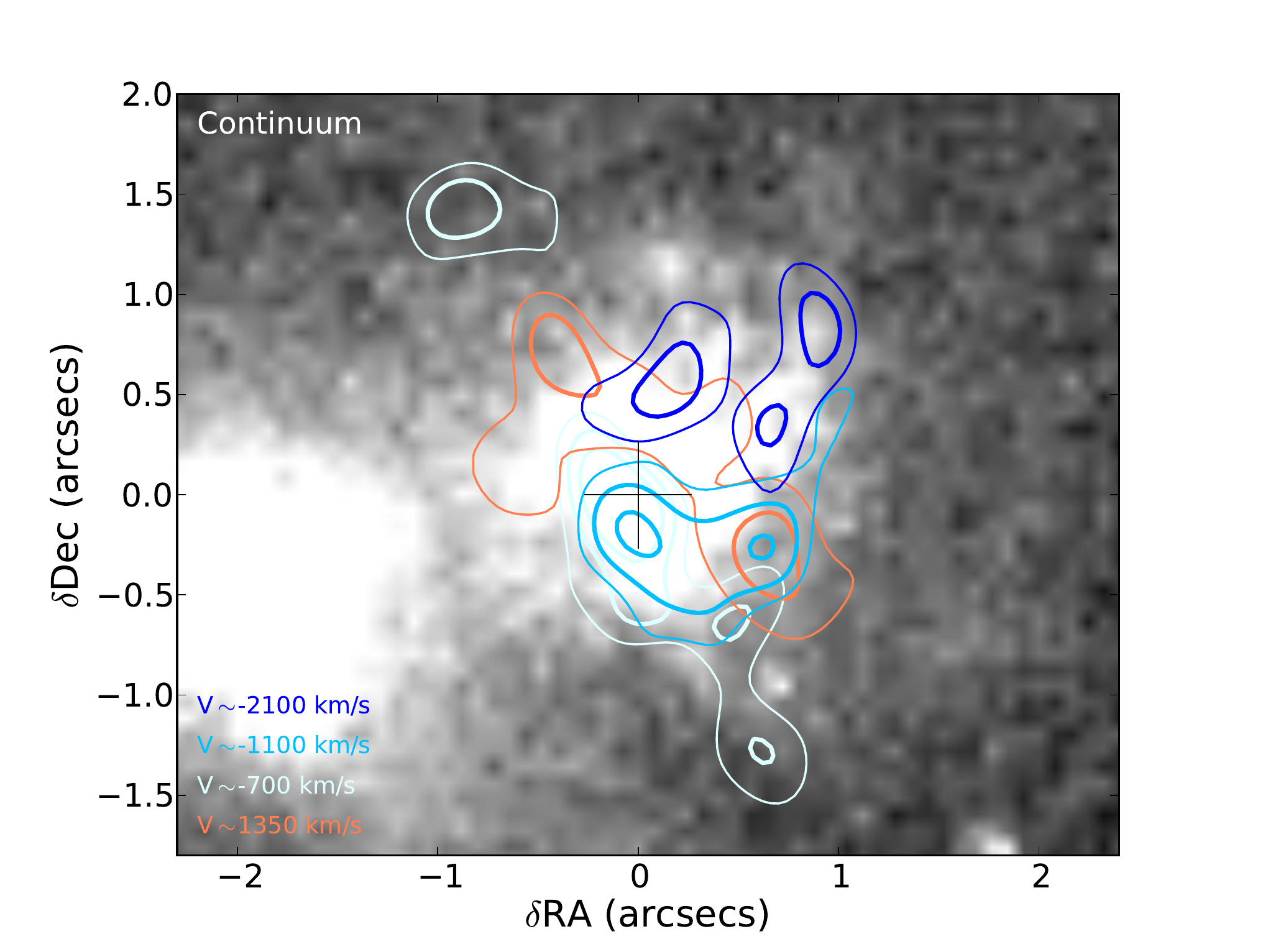}
                \caption[]{Loci of tentative detections of high-velocity molecular clouds compared to the stellar and warm, ionized gas emission. The contours are as in Fig.~\ref{fig:radial_outflows} for regions that exceed a S/N of 3. Contours of comparable velocity ranges are plotted with the same colors.}
                \label{fig:outflow_tot}
        \end{center}
\end{figure*}

In a second filament along a position angle of -30\degr\ that is designated as F2 in  Fig. \ref{fig:radial_outflows} and that is prominently seen in the stellar continuum (Fig.~\ref{fig:outflow_tot}), we again detect molecular gas. This time, we see \cooz\ emission with velocities of about -1200\kms , -2200 \kms , and with a corresponding  \htwo\ mass equal to 1.3($\pm$0.2)$\times$10$^8$\msun.  
The gas in or near this filament is counter-rotating in all CO lines. Moreover, this is the only extra-nuclear region with \coft\ detection of gas out of dynamical equilibrium; this filament is very nicely outlined in the third panel of the last row of Fig. \ref{fig:disk_detections} for gas counter-rotating with respect to the local disk velocity field.
The molecular gas mass in F1 and F2 is 1.8($\pm$0.2)$\times$10$^8$\msun .

In both filaments F1 and F2, we also observe marginal emission at positive velocities. Focusing on velocities near 1370\kms\, we see that the redshifted emission has the shape of a ring, which is centered at the nucleus but occupies the whole disk. The \htwo\ mass inside it, excluding the nuclear component contribution, is calculated using the integrated  \cooz\ flux of the area between the ellipses of the left panel of the third row of Fig. \ref{fig:radial_outflows}: namely 1.8($\pm$0.4)$\times$10$^8$\msun\ .  
To ensure that all extranuclear components are accounted for, we extracted the spectrum from a larger area around the nucleus as shown in the last panels of Fig. \ref{fig:radial_outflows}.  
Subtracting the spectrum of the nuclear wind and integrating the \cooz\ flux in the velocity range: -980\kms$<$V$<$-400\kms\, we calculate the mass of the blueshifted extended emission to be: 6.0($\pm$1.0)$\times$10$^8$\msun .

Adding the mass probed by the redshifted emission disk-wide and the blueshifted extended emission 
to the mass probed by F1 and F2, we find that the total mass of the extended wind reaches 
1.0($\pm$0.1)$\times$10$^9$\msun\ and tentatively spreads over an area of roughly 40kpc$^2$.

\begin{figure*}
        \begin{center}
                \includegraphics[width=6cm]{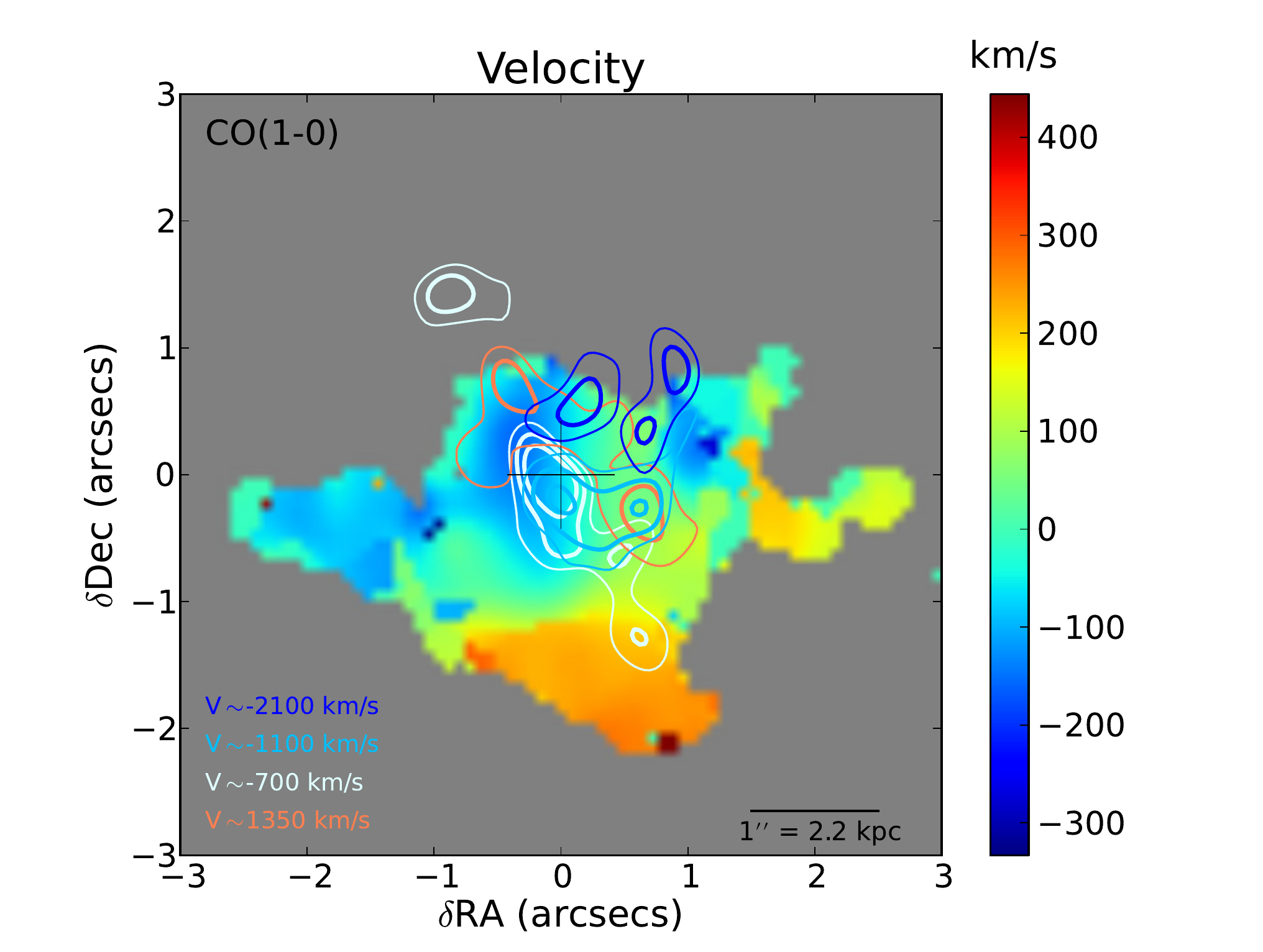}
                \includegraphics[width=6cm]{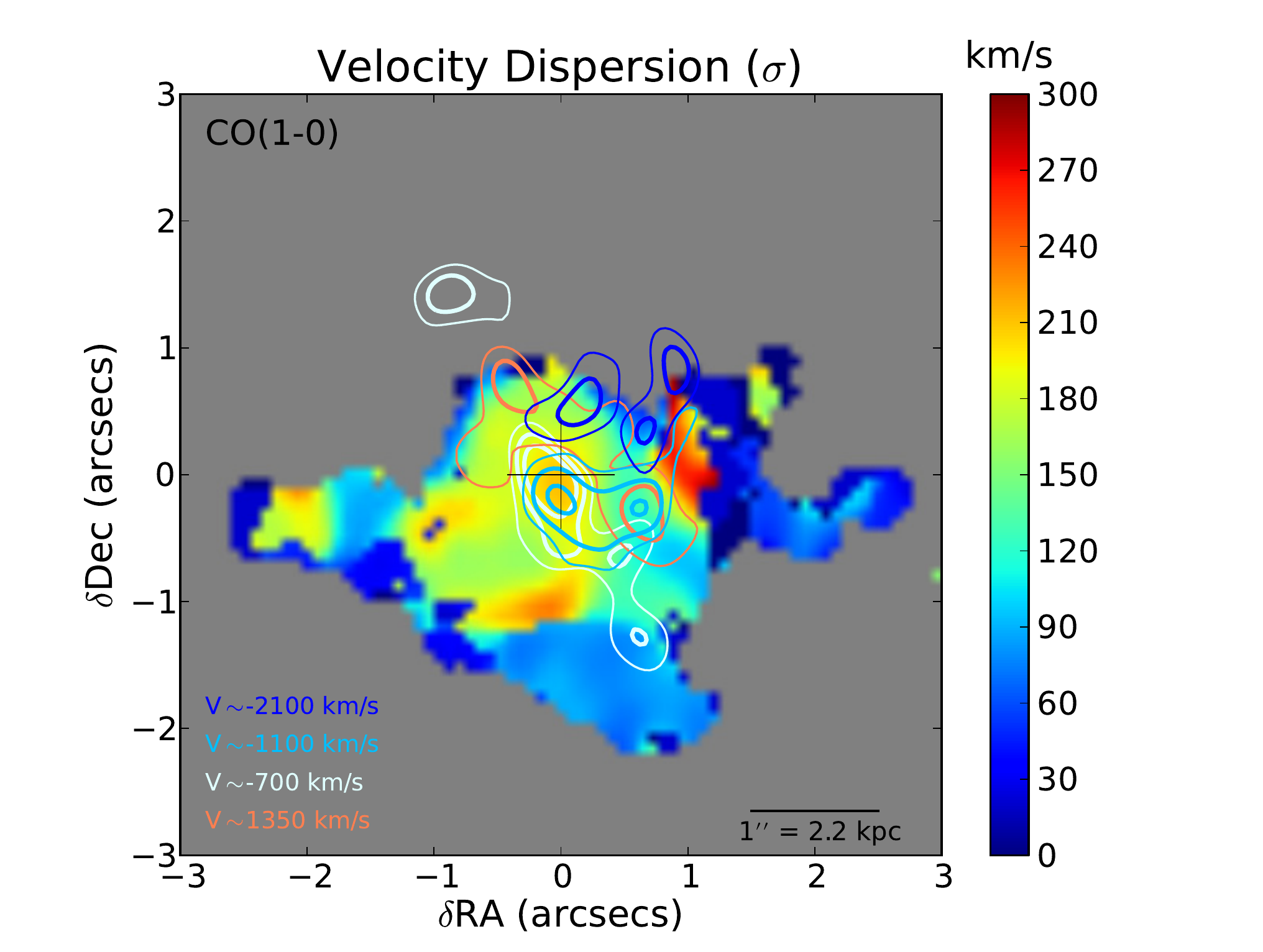}

                \includegraphics[width=6cm]{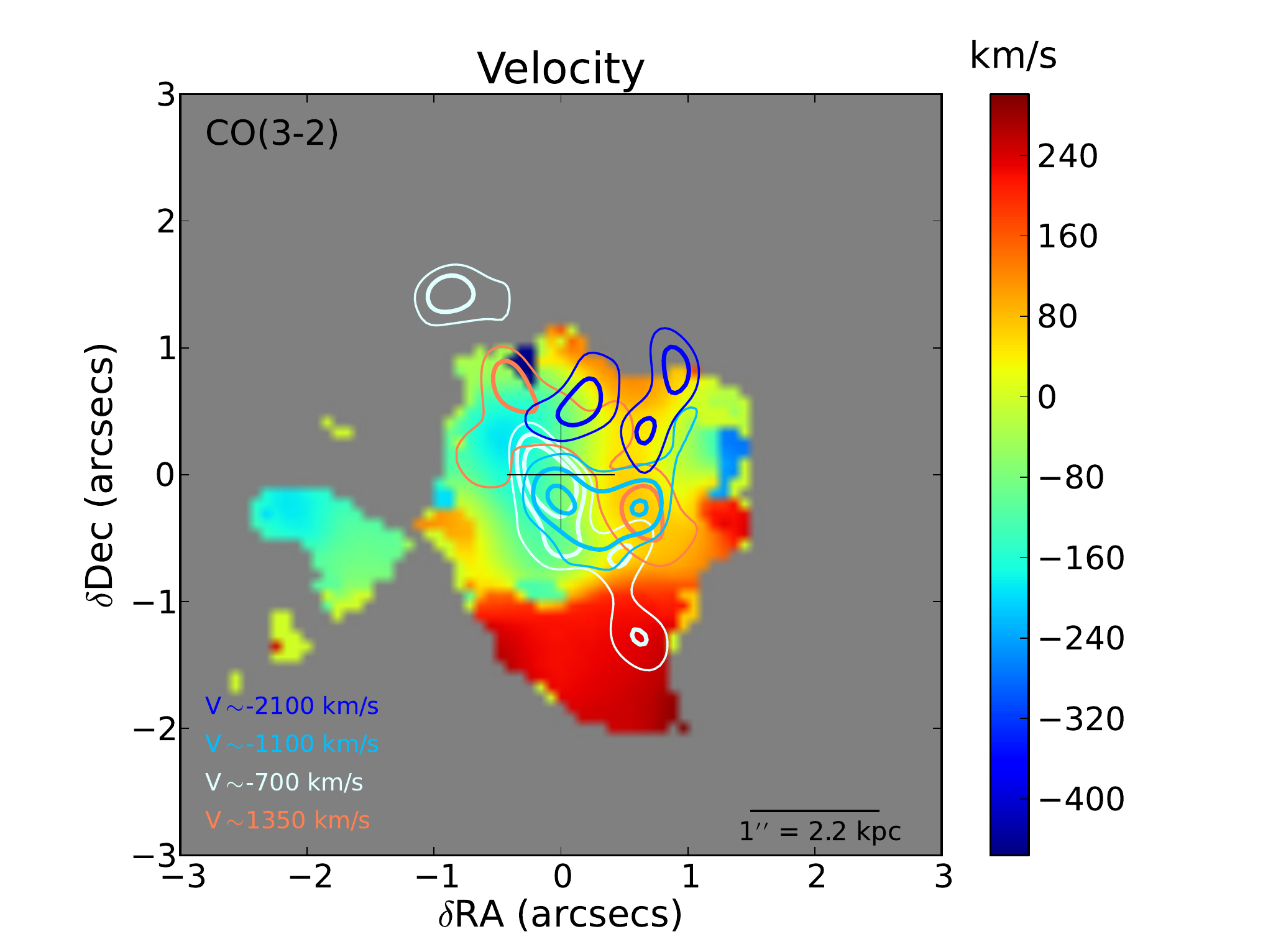}
                \includegraphics[width=6cm]{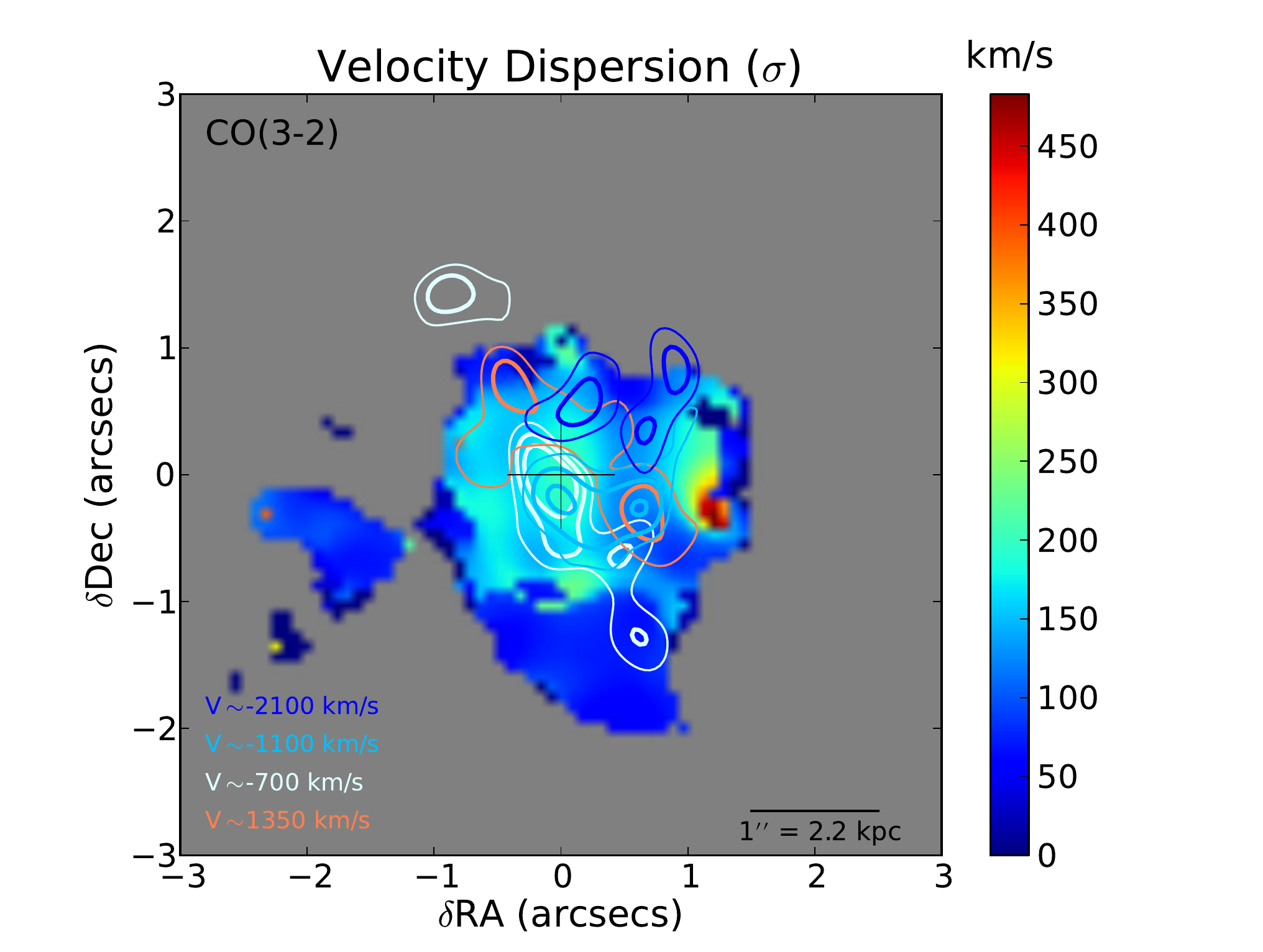}
                \caption[]{ Loci of tentative detections of high-velocity molecular clouds compared to the velocity and the velocity dispersion maps of \cooz\ and \cott\ of Fig. \ref{fig:momenta}. The contours are as in Fig.~\ref{fig:outflow_tot}.
                }                               
                \label{fig:extended_over_maps}                  
        \end{center}
\end{figure*}

Several arguments are in favor of the credibility of the wind detections in the filaments and in the extended disk-wide region despite their low S/N ratios. 
At $\sim$0.7\arcsec\ blueshifted and redshifted emission partially overlaps in the disk-wide region. The detections are located along a line that connects the main nucleus to the region of highest-velocity dispersion, as shown in Fig. \ref{fig:extended_over_maps}. This is observed in both \cooz\ and \cott\ maps.
The shapes and the loci of the high-velocity blobs F1 and F2 coincide with those of their optical filamentary counterparts (in either the \oiii\ or stellar continuum image; Fig.~\ref{fig:outflow_tot}). 
The \cooz\ emission in F1 and F2 has components in two or more velocity ranges.
Additionally, F2 is detected in both \cooz\ and \coft .
Some of the blobs also trace the excess emission that we see after the subtraction of a bulge and two disks from the \ha +continuum image (Fig.~\ref{fig:outflow_tot}). 
The detection of stars along some regions  also indicates two potential origin mechanisms for the accelerated gas: either it is related to a locally generated stellar wind or to an AGN generated nuclear wind. We evaluate the most likely scenario in Section~\ref{sec:discussion}, noting that the conclusions we draw about the origin of localized winds hold even if some detections turn out not to be real.

\subsection{Gas excitation}\label{sec:res_excitation}
\label{sec:excitation}

\begin{figure*}
        \begin{center}
                \includegraphics[width=6cm]{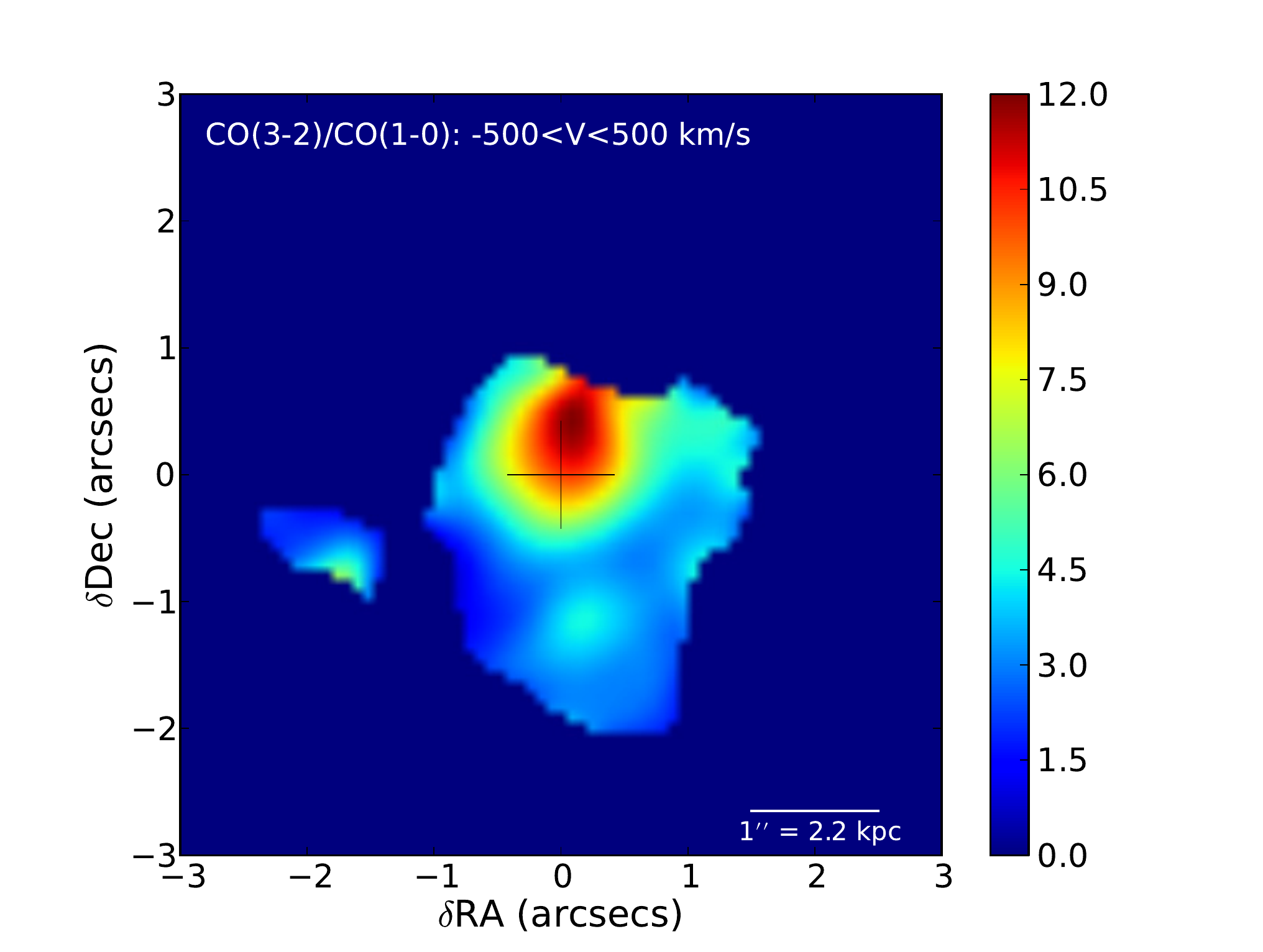}
                \includegraphics[width=6cm]{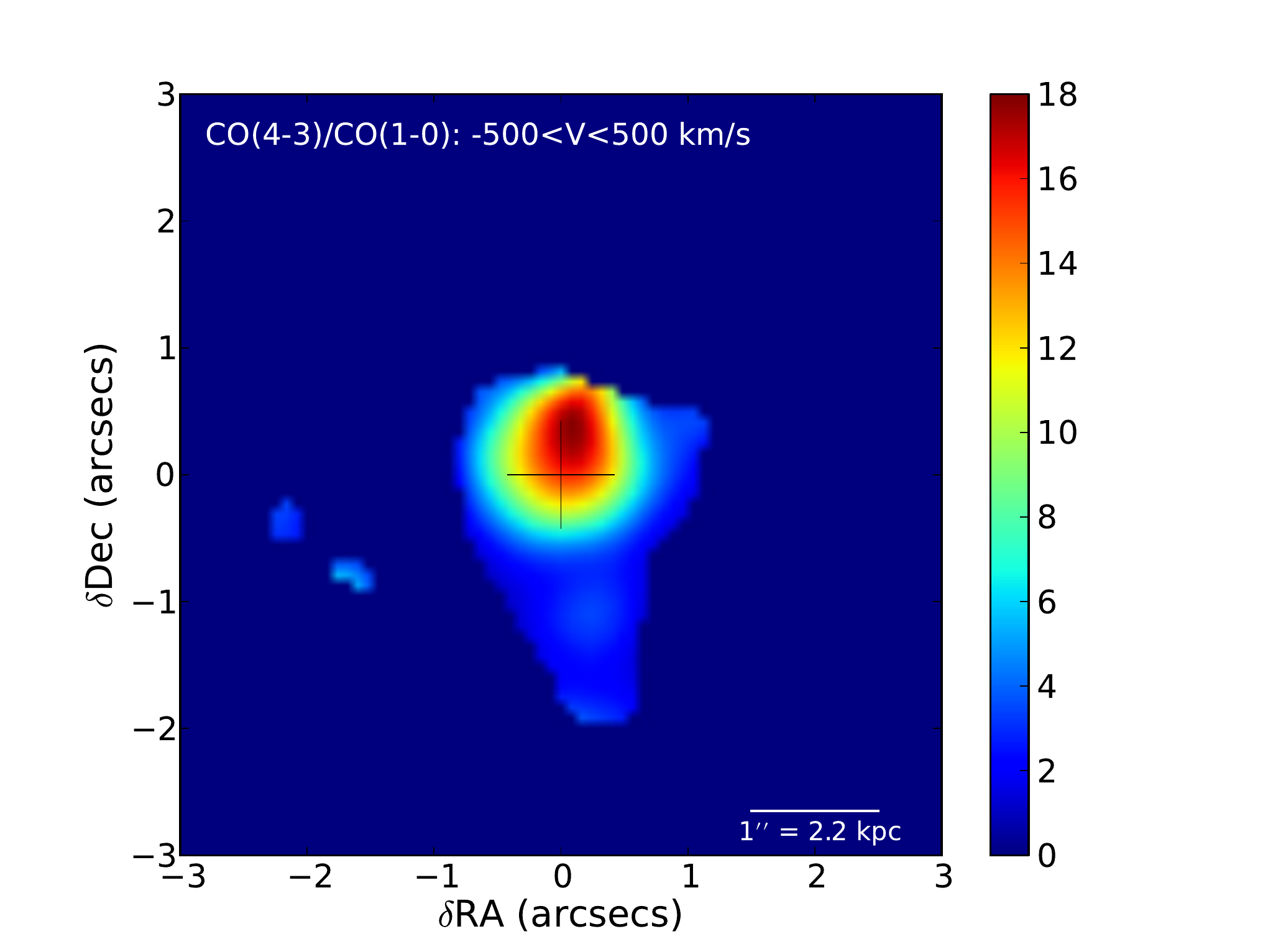}
                \includegraphics[width=6cm]{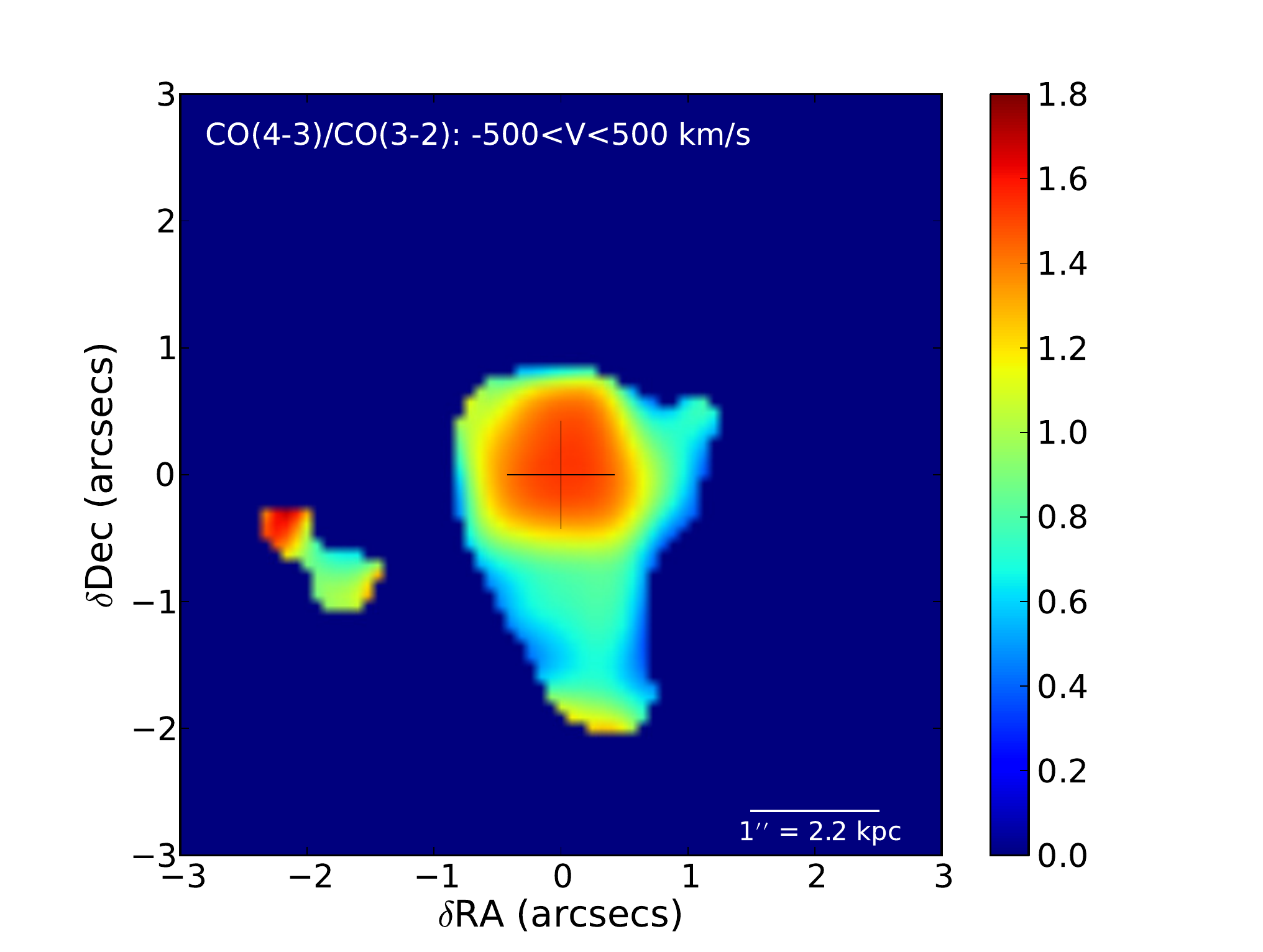}

                \includegraphics[width=6cm]{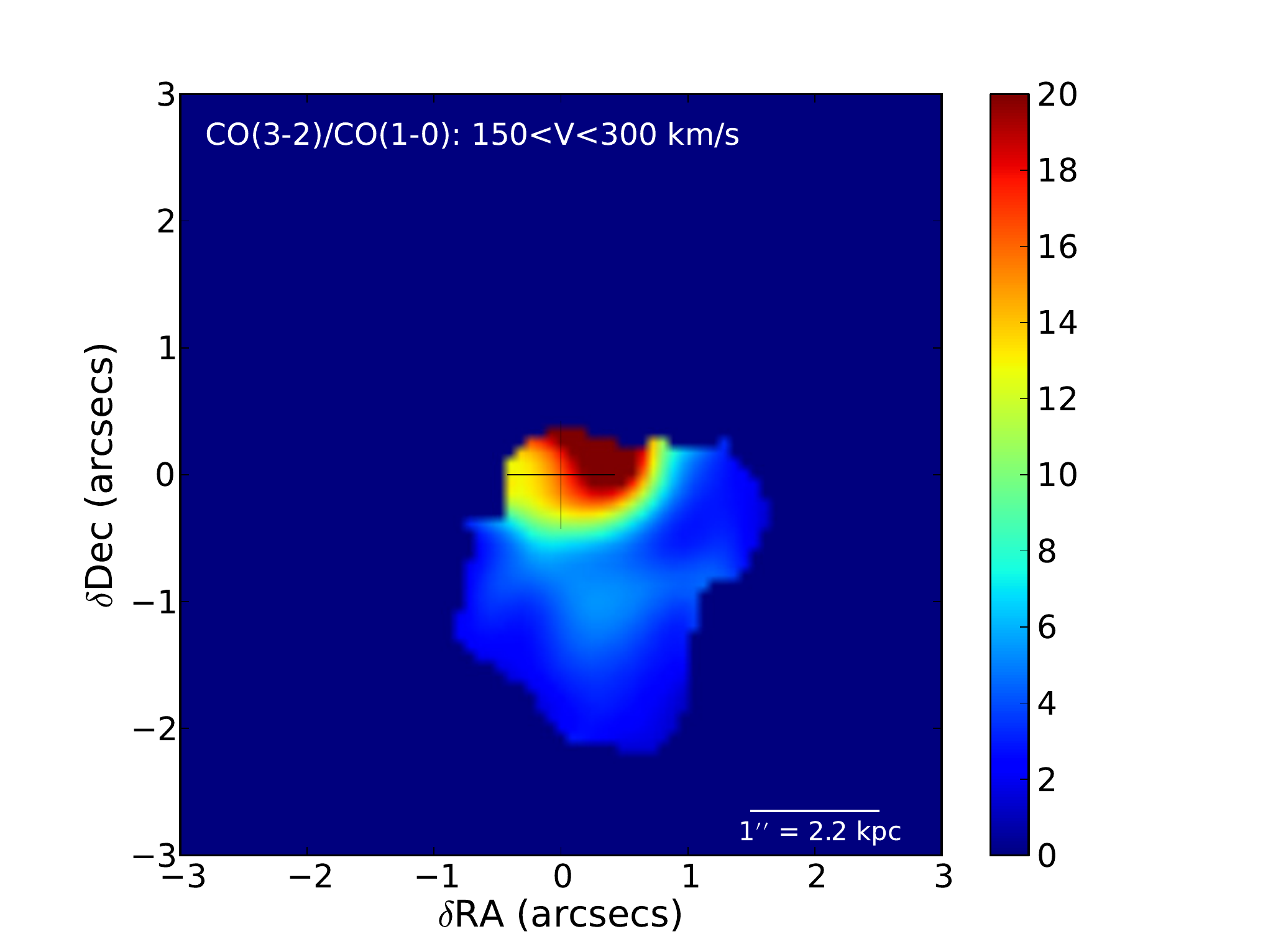}
                \includegraphics[width=6cm]{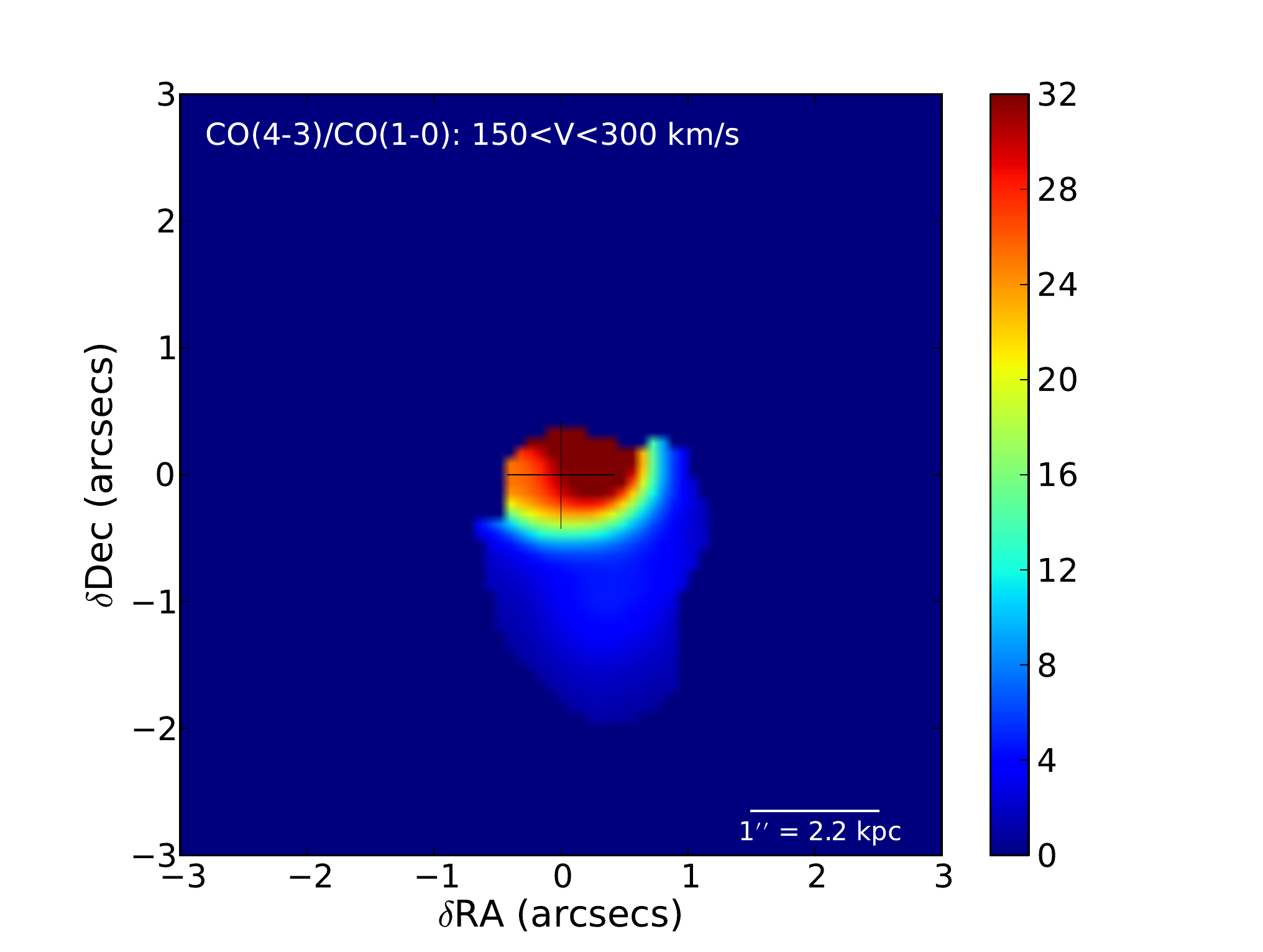}
                \includegraphics[width=6cm]{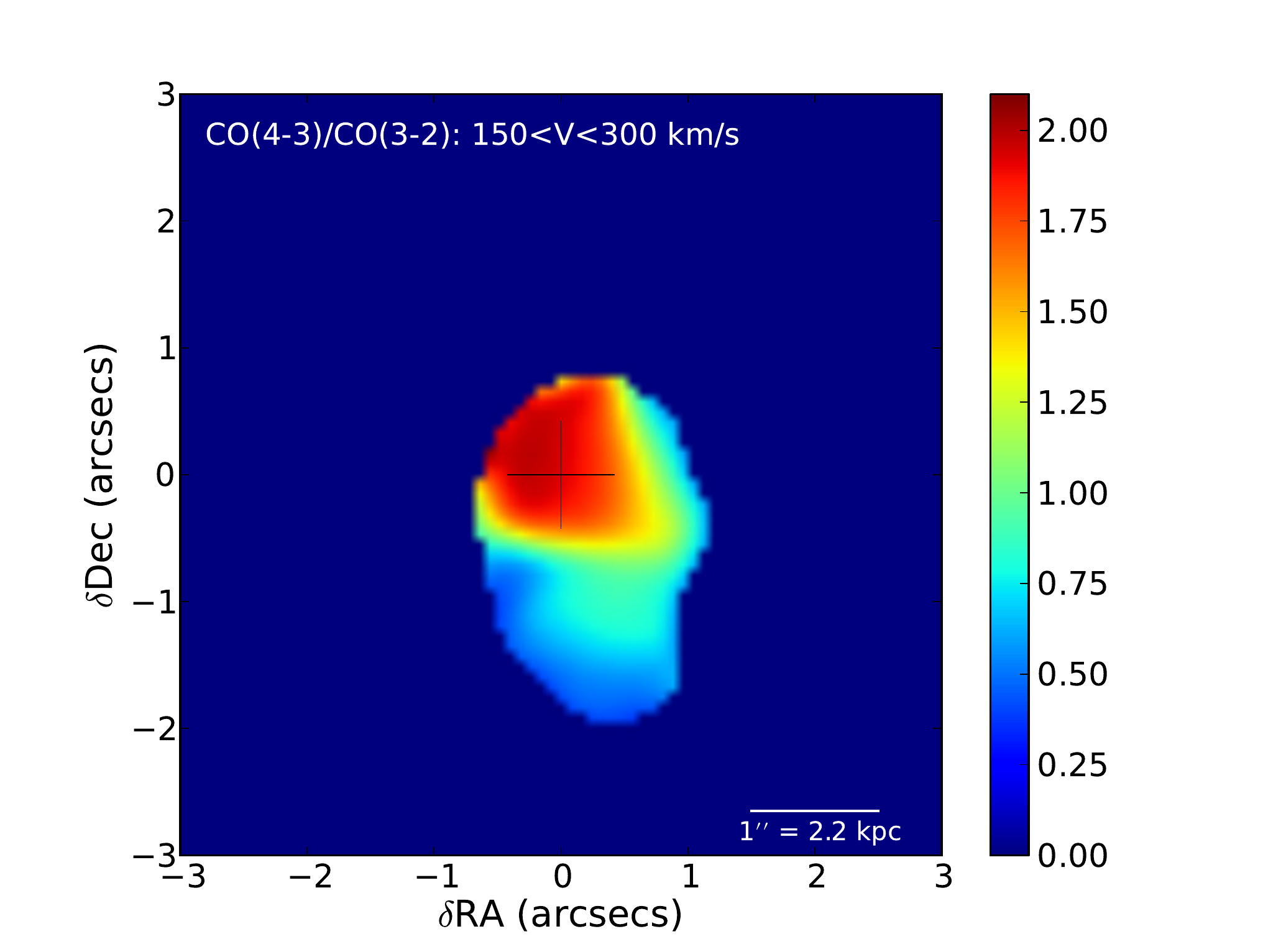}

                \includegraphics[width=6cm]{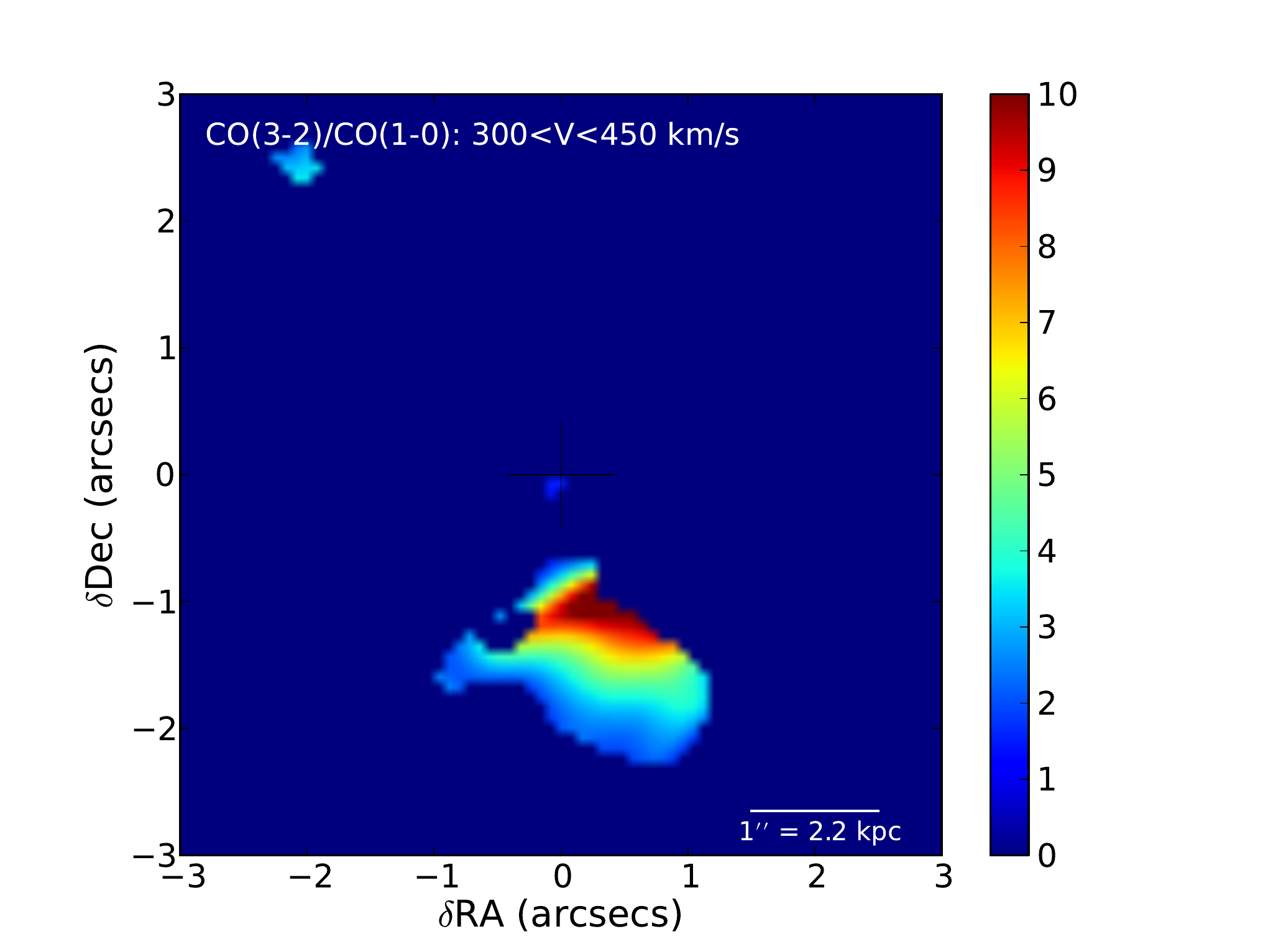}
                \includegraphics[width=6cm]{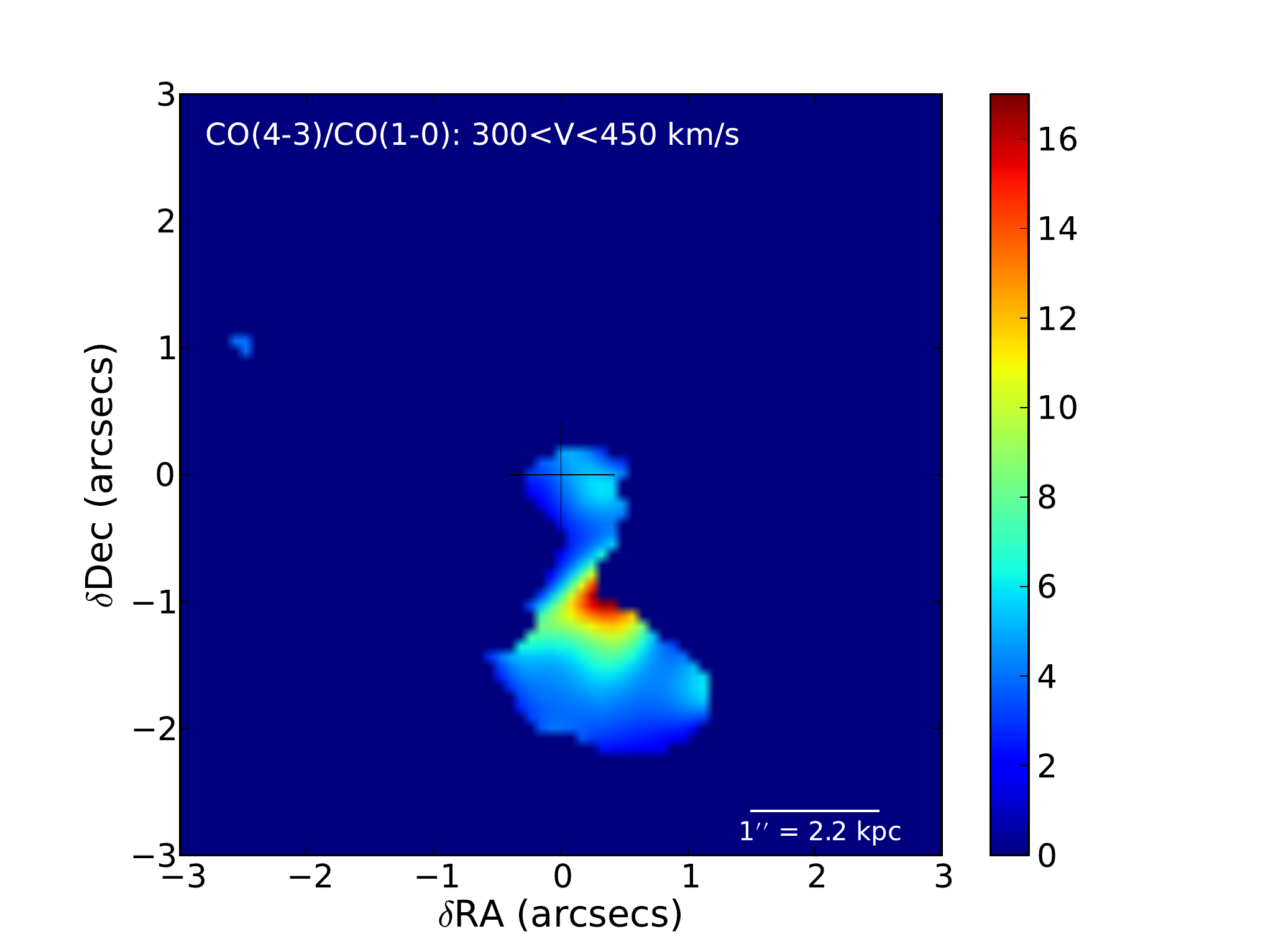}
                \includegraphics[width=6cm]{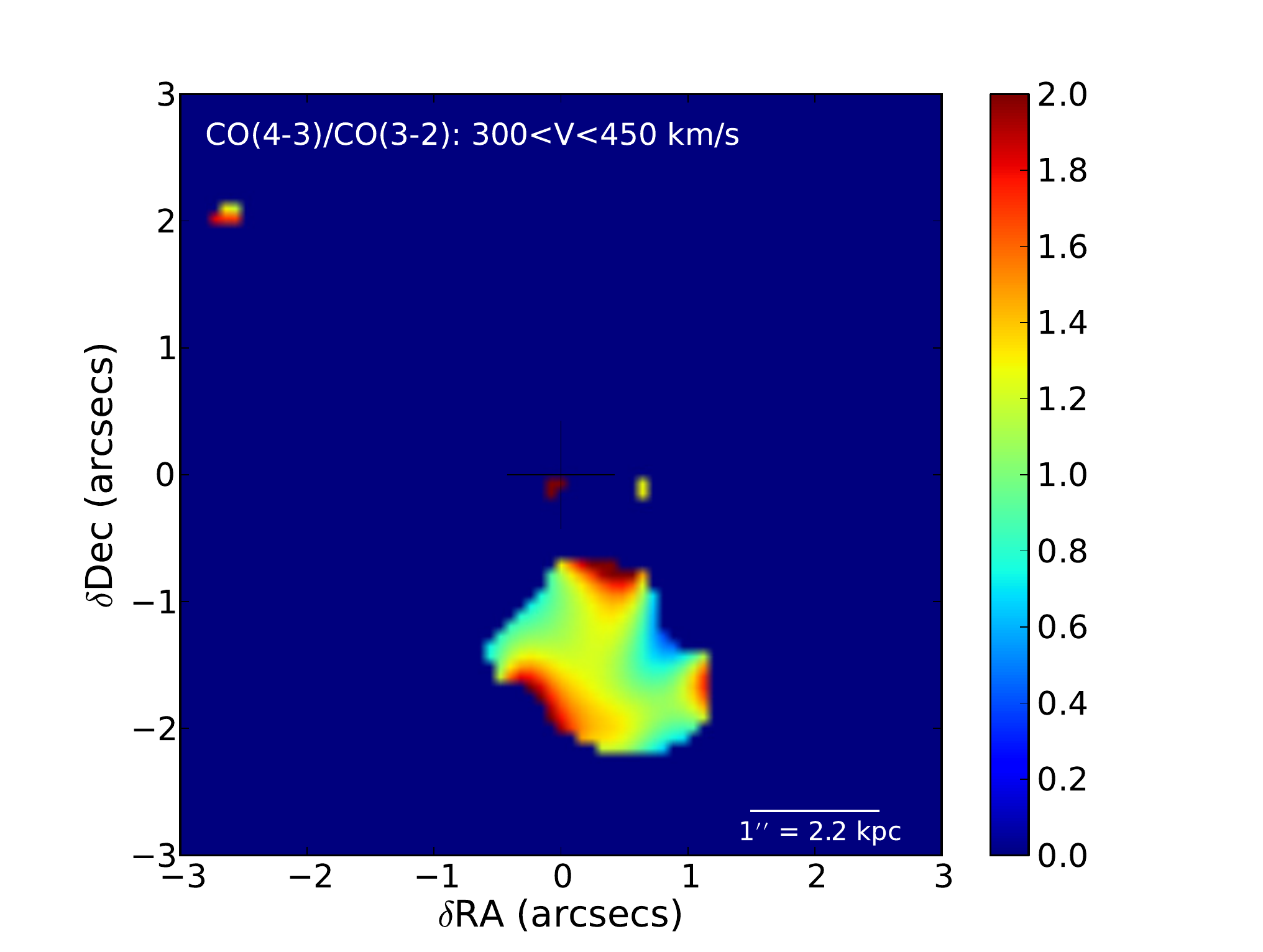}               
                \caption[]{\cott$/$\cooz  , \coft$/$\cooz\ and \coft$/$\cott\ flux ratios. Values exceeding 9, 16 and 1.78 for
                        \cott$/$\cooz\ , \coft$/$\cooz\ and \coft$/$\cott\ respectively, indicate the presence of optically thin gas.
                         First row: Flux ratios in the velocity range of the entire disk. Second row: Flux ratios in disk velocities that show high excitation close to the nucleus. Third row: Flux ratios in a sub-range of the velocities that shows excitation in the southern tail-like structure. 
                } 
                \label{fig:disk_excitation_highres}                     
        \end{center}
\end{figure*}

Because the accelerated gas can have different excitation from that of the ambient gas, CO flux ratios are examined as wind indicators. Observed flux ratios, computed from the data at the common beam resolution,  are shown in Fig.~\ref{fig:disk_excitation_highres}. Clear differences are seen between the excitation of the nucleus and that of the southern tidal structure. The gas in the southern structure is subthermally excited.  A range of excitations is seen between this region and the disk. Contrarily, higher excitation is observed near the nucleus. At positive velocities, some gas displays a \coft$/$\cott\  ratio above the value of 1.78, which is the upper limit in case of optically thick gas emission. Excitation temperatures from these maps will be presented in a forthcoming paper (Paraschos et al. 2019, in prep.).
 
Potential excitation differences can also be revealed by the comparison of multi-wavelength, multi-temperature gas probes. From previous \spi\ data, we know that in the warm  ($\sim$400$K$) \htwo\ phase, the outflow has a mass of 5.2$\times$10$^7$\msun \citep{dasyra11}, whereas the disk has a mass of 1.4$\times$10$^8$\msun \citep{dasyra11}. Therefore, the fraction of warm \htwo\ in the wind exceeds 30\%. From our new ALMA data, we find that the \htwo\ mass of the wind is as high as 1.5$\times$10$^9$\msun\, whereas that of the ambient gas is 7.7$\times$10$^{9}$\msun . Therefore, the fraction of cold \htwo\ in the wind is less than 20\%.  This confirms our past findings of heating of the accelerated gas. It is noteworthy that some of the cold gas mass in the wind may be unaccounted for due to the spatial overlap of multiple structures out of dynamical equilibrium in our ALMA data. Still, the mass loss cannot be as high as required to alter our initial conclusion. The effects of gas acceleration and heating could possibly delay star formation temporarily.

\begin{table}[h!]
\caption[]{Properties of tentative extranuclear wind detection}
\label{tab:extended_wind}
\scalebox{0.65}{
        \begin{tabular}{c c c c c c c} 
                \hline\hline 
                Region 
                & $M$ 
                & $d$ 
                & $\mid V_{mean}\mid$
                & $\dot{M}$  
                & $L_{kin}$  
                & $\dot{M}V$ \\  
                & ($\times$10$^8$\msun )
                & (\arcsec )
                & (\kms )
                & (\msun$yr^{-1}$) 
                & ($\times$10$^{44}$ $\mathrm{erg}\mathrm{s}^{-1}$)
                & ($\times$10$^{36}$ $\mathrm{erg}\mathrm{cm}^{-1}$)  \\                                  
                \hline 
                F1                              
                & 0.5($\pm$0.1)
                & 0.6
                & 2200
                & 90
                & 1.3
                & 1.2
                \\                      
                F2                              
                & 0.6($\pm$0.2)
                & 0.8
                & 2180
                & 70    
                & 1.1
                & 1.0   
                \\
                F2                              
                & 0.7($\pm$0.2)
                & 0.8
                & 1120
                & 50    
                & 0.2
                & 0.3           
                \\
                \begin{tabular}{@{}c@{}}redshifted \\ disk-wide\end{tabular}                            
                & 1.8($\pm$0.4)
                & 0.8
                & 1370
                & 140   
                & 0.9
                & 1.2
                \\
                \begin{tabular}{@{}c@{}}blueshifted \\ extended\end{tabular}                            
                & 6.0($\pm$1.0)
                & 1.2
                & 700
                & 160   
                & 0.3
                & 0.7
                \\
                \hline 
        \end{tabular}
}
\tablefoot{Using 0.1\arcsec\ as a typical error for $d$ and 10\kms\ as a typical error for $V_{mean}$, the errors in $\dot{M}$, $L_{kin}$, and $\dot{M}V$ are roughly 20\% of their values.}

\end{table}

\section{Discussion}
\label{sec:discussion}

\subsection{Energy output of the galaxy at radio and infrared wavelengths}\label{sec:energetics}

\begin{figure*}
        \begin{center}
                \includegraphics[width=10cm]{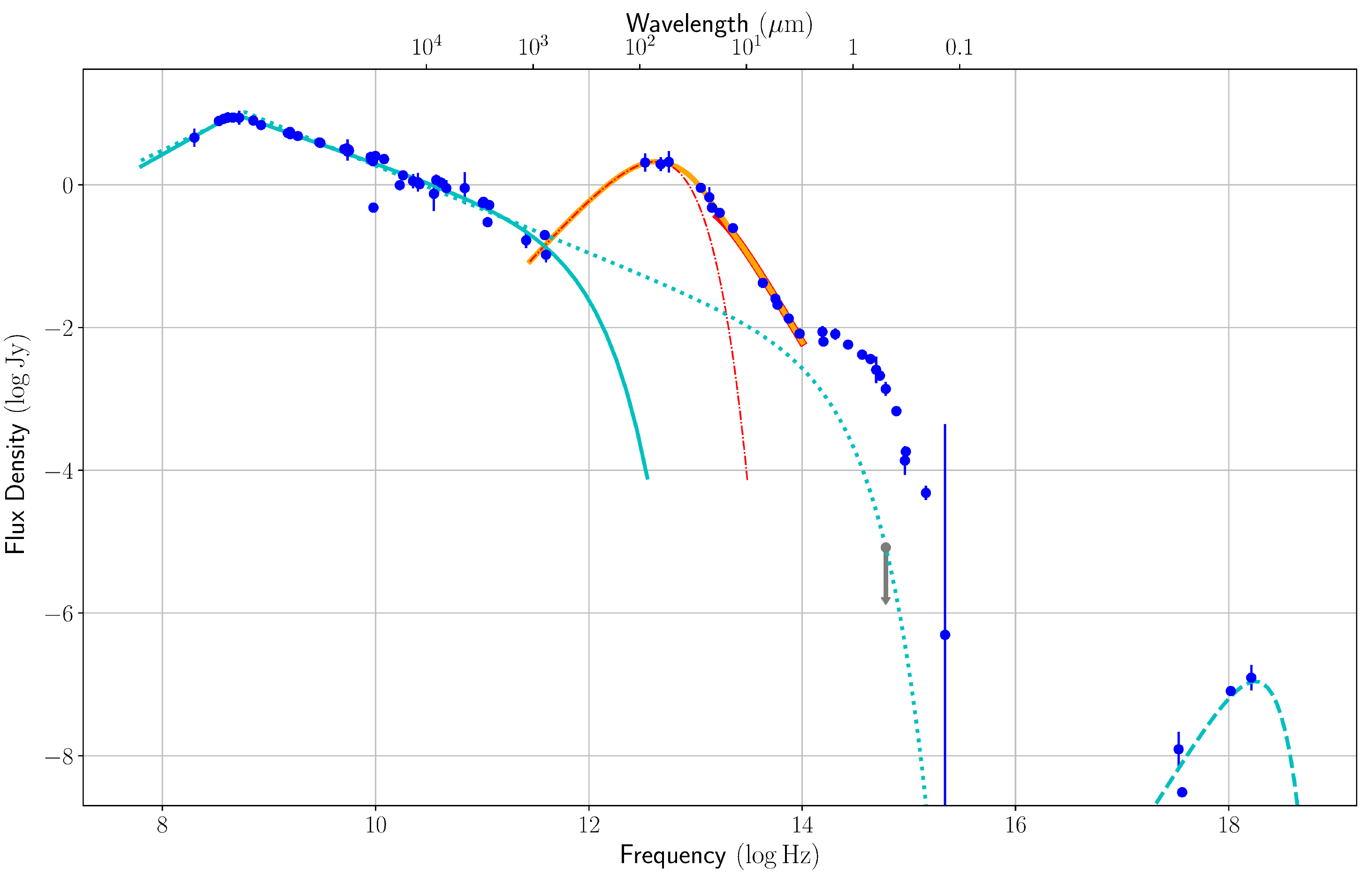}                                                                            
                \caption[]{Fit of the spectral energy distribution of 4C12.50. The infrared part of the spectrum was fit with a modified black body at long wavelengths (red dashed-dotted line) and with a power-law representing a series of modified black bodies at short wavelengths (red thick line). The total infrared model is plotted with a solid orange line. A broken power-law approximation with exponential cutoff is used for the radio emission. Minimum and maximum synchrotron flux models are given for the radio to optical range (solid cyan curve and dotted cyan curve, respectively; see the text for more details). A maximum model for the inverse Compton emission in the X-rays is shown with a dashed line. }
                \label{fig:radio_spectrum_fit}
        \end{center}
\end{figure*}

To evaluate whether the jet, the AGN radiation pressure, or the starburst can sustain winds, we first need to calculate the energy output of the galaxy at radio and infrared wavelengths. To obtain the jet power, $P_j$, we fitted the SED at radio, IR, optical, UV, and X-ray wavelengths. Because the jet and the dust can emit at common frequencies, we simultaneously fitted their SEDs using a standard $\chi ^2$ minimization method \citep{lmfit}. For the dust emission, we used a modified black-body law at long wavelengths, coupled with a power law at short wavelengths \citep{casey12}. We found the dust temperature to be $55 \, \mathrm{K}$ and \lir\ to be 2.5$\times 10^{12}$\lsun\  (in good agreement with \citealt{dasyra14}). For $P_j$, we fitted the synchrotron radiation using a broken power law with an exponential cut off \citep{ryb12}. At radio wavelengths, up to 4$\times10^{11}$Hz, this model is well determined from our data (Fig.~\ref{fig:radio_spectrum_fit}). The integral of the fit provides the lower limit for the synchrotron power, which is 9.1$\times 10^{43}\,\mathrm{erg}/\mathrm{s}$. At higher frequencies however, the dust, stellar, and AGN emission outshine the jet emission. To find an upper limit for the synchrotron power, we fitted the data under the extreme assumption that 0.6\% of the 6$\times10^{14}$Hz emission originates from the jet. The fraction 0.6\% corresponds to the part of the 5092\ang\ flux that is enclosed in a radius of 150\,pc in the  \hst\ data;  150\,pc is the distance at which the radio jet interacts with ISM clouds per \citet{morganti13b}. In this model (Fig.~\ref{fig:radio_spectrum_fit}), $P_j$ is 6.2$\times 10^{44} \,\mathrm{erg}/\mathrm{s}$. The X-rays may also be bolstered by a contribution from various Compton components related to the jet or magnetic fields \citep{finke16}. To obtain the maximum possible contribution of the jet to the X-ray emission, we again adopted a power law with a cut-off that was steeper than before due to the absence of pertinent data above $10^{19}$Hz. The maximum power output in the X-rays is 8.4$\times 10^{43} \, \mathrm{erg}/\mathrm{s}$. In summary, we find the jet power to be in the range $10^{44}-8\times10^{44}\,\mathrm{erg}/\mathrm{s}$. For comparison, \citet{guillard12} calculated the power of the radio-jet of 4C12.50 from the monochromatic 178 MHz flux, following \citet{punsly05}.
They adopted a calibration of the monochromatic flux to the bolometric output that takes into account X-rays and the plasma thermal energy in the lobes. Their approach yielded $P_j$=3$\times$10$^{45}$erg$\,s^{-1}$. 
 Likewise, using the calibration between the monochromatic 1.4 GHz luminosity and the bolometric jet luminosity following  \cite{sulentic10},  $P_j$ can be as high as 10$^{46}$erg$\,s^{-1}$ .

To obtain the force exerted on the gas due to AGN radiation pressure we assume that the AGN luminosity is well described by its \lir\ value, as the flux that was absorbed and re-emitted by the dust is higher than that seen in the optical. Following \citet{veilleux09}, we ascribe half the \lir\ to the AGN. Based on our previous analysis, the AGN-related part of the \lir\ is
4.7$\times$10$^{45}$\,erg\,s$^{-1}$. The force exerted on the gas due to the AGN radiation pressure is $L_{AGN}/c$= 1.6$\times$10$^{35}$\,erg\,cm$^{-1}$.

To evaluate whether or not the starburst can sustain winds, we need to calculate the energy released by SNe and the force exerted on the gas due to stellar radiation pressure. We assume that for every 100\msun\ formed, there is one SN. We also assume that each SN ejects material with a kinetic energy of $10^{51} erg$. Given a star formation rate (SFR) of $\sim$200\msun \,yr$^{-1}$, computed from the  \cite{kennicutt98} formula for \lir /2, the power released by the SN is 6$\times$$10^{43}$ergs$^{-1}$. Also taking into account the fraction of the total star formation that can be ascribed to any given area (as explained in Sections  \ref{sec:disc_ridge} and \ref{sec:disc_nuclear}) we calculate the power released locally by the SN. Furthermore, using the fraction of the total \lir$/$2 that can be attributed to a local starburst, we calculate the force exerted on the gas due to stellar radiation pressure.

\subsection{Extended wind energetics: gas acceleration not sustained by a local starburst}\label{sec:disc_ridge}

We begin our energetics study of the extended wind by ruling out that the low-S/N CO detections in filaments are accelerated by the local starburst. For this purpose, we attribute a part of the SFR or of \lir\ to each ellipse of F1 and F2 of Fig.~\ref{fig:radial_outflows} using the ratio of the locally enclosed \cooz\ emission over the total \cooz\ emission in our high-resolution data. The filament F1, for example, comprises roughly 8\% of the total CO emission. Based on this fraction, the local \lir\ is 3.8$\times$10$^{44}$\,erg\,s$^{-1}$ and the local SFR is $\sim$15\msun \,yr$^{-1}$. In this case, the mechanical energy from the SN is  4.8$\times$10$^{42}$erg\,s$^{-1}$ and the force due to stellar radiation pressure is 1.3$\times$10$^{34}$\,erg\,cm$^{-1}$. 
To calculate the energetics of the gas in F1, we need to use the distance $d$ of the accelerated gas from the spot of the wind generation. For this purpose, we use the mean value of the ellipse semi-axes (0.4$\arcsec$) as the distance from the star-forming region. Using the mass and the mean velocity presented in Section~\ref{sec:res_ridge}, we find that the wind kinetic luminosity,  $L_{kin}$, computed as $(1/2) MV^{3}d^{-1}$, is 2.0$\times$10$^{44}$erg\,s$^{-1}$. The wind momentum rate, $\dot{M}V$  \citep{combes13}, is 1.8$\times$10$^{36}$ \,erg\,cm$^{-1}$. In this formula,  $\dot{M}$ is the mass-flow rate of the accelerated gas, equal to $MVd^{-1}$. We find that neither the SN nor the stellar radiation pressure in the area can drive the wind along the filament, each being at least two orders of magnitude short of the required levels.

The same calculation for the filament F2 shows that it comprises roughly 9\% of the total CO emission. Therefore, the local \lir\ is  4.2$\times$10$^{44}$\,erg\,s$^{-1}$ and the local SFR is $\sim$20\msun \,yr$^{-1}$.
The mechanical energy released by the SN is 5.4$\times$10$^{42}$erg\,s$^{-1}$ and the force due to stellar radiation pressure is 1.4$\times$10$^{34}$\,erg\,cm$^{-1}$.
The distance $d$ is 0.7\arcsec and calculating the wind energetics as in the case of F1 we find that $L_{kin}$ is 1.4$\times$10$^{44}$erg\,s$^{-1}$ and $\dot{M}V$ is 1.5$\times$10$^{36}$ \,erg\,cm$^{-1}$.
In this case neither the SN nor the stellar radiation pressure in the area are adequate drivers of the wind. 

Inversely, how high does the local SFR need to be to sustain the wind?
If the SN were a local generation mechanism,  
the SFR would need to be higher than 630\msun yr$^{-1}$ and 440\msun yr$^{-1}$ in the areas of F1 and F2, respectively, in order to sustain the wind. Therefore, the local SNe cannot drive the flow by themselves in the areas of the filaments. However, given that the velocity in the filaments is the highest that we have observed, the local starburst is likely to assist in their acceleration. Still, the main driver of all wind candidates needs to be sought in a central mechanism.

\subsection{Overall wind energetics: sustainability by a central mechanism}\label{sec:disc_nuclear}

To evaluate whether or not a power source at the nucleus (i.e., within our beam) could sustain the wind, we evaluate whether the wind can be sustained via the nuclear starburst or via the jet and the AGN.
For the nuclear starburst, we compare the energy deposited by SNe to the wind kinetic luminosity and, as above, the force exerted on the gas due to stellar radiation pressure to the wind momentum rate. As explained in Section \ref{sec:energetics} the power released by the SN is 6$\times$$10^{43}$ergs$^{-1}$. Taking further into account that the nuclear SN can only be ascribed 10-20\% of the total star formation, then the total SN power is  6-12$\times$10$^{42}$ergs$^{-1}$. The fraction ascribed to nuclear star formation is computed from the fraction of \cooz\ emission within the radius of a beam (0.3$\arcsec$). 
Furthermore, taking into account that
10-20\% of \lir /2 can be ascribed to the nuclear starburst, that is, 4.7-9.4$\times$10$^{44}$\,erg\,s$^{-1}$, the force exerted on the gas due to stellar radiation pressure, $L_{SB}/c$, is then 1.6-3$\times$10$^{34}$\,erg\,cm$^{-1}$.
Alternatively, the force exerted on the gas due to the AGN radiation pressure is $L_{AGN}/c$= 1.6$\times$10$^{35}$\,erg\,cm$^{-1}$.

To find the central mechanism that could sustain the extended wind, we need to compute the respective wind kinetic luminosity and momentum rate assuming the distance of the accelerated clouds from the main nucleus. For this purpose, we compute $L_{kin}$ and  $\dot{M}V$ for each individual region of Fig. \ref{fig:radial_outflows} using its distance from the nucleus. We sum the results for the extended regions with the F1, the F2 (using the two velocity ranges of F2 as two different regions), the redshifted disk-wide and the extended blueshifted detections from Table \ref{tab:extended_wind}. We find that the mass-flow rate $\dot{M}$ carried by the extended wind is as high as  $\sim$500\msun yr$^{-1}$. Then, $L_{kin}$ (extended)  is  3.8$\times$10$^{44}$erg\,s$^{-1}$ and $\dot{M}V$(extended) is  4.4$\times$10$^{36}$ \,erg\,cm$^{-1}$. 
Therefore, the mechanical power of the SN is insufficient to drive the flow, considering that it is low and that only a part of it is deposited in the ISM.
Likewise, the stellar radiation pressure within the central 0.3$\arcsec$ cannot drive the wind,
since the momentum rate of the wind is two orders of magnitude higher than the force exerted on gas by the stellar radiation pressure. As this momentum rate is also an order of magnitude higher than  $L_{AGN}/c$, the AGN radiation pressure is unable to drive this wind, unless an energy-conserving expansion has significantly boosted the momentum rate of the gas (i.e., by a factor of 30). Such a boost is high, but not impossible, as shown by \cite{cicone14}.

To add to the above numbers those for the circumnuclear wind, we set $d$ equal to 0.3$\arcsec$, which is approximately half of the semi-major axis of the ellipse in Fig. \ref{fig:cen_wind_detection}. 
The mass of the circumnuclear wind is 5.3($\pm$0.7)$\times$10$^8$\msun , and the mean velocity of either the blueshifted or the redshifted component is $\sim$800\kms. Using these numbers and assuming that the \htwo\ clouds and the CO clouds have similar spatial and velocity distributions, we find that the kinetic luminosity of the nuclear wind is  1.3$\times$10$^{44}$\,ergs$^{-1}$ and the momentum rate is   3.3$\times$10$^{36}$\,erg\,cm$^{-1}$.
Adding these numbers to the values we calculated for the extended wind, the total kinetic luminosity and the total momentum rate of the outflow reach  5.1$\times$10$^{44}$\,ergs$^{-1}$ and 
7.7$\times$10$^{36}$\,erg\,cm$^{-1}$, respectively. These numbers may yet increase, considering that part of the circumnuclear wind is unresolved and thus the used radius may be overestimated.

Overall, a combination of the jet, radiation pressure, and SN mechanical power needs to be invoked. The jet is the most likely generation mechanism given that it carries the most power.
The jet alone could have driven the flow if a past event, which has been recorded for this system, were found to have deposited enough energy onto the ISM. In this case we are observing a fossil outflow which expands into an already carved cavity \citep{fluetsch19}. This could explain the radial symmetry of the wind even though the radiation pressure is equally well-suited to explain it. 
However, the force due to the radiation pressure cannot drive the outflow by itself, but could assist it. The same applies to the local generation mechanism along the filaments: a local starburst could further accelerate the gas that has already been accelerated.

\section{Conclusions}
\label{sec:conclusions}
We obtained ALMA data of the radio galaxy 4C12.50 in order to determine the millimeter properties of its outflow, which was previously known to exist from large- and small-angular-resolution observations at other wavelengths. We mapped the CO distribution, kinematics, and excitation with ALMA at a resolution of $\sim$0.5\arcsec\ and found the following results.
\begin{itemize}
        \item
        The main gaseous disk is rather compact, extending to radii of $\sim$6 kpc in CO (1$-$0) and $\sim$2 kpc in CO (4$-$3).  Some co-rotating CO emission is also seen in the secondary nucleus and in a bridge connecting the nuclei. 
        \item
        A shock-tracer, \water , and a hard-ionization-field tracer, \hcop\, are seen in the main nucleus.
        \item
        Several extranuclear structures are seen. These include a prominent tidal tail south of the main nucleus with gas seen in negative and positive velocities. 
        \item
        The CO emission is abruptly cut along an \oiii\ ridge north of the main nucleus, where no CO is detected. This indicates a phase transition of the molecular gas.
        \item
        Extended wind components are tentatively seen in the \cooz\ line. The detections are also seen in regions with \oiii\ and stellar continuum emission. They include high-velocity (-2000 \kms) filaments that extend radially from the nucleus to the \oiii\ ridge, a redshifted disk-wide emission (V$\sim$1350 \kms) and an extended blueshifted emission (V$\sim$-700 \kms). The total mass is 1.0($\pm$0.1)$\times$10$^9$\msun . 
        \item
        A circumnuclear wind candidate is detected in emission, in \cooz , through broad blueshifted and redshifted line wings. The emission peaks within 200 pc from the radio core, and it coincides with a region of high CO excitation (within the velocity range of the disk), as indicated by its \coft / \cott\ line ratio. The mass of the circumnuclear wind is equal to 5.3($\pm$0.7)$\times$10$^8$\msun .
        \item
        The total mass of the wind is as high as 1.5($\pm$0.1)$\times$10$^9$\msun .
    \item
        Both at the nucleus and at the extra-nuclear regions, the wind can be sustained mainly by the jet. The radiation pressure of the AGN can help, in particular near the nucleus. It is plausible that the starburst also contributes, in particular for any extended wind components. However, the starburst cannot drive the wind alone based on its radiation pressure or its energy release by SN remnants.

\end{itemize}

\begin{acknowledgements}
        This paper makes use of the ALMA data ADS/JAO.ALMA 2013.1.00180.S, 2013.1.00976.S, 2012.1.00797.S. 
        ALMA is a partnership of ESO (representing its member states), NSF (USA) and NINS (Japan), together with NRC (Canada) and NSC 
        and ASIAA (Taiwan) and KASI (Republic of Korea), in cooperation with the Republic of Chile. The Joint ALMA 
        Observatory is operated by ESO, AUI/NRAO and NAOJ.
        K. M. Dasyra acknowledges financial support by the Hellenic Foundation for Research and Innovation (HFRI) and the General Secretariat For Research and Technology, under grant number 1882. 
        We would like to thank the referee, B. Emonts, for detailed comments which led to a significant improvement of the paper.
\end{acknowledgements}

\bibliographystyle{aa} 
\bibliography{AA_34416}
\newpage
\begin{figure*}
        \begin{center}
                \includegraphics[width=6cm]{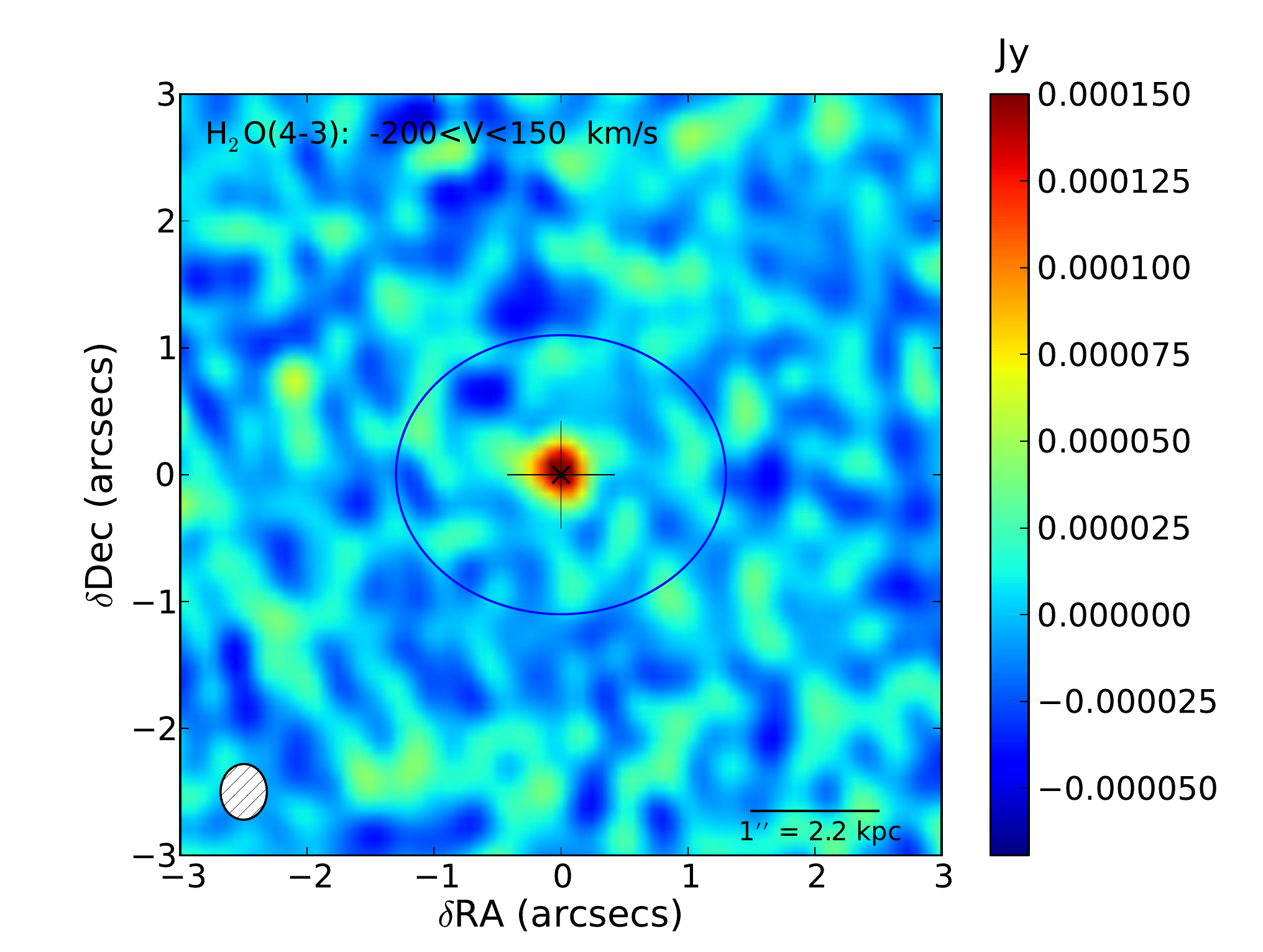}
                \includegraphics[width=6cm,height=4.5cm]{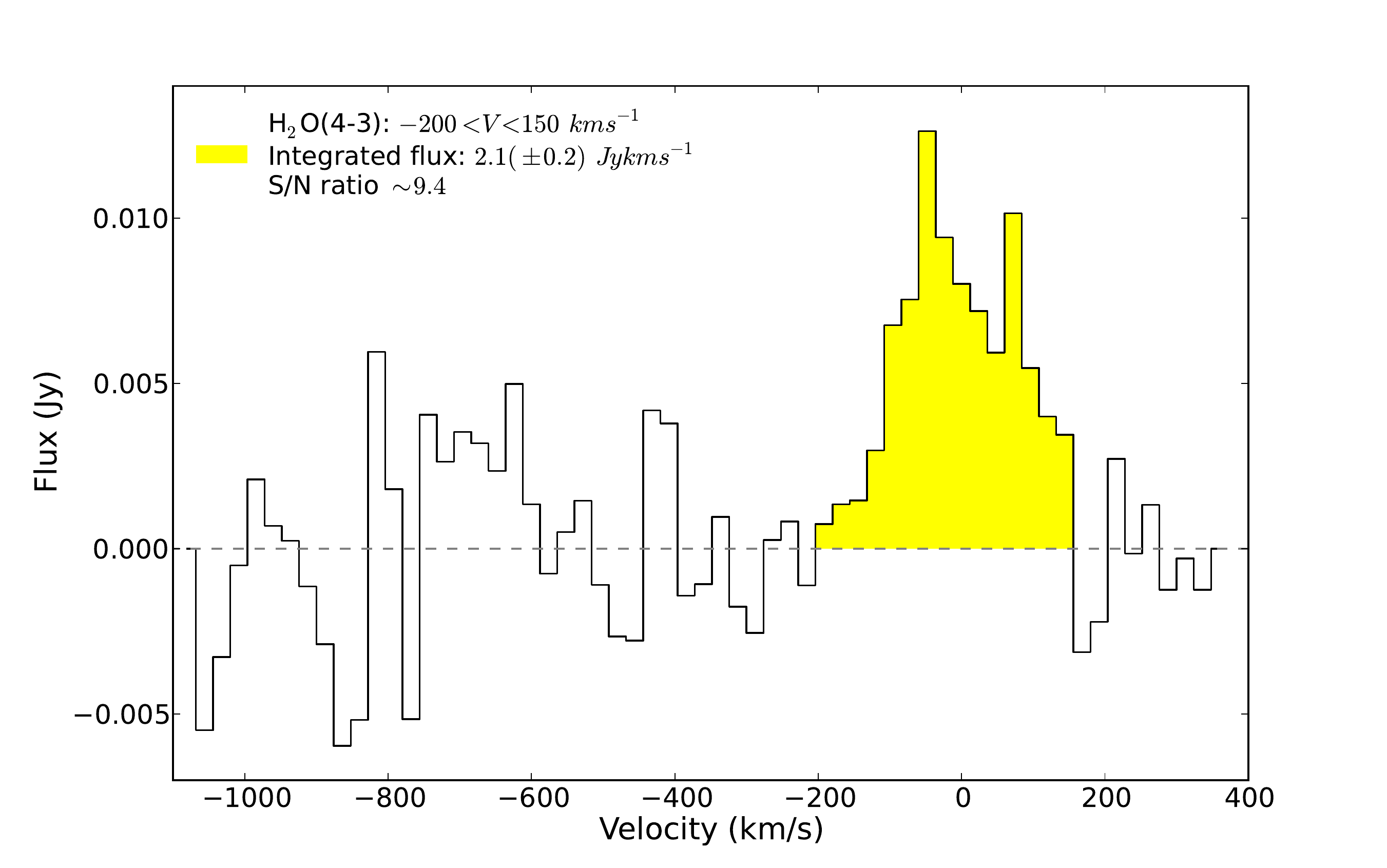}
                \includegraphics[width=6cm]{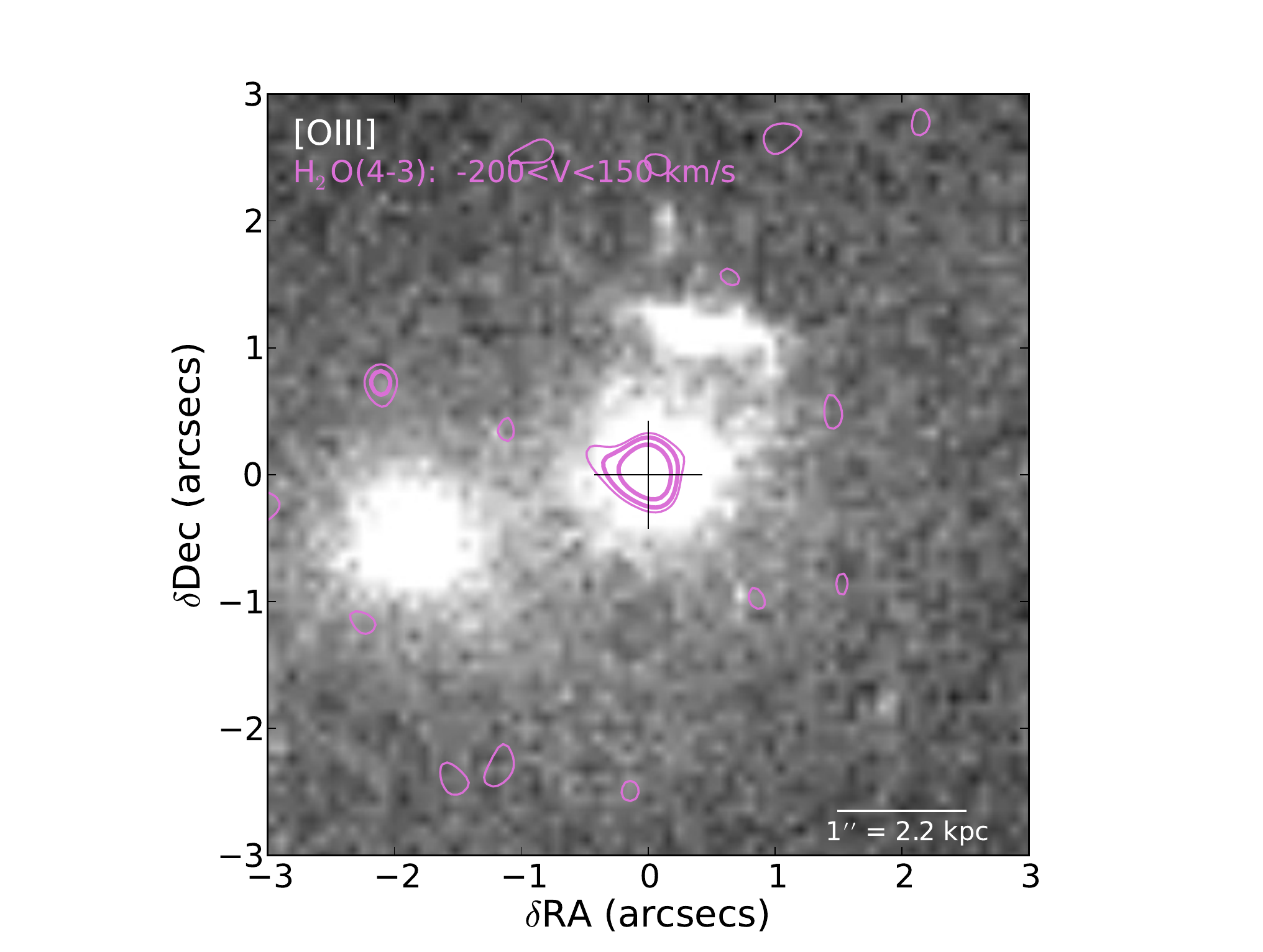}             
        \end{center}
        \caption[]{\water(4$-$3) in 4C12.50. The left panel shows the ALMA image, collapsed in the indicated velocity range. The beam is shown with a shaded ellipse. The open ellipse corresponds to the area used for the extraction of the spectrum in the middle panel. In the right panel we plot the CO contours (levels: 2$\sigma$,3$\sigma$ and 5$\sigma$) over the \hst\ \oiii\ image of Fig.\ref{fig:optical}. Crosses are as in Fig~\ref{fig:optical}. The same applies to all following figures with similar layout.}
        \label{fig:disk_H2O}
\end{figure*}
\begin{figure*}
        \begin{center}
                \includegraphics[width=6cm]{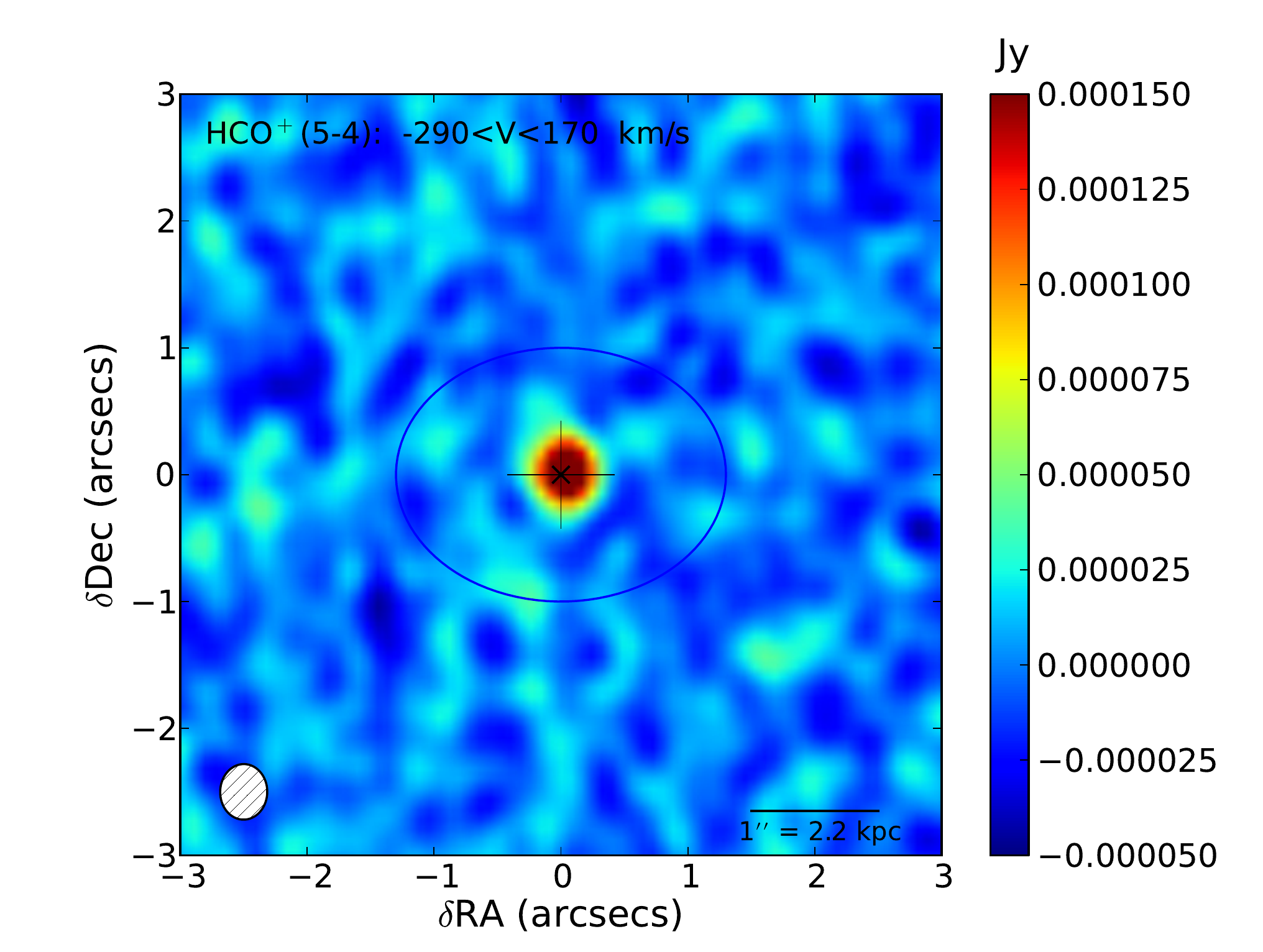}
                \includegraphics[width=6cm,height=4.5cm]{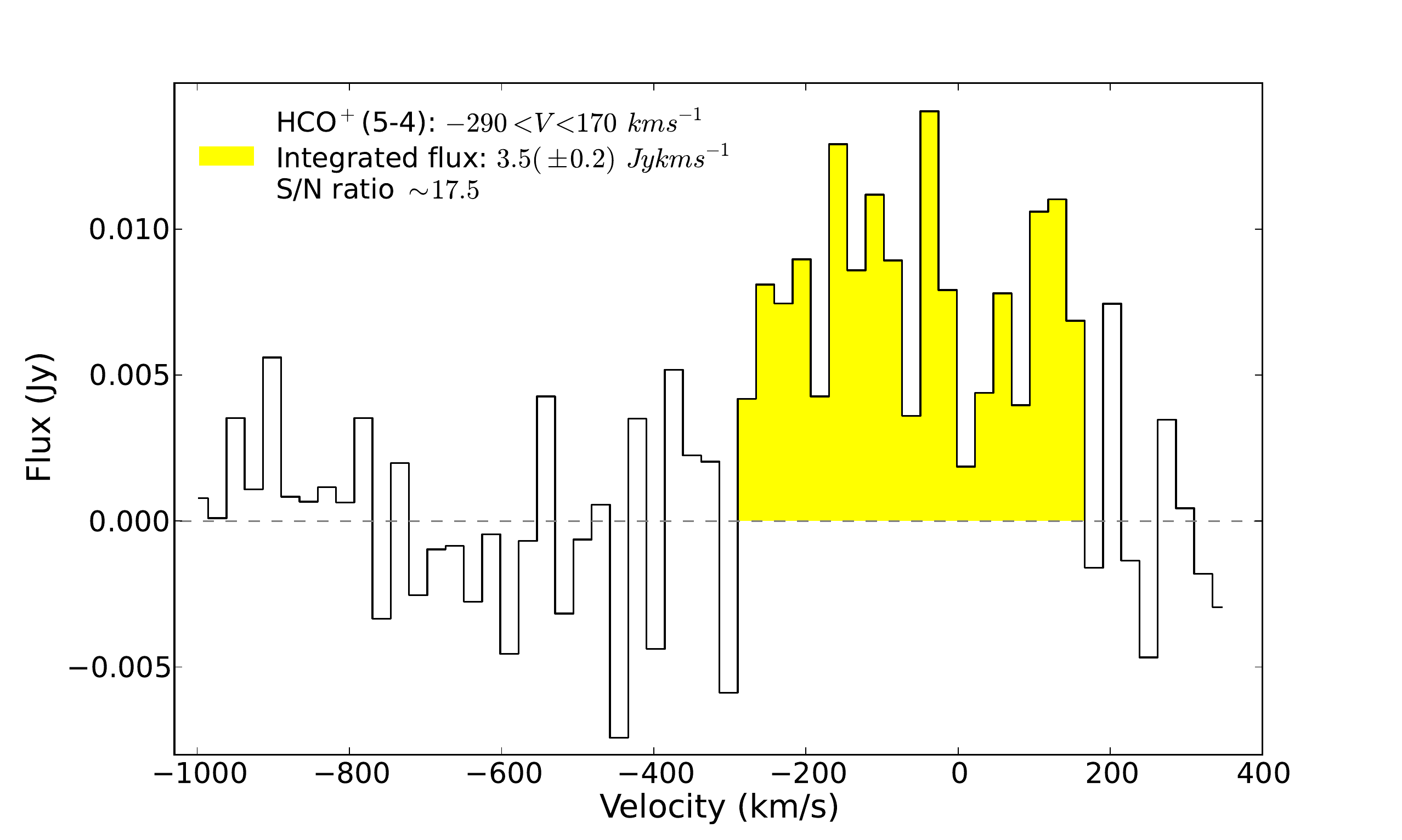}
                \includegraphics[width=6cm]{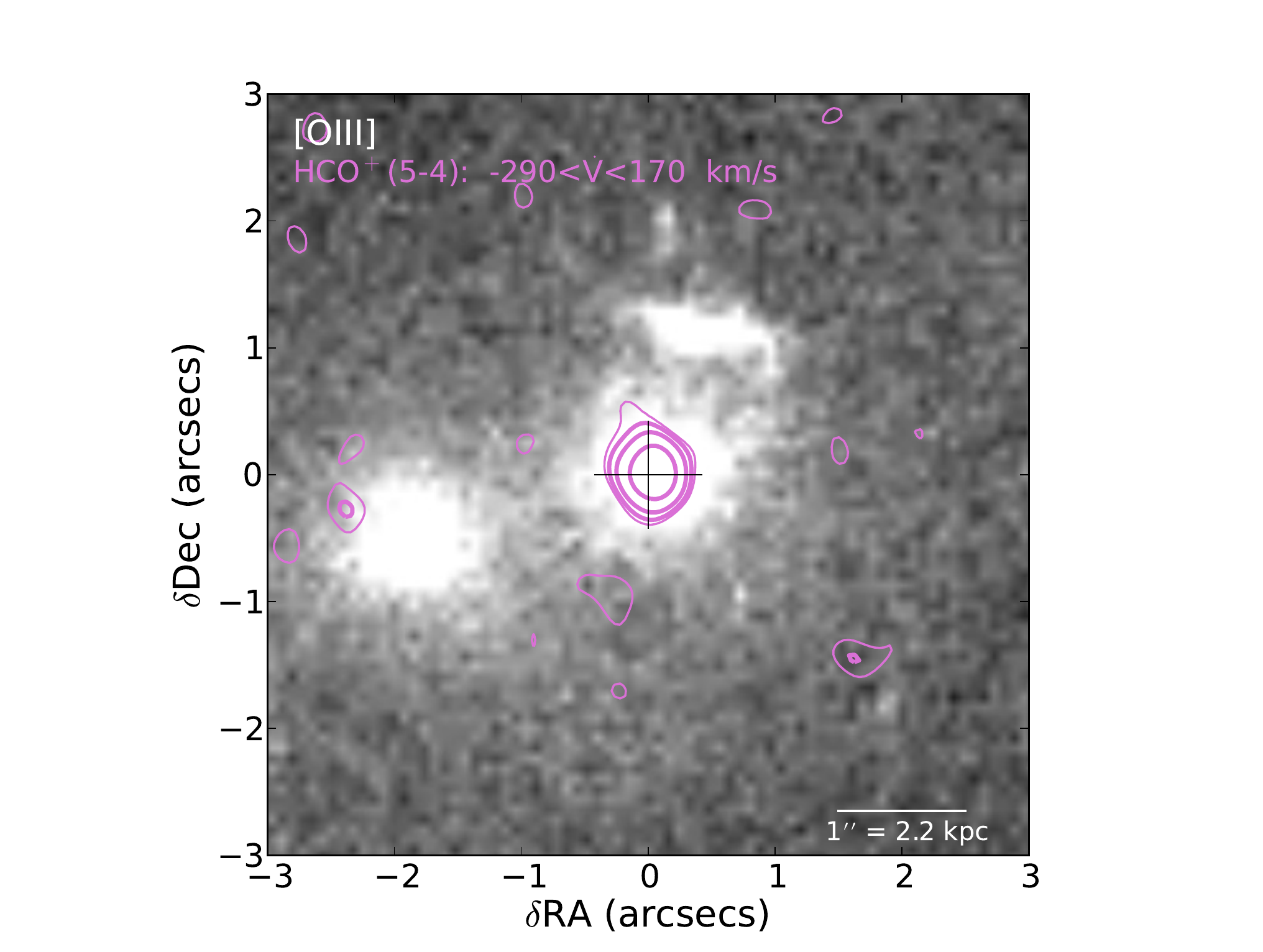}
        \end{center}
        \caption[]{\hcop\  in 4C12.50. Contour levels are at  2$\sigma$,3$\sigma$,5$\sigma$ and 10$\sigma$.
        }
        \label{fig:disk_HCO}
\end{figure*}
\clearpage

\end{document}